\documentclass[11pt]{article}
\pdfoutput=1 
\usepackage{mathrsfs}
\usepackage{xcolor}
\usepackage{jheppub}
 \usepackage{float}
\usepackage{booktabs}
\usepackage{amssymb}
\usepackage{bbm}
\usepackage{orcidlink}
\usepackage[most]{tcolorbox}
\usepackage{amsmath}
\usepackage{caption}
\usepackage{subcaption,longtable,stmaryrd}
\usepackage{slashed}
\usepackage{bigints}
\usepackage{cancel}
\usepackage{tikz}
\usepackage{enumitem}
\usepackage[customcolors,shade]{hf-tikz}
\usepackage{graphicx}
\usepackage{cancel}

\DeclareMathAlphabet\mathbfcal{OMS}{cmsy}{b}{n}
\DeclareSymbolFont{usualmathcal}{OMS}{cmsy}{m}{n}
\DeclareSymbolFontAlphabet{\mathcal}{usualmathcal}
\DeclareSymbolFont{rmlargesymbols}{OMX}{mdbch}{m}{n}
\DeclareMathSymbol{\rmintop}{\mathop}{rmlargesymbols}{82}
\DeclareMathSymbol{\rmointop}{\mathop}{rmlargesymbols}{72}

\definecolor{mygray}{gray}{0.5}

\definecolor{mygray}{gray}{0.5}
\definecolor{refonecolor}{RGB}{0,90,170}
\definecolor{reftwocolor}{RGB}{176,80,0}
\definecolor{refthreecolor}{RGB}{0,120,86}
\makeatletter
\long\def\textcolor#1#2{{\color{#1}#2}}
\makeatother

\author[a]{Arpan Bhattacharyya,}
\author[a,b]{Saptaswa Ghosh,}
\author[a]{Sounak Pal,}
\author[a]{Jagannath Santara}
\affiliation[a]{\it Department of Physics, Indian Institute of Technology, Gandhinagar, Gujarat-382055, India}
\affiliation[b]{Centre for High Energy Physics, Indian Institute of Science,\\ Bangalore 560012, India.}
\emailAdd{abhattacharyya@iitgn.ac.in}
\emailAdd{saptaswag@iisc.ac.in}
\emailAdd{palsounak@iitgn.ac.in}
\emailAdd{jagannath.santara@iitgn.ac.in}

\abstract{\textcolor{black}{We explore several aspects of categorical symmetry‑resolved entanglement entropy (SREE) directly within two‑dimensional rational conformal field theory (RCFT) (without invoking any SymTFT construction \cite{Choi:2024wfm}). We derive a general formula applicable whenever the action of the relevant topological defect lines on the annulus Hilbert space is known. This framework accommodates weakly and strongly symmetric boundaries, cloaking states, and fusion rings with multiplicities. We verify the formula in a range of diagonal unitary and non‑unitary examples, including theories with generalized Haagerup–Izumi modular data. Furthermore, we extend the analysis to non‑diagonal RCFTs. The $\frac{1}{2}E_6$ example demonstrates that closed‑channel modular data and NIM‑rep multiplicities alone do not suffice to determine the defect action on the complete open‑channel Hilbert space.}

}
\title{\boldmath{On the resolution of  categorical symmetries in (Non-) Unitary Rational CFTs }}
\begin{document}
\maketitle
\flushbottom
\section{Introduction}


Symmetries are omnipresent in nature. In physics, symmetry plays a crucial role in understanding the dynamics of quantum field theory and provides a framework for going beyond. {According to Wigner \cite{Wigner:1959}, ordinary quantum symmetries are implemented by unitary or anti-unitary operators; when such an operator has an inverse, the symmetry is group-like.} Gaiotto et al. \cite{Gaiotto:2014kfa} showed that the generalized global symmetry of a theory is associated with a topological defect operator, and there are plenty of examples found in condensed matter systems that realize this generalized global symmetry \cite{McGreevy:2022oyu}. One of the classes of these generalized symmetries is famously known as \textit{non-invertible symmetry} \cite{Bhardwaj:2017xup, Chang:2018iay, Thorngren:2019iar,Heidenreich:2021xpr,Thorngren:2021yso,Choi:2022jqy, GarciaEtxebarria:2022vzq, Cordova:2022ieu, Cordova:2022ruw, Choi:2022fgx,Choi:2022rfe,Apruzzi:2022rei,Shao:2023gho,Seiberg:2023cdc, Schafer-Nameki:2023jdn, Bhardwaj:2023wzd, Bhardwaj:2023ayw, Bhardwaj:2023fca, Choi:2023pdp, Lin:2023uvm,  Iqbal:2024pee, Choi:2024rjm, Seiberg:2024gek,Bhardwaj:2024wlr, Bhardwaj:2024kvy,Putrov:2024uor,Shao:2025mfj,Shao:2025qvf} \footnote{The references mentioned here are by no means exhaustive. Interested readers are referred to some of these reviews and lecture notes and their citations \cite{McGreevy:2022oyu,Shao:2023gho, Schafer-Nameki:2023jdn} for more references.}, which can not be described by a unitary operation, more specifically, it does not have an inverse. In mathematics literature, it is known as a fusion category \cite{Bartsch:2023wvv, Costa:2024wks}. For these kinds of categorical (non-invertible) symmetries, associated topological operators $\{\mathcal{L}_a\}$ obey fusion rules, as shown below, 
\begin{align}
    \mathcal{L}_a\times \mathcal{L}_b={\bigoplus}_{c} \mathcal{N}_{ab}^{\,\,\,c}\,\mathcal{L}_{c}
\end{align}
where,  {$\mathcal{N}_{ab}^{\,\,c}\in \mathbb{Z}_{\ge0}$ are the fusion coefficients.} These defect lines usually do not have a formal inverse, i.e., there exists no $\mathcal{L}^{-1}$ such that $\mathcal{L}\times\mathcal{L}^{-1}=\mathbb{I}$.
There are numerous examples that demonstrate this kind of \emph{categorical} symmetry exists in various branches of physics, including high-energy theory, mathematical physics, and condensed matter physics, particularly in low-dimensional quantum systems that are useful for quantum computation \cite{Preskill:1999he, Kitaev:2005dm, Kitaev:1997wr, PhysRevLett.96.110405, Aharonov:2005nkd, Feiguin:2006ydp, Dennis:2001nw, McGreevy:2022oyu, PRXQuantum.5.037001, Okada:2024qmk, Li:2024gwx, Fukusumi:2024cnl,Bhardwaj:2024ydc, PhysRevX.15.011058, Li:2025bgo,KNBalasubramanian:2025vum}. 

\paragraph{Categorical symmetry–resolved entanglement.}
Parallel to recent advances in the characterization of (non-)invertible symmetries, a complementary line of research has focused on the role of symmetries in shaping entanglement structures in quantum field theories (QFTs). Entanglement serves as a fundamental measure of quantum correlations and phases of matter, and provides a particularly sensitive probe of underlying symmetry principles, including generalized and non-invertible symmetries. In this context, symmetry-resolved entanglement entropy (SREE)~\cite{Goldstein:2017bua,german3, Belin:2013uta, Islam:2015mom, doi:10.1126/science.aau0818,FG,Murciano:2019wdl,Murciano:2020lqq,Parez:2021pgq,Murciano:2020vgh,Zhao:2020qmn,ourPartI,Horvath:2021rjd,Weisenberger:2021eby,Magan:2021myk,Ares:2022gjb,DiGiulio:2022jjd,Zhao:2022wnp,Gaur:2023yru,Kusuki:2023bsp,Northe:2023khz,Gaur:2024vdh,Banerjee:2024ldl}\footnote{For more references and details, interested readers are referred to the following review of symmetry-resolved entanglement measure \cite{Castro-Alvaredo:2024azg}.}  has emerged as a refined diagnostic that decomposes the total entanglement into contributions associated with individual symmetry sectors, thereby elucidating how symmetry structures govern the distribution of quantum information. Recent developments further demonstrate that non-invertible symmetries can impose nontrivial constraints on entanglement patterns, influence topological order, and enforce selection rules that transcend those encountered in the setting of ordinary invertible symmetries \cite{Casini:2019kex, Lin:2022dhv, Copetti:2024dcz, AliAhmad:2025bnd}.
Motivated by these developments, this paper examines how non-invertible (categorical) symmetries modify entanglement measures in certain two-dimensional statistical systems at criticality. For global invertible symmetries, SREE has been extensively studied, allowing the entanglement to be resolved sector by sector. Readers are referred to these reviews ~\cite{Castro-Alvaredo:2024azg}. In contrast, symmetry resolution for \emph{categorical} symmetries~\cite{Sierra_2024,Benedetti:2024dku, Das:2024qdx, Heymann:2024vvf, Choi:2024wfm,Das:2025xyz} remains far less explored. In $(1{+}1)$-dimensional CFTs with a categorical symmetry encoded by a fusion category~$\mathcal{C}$, symmetry resolution can be formulated in terms of boundary CFT (BCFT)  \cite{DiGiulio:2022jjd, Kusuki:2023bsp, Northe:2023khz} through projections onto categorical symmetry sectors. 
Now, on the replica geometry \cite{Calabrese:2009qy}, inserting compatible defect endpoints along the branch cut yields well-defined sector probabilities and sector-resolved Rényi and von Neumann entropies. This construction is valid whenever the categorical symmetry is non-anomalous, i.e., \ free of obstructions to gauging \cite{Choi:2023PRD, Seifnashri:2025fgd}. Alternatively, there is a twist-field approach that computes directly symmetry-resolved entanglement entropy using charged moments, and this method is applied to compute SREE for rational CFT with Lie group symmetry by \cite{Calabrese:2021wvi} and for certain non-invertible symmetry \cite{Das:2025xyz}.\footnote{We were informed about \cite{Das:2025xyz} by a  private communication from Arpit Das.}\par 
Interestingly, for $1+1$D CFTs with global internal abelian symmetry, SREE shows universal results in the sense of \textit{equipartition of entanglement} among all charge sectors \cite{german3, Castro-Alvaredo:2024azg}\footnote{One may note that for abelian symmetry the entanglement equipartition does not hold for all orders of UV cutoff, see \cite{Northe:2023khz,  Northe:2025qcv}.}.  On the other hand, in the context of rational CFTs (RCFTs)\footnote{It turns out that almost all rational CFTs possess non-invertible symmetry.}, for non-invertible categorical symmetry-resolved entanglement entropy, this {equipartition of the entanglement is broken \cite{Heymann:2024vvf,Choi:2024wfm}}. These deviations are controlled by the fusion data of~$\mathcal{C}$ and by the choice of symmetric boundary condition. Note that computation of the entanglement requires partitioning the Hilbert space, hence the boundary state placed at the entangling surface plays a crucial role in refining the entropy~\cite{Ohmori:2014eia, Cardy_2016}. 
\textcolor{black}{In contrast to the group-like setting, non-invertible symmetries generally admit multiple, and often inequivalent, choices of boundary conditions. {Although this issue has been noted in the literature
\cite{Choi:2024wfm,Choi:2023PRD,Choi:2024tri,Lin:2022dhv,
Heymann:2024vvf,Das:2024qdx}, previous discussions have often focused on
weakly or strongly symmetric boundary conditions, typically in
multiplicity-free settings. For invertible symmetries, these notions coincide
whenever the boundary preserves the symmetry, since the corresponding
topological lines have quantum dimension one. However, the entangling boundary
is not determined by the bulk symmetry alone; in a microscopic realization,
such as a lattice model with a specified entanglement cut, it requires additional
physical data. Accordingly, the entanglement spectrum may be viewed as a
boundary-CFT spectrum with one boundary induced by the cut
\cite{Roy:2025hew}. For non-invertible symmetries, this boundary dependence is especially important, since strongly and weakly symmetric boundaries are distinct, and different compatible boundaries can yield different categorical sector resolutions. In most of our non-invertible examples, however, an explicit microscopic realisation is not yet available; we therefore treat the boundary condition as an additional assumption of the RCFT/BCFT construction. We illustrate this through examples with weakly symmetric, strongly symmetric, and cloaking boundaries,
including fusion rings with non-trivial multiplicities.
} In this work, we revisit this question through a series of concrete examples that clarify how different boundary conditions naturally arise and how they affect the symmetry resolution. In particular, we highlight situations in which the weakly symmetric boundary state is absent and motivate the use of cloaking boundary conditions (or states) \cite{Brehm:2022JPA,Brehm:2024arXiv}, which, despite not corresponding to physical Cardy states, play a crucial role in implementing symmetry resolution in non-invertible symmetries consistently.}
\par

\paragraph{Non-unitarity and symmetry resolution.} 
\noindent
Most developments on symmetry resolution have focused on \emph{unitary} settings, where the physical Hilbert space and defect operations admit a positive-definite inner product, with positive central charge. In contrast, symmetry resolution for categorical symmetries in non-unitary setting remains comparatively unexplored, despite clear motivation \cite{Fossati:2023zyz,Fukusumi:2025xrj}. By the bulk–boundary correspondence, when a $(2{+}1)$-dimensional topological phase is described by a unitary TQFT, the universal properties of its gapless boundary are captured by a two-dimensional unitary RCFT \cite{Witten:1988hf, Moore:1988qv, Moore:1989vd, Dijkgraaf:1989pz, Verlinde:1988sn, Fuchs:2009iz}. Extending this logic beyond unitarity is natural \cite{Gang:2023rei}. It has been argued that a family of \emph{non-unitary} TQFTs arises from topological twists of certain three-dimensional $\mathcal{N}=4$ rank-zero (S-fold) SCFTs, producing modular data that generalize the \textit{non-unitary Haagerup–Izumi modular data} \cite{Gang:2023ggt, Gang:2022kpe}. This, in turn, leads to the question of whether there exist boundary RCFTs compatible with such generalized Haagerup modular data \cite{Haagerup1994, Evans:2010yr}. The answer is affirmative: using the so-called Riemann–Hilbert approach, conformal characters consistent with the proposed modular data have been explicitly constructed~\cite{Gang:2023ggt}. These developments supply a concrete arena in which to formulate and study \emph{categorical symmetry resolved entanglement} for non-unitary RCFTs, thereby extending symmetry resolution beyond the unitary paradigm. At this point, note that there are recent studies aimed at constructing critical lattice models in two dimensions, realizing the symmetry associated with the \textit{unitary} Haagerup fusion category $H_3$ \cite{Huang:2021nvb, Vanhove:2021zop,Jia:2024wnu, Bottini:2025hri} and extracted the central charge (and operator spectra) which turns out to be $c\sim 2$. Then it was further conjectured that this two-dimensional CFT can be constructed by a taking coset of a Haagerup rational CFT characterized central charge multiple of $8$ \cite{Evans:2010yr}, even though there is a puzzle regarding the existence of such $c=8$ unitary rational CFT with modular tensor category (MTC) $\mathcal{Z}(H_3)$ (for more details, interested readers are refered to T. Gannon's talk \cite{Gannon:2025birs})\footnote{Recently, entanglement asymmetry for $\mathcal{Z}(H_3)$ has been computed \cite{Benini:2025lav}.}.  Nonetheless, we draw motivations from these studies and focus on certain non-unitary generalized Haagerup fusion categories \cite{Gang:2023ggt}(some of them are tensor products of certain non-unitary minimal models) and discuss symmetry-resolution of entanglement entropy.
\noindent
\paragraph{Goal and brief takeaway.} Besides the mentioned developments in the computation of SREE, the explicit resolution in non-diagonal and non-unitary theories remains immensely unexplored. In this paper, we take a step toward filling this gap.  \textcolor{black}{Our main result is a direct RCFT based derivation of SREE from annulus characters, open-channel defect actions using open-closed duality. The resulting formula applies whenever the relevant defect action on the annulus Hilbert space is known, including diagonal unitary and non-unitary examples, different entangling boundaries, and fusion multiplicities. This provides a novel and directly usable method for explicit computations of SREE using only the standard RCFT data. By contrast, the general TQFT prescription remains largely formal until the additional theory-dependent boundary data needed to evaluate its projectors are supplied \cite{Choi:2024wfm}. We test our general formula against several \textit{explicit} examples, including unitary diagonal models such as Ising CFT, tetracritical Ising CFT, and non-unitary models realized by \textit{generalized} Haagerup-Izumi modular data.   Furthermore, the half-$E_6$ example shows that, in a non-diagonal RCFT, the defect action need not be determined by the modular invariant and the non-negative-integer matrix representation (NIM-rep) of the fusion algebra, which encodes the annulus multiplicities between boundary conditions.}

\noindent
Our paper is organised as follows. In Section~\eqref{sec2} we review the essential ingredients and terminology used throughout the work. We outline the different types of boundary conditions relevant for our analysis. 
Section~\eqref{sec3} presents the general strategy for computing symmetry-resolved entanglement entropy (SREE) by decomposing the entanglement into contributions from individual representations. We examine how different choices of boundary states inserted at the entangling surface of a bipartite system influence the resolution of entanglement. In this context, motivated by the SymTFT construction of \cite{Choi:2024wfm}, \textcolor{black}{we provide a direct RCFT based derivation of SREE using the annulus Hilbert space, modular transformations, and the open-channel action of the TDLs. The formula applies whenever this action is specified.}
\textcolor{black}{In Section~\eqref{sec4} we test the formula in several unitary examples and study the role of different entangling boundaries. We also discuss the non-diagonal $SU(2)_{10}$ theory with the $E_6$ modular invariant. In this example, the available closed-channel defect data do not determine the defect action on every open-channel module. We contrast this with the $\mathbb Z_3$ resolution of the three-state Potts model, where the required open-channel action is known}. Finally, in Section~\eqref{sec5} we turn to several non-unitary models and show that the modified Affleck--Ludwig formula continues to hold in these cases as well, thereby extending our analysis beyond the domain of unitary RCFTs. We conclude with a discussion in Section~\eqref{Sec6}.




\section{Background and basic setup}\label{sec2}
We begin by reviewing the fundamental principles and structural properties of Rational Conformal Field Theories (RCFTs). An RCFT is defined by its modular‑invariant torus partition function and possesses a finite set of primary fields with respect to an extended chiral algebra, within which the Virasoro algebra appears as a subalgebra. This finiteness condition ensures that both the central charge and the conformal dimensions take rational values. In general, the torus partition function of an RCFT takes the form \cite{DiFrancesco:1997nk},
\begin{align}
{Z}(\tau,\,\bar \tau)=\sum_{i,j}M_{ij}\,\chi_i(\tau)\bar{\chi}_j(\bar \tau)\,\,\,\,\,\,\,\,\,\,\,\,\,i,j\in \textrm{Primary spectrum}\label{2.1q}
\end{align}
where, the characters $\chi_i(\tau)$ are defined as,
\begin{align}
\chi_{i}(\tau)=\text{Tr}_{\mathcal{H}_{R_i}}\,q^{L_0-\frac{c}{24}},\, \textrm{with,}\,\,\, q=e^{2\pi i \tau}\,.
\end{align}
Now the degeneracy matrix $M_{ij}$ in \eqref{2.1q} need not be diagonal always. When $M_{ij}=\delta_{ij} $, we call the RCFT a diagonal one; otherwise, we call it a non-diagonal RCFT. In the next subsection, we will briefly introduce the notion of boundaries and defects, which will be relevant for our computation of SREE in the subsequent sections.


\subsection*{Introducing defects and conformal boundary conditions in RCFT}

\begin{figure}[htb!]
    \centering
    \includegraphics[width=0.38\linewidth]{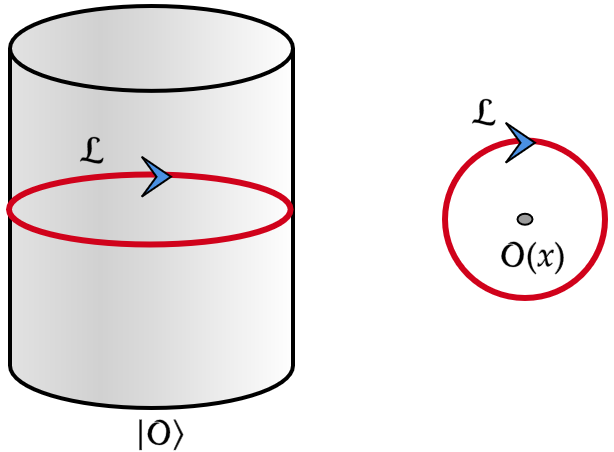}
    \caption{ The topological defect line($\mathcal{L}$) acts on state $| O\rangle $, which, after the conformal transformation (from cylinder to plane), acts on the local operator (right) using state-operator correspondence.}
\label{fig:1}
\end{figure}

A conformal defect can be understood as an extended operator that extends along the Euclidean time direction while remaining localized in the spatial dimensions \cite{Shao:2023gho,Schafer-Nameki:2023jdn} as shown in Fig.~(\ref{fig:1}). For non-invertible defects, the corresponding defect lines are known to be equivalent to the charge operators in the associated  theory. This correspondence offers a conceptual framework for computing the SREE through the evaluation of charge moments \cite{Goldstein:2017bua,german3,Bonsignori:2020laa}. In contrast, our work adopts a different strategy: we employ the established BCFT formalism to determine the SREE \cite{Kusuki:2023bsp}. Regardless of the chosen approach, the SREE can ultimately be expressed in terms of the quantum dimensions of the defect lines contributing to each symmetry sector. {In the context of BCFT, the boundary states are naturally chosen as Cardy states (for diagonal RCFTs). For non-diagonal theories, the relevant boundary labels are graph/NIM-rep labels, which we will discuss in detail later.}  These {Cardy} states are the linear combination of the Ishibashi states with specific coefficients.  Ishibashi states are solutions of the Ishibashi condition given by the construction of BCFT as \cite{Ishibashi:1988kg, Cardy:1986gw, Cardy:1989ir, Cardy:1989vyr, Cardy:1991tv, Cardy:2004hm} \footnote{For more details, we suggest the reader the following review \cite{Northe:2024tnm} and the book \cite{Recknagel:2013uja}. },
\begin{align}
    (T(z)-\bar T(\bar z))|_{\rm{bdy}}=0\,.
\end{align}
Now, using {the} Virasoro mode decomposition, $T(z)=\sum\limits_{n\in\mathbb{Z}} \frac{L_n}{z^{n+2}}$ for the Ishibashi condition, we get the following equation,
\begin{align}
    (L_{n}-\bar{L}_{-n})| \alpha\rangle\!\rangle=0\, \quad n \in \mathbb{Z}
\end{align}
where, {$\{L_n\}$} are the Virasoro generators \footnote{For extended chiral algebra, one will also need extra gluing conditions for additional conserved currents. This will not be explicitly required for our study.}.\\\\
 \textbf{Strong and Weakly symmetric boundary:}
We call a boundary {is weakly symmetric} if the symmetry defects can terminate topologically on it, leading to conserved operators that commute with the annulus Hamiltonian in the open-string channel. Similarly, for a boundary becoming strongly symmetric, the relevant boundary state is an eigenstate of the symmetry operators (closed string channel). These two notions are identical for invertible symmetries but quite different for non-invertible symmetries.
 We define them as follows. A boundary state $|\!|B\rangle\!\rangle$ is called \textit{strongly symmetric} if for every defect line $\mathcal{L}_a$ for $a\in \mathcal{C}$ it satisfies the following condition \cite{Choi:2023PRD, Choi:2024tri},
\begin{align}
\mathcal{L}_a|\!|{B}\rangle\!\rangle=\langle\mathcal{L}_a \rangle|\!|{B}\rangle \!\rangle\hspace{3 cm} \forall a\in \mathcal{C}
\end{align}
where $\langle\mathcal{L}_a \rangle$ is the quantum dimension of the defect line $\mathcal{L}_a$ and similarly a boundary is called \textit{weakly symmetric} if the following equation holds,
\begin{align}
\mathcal{L}_a|\!|{B}\rangle\!\rangle=|\!|{B}\rangle\!\rangle+ \cdots \hspace{3 cm} \forall a\in \mathcal{C}
\end{align}
where the ellipsis denotes the sum of other Cardy boundary states {(for non diagonal theory graph-boundary states/NIM-rep boundary states)} present in the theory. If there is a defect line with a non-integer quantum dimension, it turns out that one can not find any strongly symmetric boundary state \cite{Choi:2023PRD, Heymann:2024vvf}.\\
\paragraph{Cloaking boundary.}
Apart from the standard Cardy boundary conditions, we also consider a special class of boundary states known as \emph{cloaking boundary states}~\cite{Donnelly:2018ppr, Hung:2019bnq, Brehm:2022JPA, Brehm:2024arXiv}. These boundary conditions play a central role in symmetry resolution because they are constructed to be invisible to a chosen set of topological defect lines. Let the defect lines of the category $\mathcal{C}$ be labeled by $\mathcal{I}$. We want to construct a cloaking boundary condition for every subset $\mathcal{J}\subset \mathcal{I}$ that closes under fusion.

A boundary condition is said to be \emph{cloaked} with respect to $\mathcal{J}$ if its Ishibashi components are invisible to the action of every defect line in $\mathcal{J}$, i.e., the defect line passes through the boundary, unlike the Cardy states, where the lines end. This invisibility condition is imposed by requiring that the row of the modular $S$-matrix indexed by $x$ agrees with the vacuum row when restricted to the columns labeled by $\mathcal{J}$. Explicitly, the allowed Ishibashi labels $x \in \mathcal{I}$ satisfy
\begin{equation}
\mathcal{I}_{\mathcal{J}}=\left\{
x \in \mathcal{I}\,\Bigg|\,
\frac{S_{xj}}{S_{x1}} = \frac{S_{\mathbf{1}j}}{S_{11}}
\ \ \forall\, j \in \mathcal{J}
\right\}.
\label{2.7}
\end{equation}
\noindent
\\
Here $S_{ij}$ denotes the elements of the modular S-matrix.  Whenever the above condition holds, the Ishibashi state $|x\rangle\!\rangle$ contributes to the cloaking boundary. \textcolor{black}{The associated cloaking boundary state is 
\begin{equation}
\bigl\| \gamma_{\mathcal{J}}\bigr\rangle\!\bigr\rangle
= \delta_{\mathcal{J}}
\sum_{x \in \mathcal{I}} \sqrt{S_{x1}}\, |x\rangle\!\rangle ,
\label{2.8}
\end{equation}
In general, $\bigl\|\gamma_{\mathcal{J}}\bigr\rangle\!\bigr\rangle$ is not a Cardy boundary state. We use it as a defect-transparent state at the entangling cut. For the calculations below, the known action of the TDLs on $\bigl\|\gamma_{\mathcal{J}}\bigr\rangle\!\bigr\rangle$ is sufficient to determine the normalized sector weights. A complete RCFT realization of this entangling boundary would additionally require fixing its normalization $\delta_{\mathcal{J}}$ and
specifying how the defect lines meet the cut. We leave these questions for future investigation; see \cite{Donnelly:2018ppr, Brehm:2022JPA, Brehm:2024arXiv} for related discussions\footnote{\textcolor{black}{While calculating the probabilities of each symmetry resolution sector, this factor of $\delta_{\mathcal{J}}$ will not affect the probabilities, while it will shift the boundary entropy constant.}}.} \par


\vspace{0.75cm}
Before we end this section, we note that, for unitary diagonal theories, the vacuum and the ground state coincide, so the usual
modular quantum dimension \(S_{a1}/S_{11}\) agrees with the
Perron-Frobenius quantum dimension (see \cite{2004math......1246C} for more details). In non-unitary theories, we distinguish
these two notions. Throughout the entropy formulas below, we use the
ground-state, or Perron-Frobenius, quantum dimension
\[
d_a \equiv d_a^{\rm PF}=\frac{S_{a\Omega}}{S_{\mathbf{1}\Omega}},
\]
where \(\Omega\) labels the ground-state of the theory. The vacuum-row quantity \(S_{a1}/S_{11}\) enters only in the cloaking condition itself, where one asks whether the defect acts on the corresponding Ishibashi states as it acts on the vacuum sector. However, in the subsequent sections, we use the notation $d_a$ for both unitary and non-unitary cases, and the interpretations are implied.

\section{Symmetry resolved entanglement entropy (SREE)}\label{sec3}
\begin{figure}[htb!]
    \centering
\includegraphics[width=0.56\linewidth]{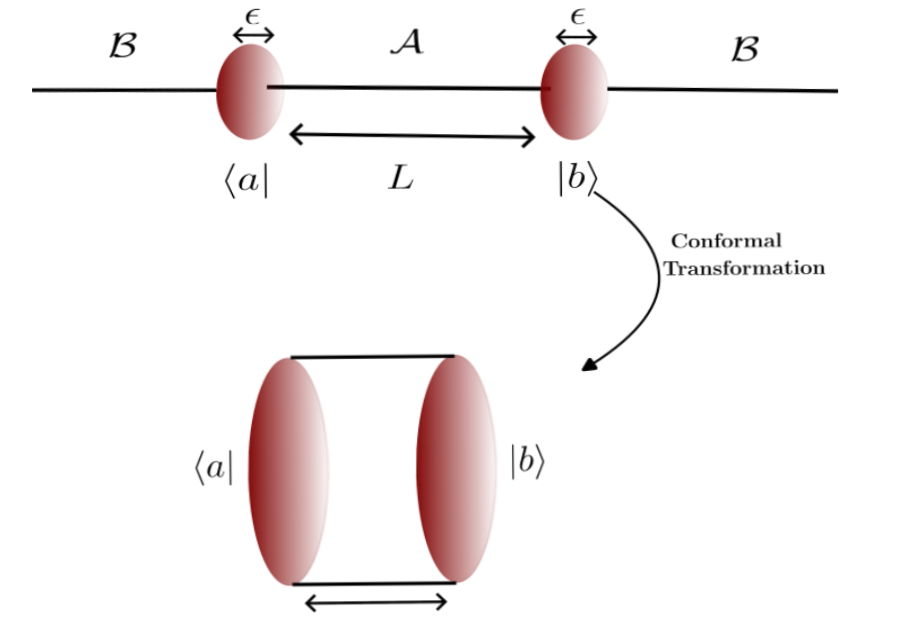}
    \caption{For a set of intervals A and B, as the Hilbert space is not a direct product, we have put conformal boundary conditions with two small $\epsilon$ circles at the entangling surface, which turn into two different Cardy boundary states {(or more generally for non-diagonal theories graph fusion states/ NIM-rep states )} after conformal mapping to an annulus.}
    \label{fig:2}
\end{figure}
We consider a $(1+1)$ dimensional quantum system endowed with an internal symmetry and partition it into two complementary subsystems, $A$ and $B$. In order to compute the entanglement entropy of subsystem A of length $L$ with B, we have traced out the subsystem B, and this configuration is conformally equivalent to an annulus of width $W = 2\log(\frac{L}{\epsilon})$, see Fig.~(\ref{fig:2}). {For abelian (invertible) symmetries}, let $\mathcal{Q}$ denote the conserved charge operator generating the symmetry. \textcolor{black}{We assume that the total charge decomposes additively as}
\begin{equation}
    \mathcal{Q} = \mathcal{Q}_A \oplus \mathcal{Q}_B .
\end{equation}
If the global state of the system is an eigenstate of $\mathcal{Q}$, it follows that
\begin{equation}
    [\widehat\rho, \mathcal{Q}] = 0 ,
\end{equation}
where $\widehat\rho$ is the density matrix of the full system. By tracing over the degrees of freedom of $B$, we obtain the reduced density matrix $\widehat\rho_A = \mathrm{Tr}_B\,\widehat\rho$, which commutes with the subsystem charge operator $\mathcal{Q}_A$. 
Consequently, $\widehat\rho_A$ has a block-diagonal structure (where each block is denoted by $\mathcal{R}_\rho$), with each block labeled by an eigenvalue ${q_\rho}$ of $\mathcal{Q}_A$:
\begin{equation}
   \widehat \rho_A = \bigoplus_\rho\widehat\rho_A(q_\rho)\,. \label{eq:block}
\end{equation}
\noindent
Each block $\widehat\rho_A(q_\rho)$ is obtained by projecting $\widehat\rho_A$ onto the eigenspace of $\mathcal{Q}_A$ with corresponding eigenvalue. 
Through this decomposition, one can define symmetry-resolved entanglement measures. {
The non-invertible analogue of charge resolution is not a decomposition into eigenspaces of an additive subsystem charge. For an ordinary invertible
symmetry, one can often split the conserved charge as \(\mathcal{Q}=\mathcal{Q}_A+\mathcal{Q}_B\), so that
\(\widehat\rho_A\) decomposes into sectors labelled by the eigenvalues of \(\mathcal{Q}_A\).
For a non-invertible symmetry, however, there is, in general, no such local
charge operator. The symmetry is instead generated by topological defect lines
\(\mathcal L_a\), labelled by simple objects \(a\), with fusion
\begin{align}
    \mathcal L_a\times \mathcal L_b
    =
    \sum_c \widehat{\mathcal N}_{ab}^{\;\;c}\,\mathcal L_c\,,
\end{align}
where \(\widehat{\mathcal N}_{ab}^{\;\;c}\) are fusion multiplicities.
}
{
When the entangling boundary condition is compatible with these defects, the defect lines can be moved topologically to the boundary. In the open-string channel they are represented by operators \(\widehat{\mathcal L}_a\), which
commute with the annulus Hamiltonian,
\begin{align}
    [H_{\rm annulus},\widehat{\mathcal L}_a]=0\,,
\end{align}
and 
 also commutes with the reduced density matrix 
\begin{align}
    [\widehat\rho_A,\widehat{\mathcal L}_a]=0,
    \qquad \forall a \,,
\end{align}
then the appropriate analogue of symmetry resolution is the
decomposition of the open-string Hilbert space,
\begin{align}
    \mathcal H_A
    =
    \bigoplus_\mu V_\mu\otimes \mathcal M_\mu^{A}.
\end{align}
Here \(V_\mu\) is an irreducible module of the relevant defect algebra, while
\(\mathcal M_\mu^{A}\) is the corresponding multiplicity space. By Schur's
lemma, \(\widehat\rho_A\) acts trivially on \(V_\mu\) and only nontrivially on the
multiplicity factor:
\begin{align}
    \widehat\rho_A
    =
    \bigoplus_\mu
    \mathbf 1_{V_\mu}\otimes \widehat\rho_A^{\,(\mu)} .
\end{align}
Thus, the labels \(\mu\) play the role of generalized charge sectors. They are not eigenvalues of an additive charge operator, but categorical sectors associated with irreducible representations of the non-invertible defect algebra.
}
Following \cite{Goldstein:2017bua, german3, Bonsignori:2019naz, DiGiulio:2022jjd}, the symmetry-resolved Rényi entropy, which quantifies the entanglement between subsystems \(A\) and \(B\) within a fixed charge sector \(\mathcal{R}_\rho\), is introduced as, 
\begin{align}
    S_{n}(\mathcal{R}_\rho)
    =\frac{1}{(1-n)} \log{\rm Tr}( \widehat{\rho}_A(q_\rho)^{n}) 
    = \frac{1}{1-n} \log\!\left( \frac{Z_{B_1|B_2}(q^n,\mathcal{R}_\rho)}{Z_{B_1|B_2}^n(q,\mathcal{R}_\rho)} \right). \label{eq3}
\end{align}
Here, {\(Z_{B_1|B_2}(q)\)} denotes the full annulus partition function in the open-string channel and closed-string channel related by modular $S$ transformation \cite{Cardy:2004hm,Petkova:2000ip}, expressed as \footnote{The closed-string amplitude contains both chiralities. It reduces to one chiral character because of the left-right level matching in an Ishibashi state.},
\begin{align}
    Z_{B_1|B_2}(q) =\sum_{\rho} Z_{B_1|B_2}(q,\mathcal{R}_{\rho})= \sum_{i} n_{iB_1}^{\,B_2} \, \chi_i(q)
    {=} \langle\!\langle B_1|\!|\tilde{q}^{{\color{black}{\frac{1}{2}(\,L_{0}+\bar{L}_0-\frac{c}{12})}}}|\!|B_2\rangle\!\rangle \label{eq4}
\end{align}
where,  we use $q=e^{-\frac{2\pi^2}{W}}, \tilde{q} = e^{-2W}$, \(B_1\) and \(B_2\) label the conformal boundary states. {For diagonal theories, these can be Cardy states, while in non-diagonal examples, they should be understood as graph/NIM-rep boundary labels.} Now, \eqref{eq4} can be justified by the following definition in the open-string channel as,
\begin{align}
\begin{split}
    Z_{B_1|B_2}(q)&=\,\text{Tr}_{\mathcal{H}_{B_1|B_2}}(q^{L_0-c/24})
    =\text{Tr}_{\oplus_{\rho}\mathcal{H}_{B_1|B_2}^{\rho}}(q^{L_0-c/24})
    =\sum_{\rho}Z_{B_1|B_2}(q,\mathcal{R}_{\rho})
    \end{split}
\end{align}

\textcolor{black}{The sector Hilbert space is the image of a projector ($P^\rho_{B_1|B_2}$) acting on the full
annulus Hilbert space:
\[
\mathcal H^\rho_{B_1|B_2}
=
P^\rho_{B_1|B_2}\mathcal H_{B_1|B_2},
\qquad
Z_{B_1|B_2}(q,R_\rho)
=
\operatorname{Tr}_{\mathcal H_{B_1|B_2}}
\left[
P^\rho_{B_1|B_2}q^{L_0-c/24}
\right].
\]
This projected annulus trace is also the observable considered in \cite{Choi:2024wfm}. In that work, the projector is constructed using TQFT boundary data, whereas here it is obtained directly from the action of the RCFT topological defect lines on the open-channel states. For a diagonal unitary  RCFT, the two descriptions agree after matching the sector labels and defect orientations.\footnote{\textcolor{black}{In the diagonal  case, the quantum dimensions and boundary factors appearing in \cite{Choi:2024wfm} reduce to the corresponding RCFT quantities, while its boundary fusion multiplicity is identified with the
annulus multiplicity, $N^{\underline B_1}_{i\underline B_2}
=n^{B_2}_{i B_1}$. With this identification, the subsequent equation (\ref{3.15k}) is the diagonal-unitary RCFT realization of Eq.~(19) of \cite{Choi:2024wfm}. For a general non-diagonal RCFT, the same identification additionally requires the action of the projector on all primaries in the open channel and their multiplicity spaces. This depends on further boundary and junction data, including boundary $\widetilde F$-symbols. The explicit examples computed in \cite{Choi:2024wfm} are cases for which these data are already known or readily obtainable.}}}

The coefficients \(n_{iB_1}^{\,B_2}\) denote annulus multiplicities: in a diagonal case, they reduce to the usual Verlinde coefficients \(N_{iB_1}^{\,B_2}\), while for non-diagonal theories, they are the relevant NIM-rep coefficients. Throughout this work, we adopt the notation \(Z(q)\) in place of \(Z_{B_1|B_2}(q)\). In subsequent sections, the structure and interpretation of the coefficients \(n_{iB_1}^{\,B_2}\) will be elaborated in detail. Accordingly, \(Z(q, \mathcal{R}_\rho)\) refers to the sector of the full partition function corresponding to the symmetry representation denoted by \(\mathcal{R}_\rho\). {For $1+1$D RCFT, the global symmetry is generated by topological defect lines (TDLs), and they form a fusion category. In order to resolve the entanglement Rényi entropy with respect to the fusion category of the underlying TDLs, the partition functions of each irreducible sectors can be extracted from the full annulus partition function by looking at the action of the TDLs on the primary fields rather than computing charged moments \cite{DiGiulio:2022jjd}}\footnote{For the computation of charged moments, we have to compute the projectors in terms of the $F$-symbols. For generic RCFT, it is very hard to compute these symbols. For more detail about the $F$-symbols upto rank $7$, see \cite{Ardonne_2010, vercleyen2023lowrankfusionrings}.}.

\subsection*{Boundary conditions and SREE}
In this section, we briefly review some subtleties in the choice of boundary states and their consequences for the symmetry-resolved Rényi entropies (SRRE). 
For illustration, suppose there are {$\mathscr{N}$} irreducible blocks; the Rényi entropy for the $\rho$'th block admits the following asymptotic behavior using \eqref{eq4}, in the limit $q\to 1$ or equivalently $\tilde{q}\to 0$ ,

\begin{align}
\begin{split}
    S_{n}(\mathcal{R}_\rho)
    &=\frac{1}{1-n}\log\!\left(\frac{Z(q^n,\mathcal{R}_\rho)}{Z^n(q,\mathcal{R}_\rho)}\right),
\end{split}
\end{align}
{with the following definition of the partition function and probability,
    \begin{align}
        \begin{split}
    \frac{Z(q^n,\mathcal{R}_\rho)}{Z^n(q,\mathcal{R}_\rho)}
    \xrightarrow{\tilde q\to 0}\frac{
    p_\rho^{\,1-n}\,\tilde q^{\big(n-\frac{1}{n}\big)\frac{c_{\rm eff}}{24}}}{\left(\sum_{\rho}n_{\rho B_1 }^{\,\,B_2}S_{\rho \Omega}\right)^{n-1}}, \textrm{with,\,}\,p_{\rho}=\frac{Z(q,\mathcal{R}_{\rho})}{Z({q})}.
    \label{3.11b}
\end{split}
\end{align}}
where $c_{\rm eff} = c-24\Delta_{\Omega}$ and  $\Delta_{\Omega}$ is the conformal dimension of the ground state (for unitary theories the ground state is the vacuum state with $\Delta_{\Omega}=0$, but non-zero for non-unitary theory), $Z({q,\mathcal{R}_{\rho}})$ and $Z(q)$ are the sector partition function and full partition function respectively, {for usual Cardy boundary state, the coefficient $n_{\rho B_1}^{\,\,B_2}$ is the Verlinde coefficient (we will comment on this in detail in the subsequent subsections) and is denoted as $N_{\rho B_1}^{\,\,B_2}$} , and $\Omega$ denotes the entry of S-matrix corresponding to the ground state. In this limit, the SREE for the irreducible block becomes
\begin{align}
    S_{n}(\mathcal{R}_\rho)
    =\frac{c_{\rm eff}}{6}\left(\frac{n+1}{n}\right)\log\!\left(\frac{L}{\epsilon}\right)+\log\!\big(p_\rho\big)+\log\left(\sum_{\rho}n_{\rho B_1 }^{\,\,\,B_2}S_{\rho \Omega}\right)+\cdots\,.\label{3.7j}
\end{align}
{
It is straightforward to show that the last term is the boundary entropy  and can be written as,
\begin{align}
    S_{b}=\log(g_1 g_2), \,\,\textrm{where,}\,\,g_1g_2=\langle\!\langle B_1|\!|\Omega\rangle ^*\langle\!\langle B_2|\!|\Omega\rangle \,.
\end{align}
{
\textcolor{black}{The  boundary states in general can be expanded in the Ishibashi basis as
\begin{align}
\begin{split}
    |\!|B_a\rangle\!\rangle
    =
    \sum_{j\in \mathrm{Exp}(G)}
    B_a^j |j\rangle\!\rangle\,,
    \qquad
    B_a^j
    =
    \frac{\psi_a^j}{\sqrt{S_{\mathbf 1 j}}}\,.
\end{split}
\end{align}
Here \(\mathrm{Exp}(G)\) denotes the set of Ishibashi, or exponent, labels, and
\(\psi_a^j\) is the eigenvector matrix diagonalizing the NIM-rep. Now the boundary term can be explicitly computed,
\begin{align}
\begin{split}
   \langle\!\langle B_1|\!|\Omega\rangle=  g_1 g_2
    &=
    \left(B_{B_1}^{\Omega}\right)^{*}
    B_{B_2}^{\Omega}  =
    \frac{
    \left(\psi_{B_1}^{\Omega}\right)^{*}
    \psi_{B_2}^{\Omega}
    }{
    S_{\mathbf 1\Omega}
    } \,.
\end{split}
\end{align}
The NIM-rep matrices are diagonalized by the same \(\psi\)-matrix:
\begin{align}
\begin{split}
    (n_\rho)_{B_1}^{\;\;B_2}
    =
    \sum_{j\in \mathrm{Exp}(G)}
    \frac{
    S_{\rho j}
    }{
    S_{\mathbf 1 j}
    }
    \left(\psi_{B_1}^{j}\right)^{*}
    \psi_{B_2}^{j}.
\end{split}
\end{align}
Now consider the following expression,
\begin{align}
\begin{split}
    \sum_{\rho} (n_\rho)_{B_1}^{\;\;B_2} S_{\rho\Omega}
    &=
    \sum_{\rho}
    \sum_{j\in \mathrm{Exp}(G)}
    \frac{
    S_{\rho j}
    }{
    S_{\mathbf 1 j}
    }
    \left(\psi_{B_2}^{j}\right)^{*}
    \psi_{B_1}^{j}
    S_{\rho\Omega}\,,  \\
    &=
    \sum_{j\in \mathrm{Exp}(G)}
    \frac{
    \left(\psi_{B_2}^{j}\right)^{*}
    \psi_{B_1}^{j}
    }{
    S_{\mathbf 1 j}
    }
    \left(
    \sum_{\rho}S_{\rho j}S_{\rho\Omega}
    \right)\,, \\
    &=
    \sum_{j\in \mathrm{Exp}(G)}
    \frac{
    \left(\psi_{B_2}^{j}\right)^{*}
    \psi_{B_1}^{j}
    }{
    S_{\mathbf 1 j}
    }
    \delta_{j\bar\Omega} =    \frac{
    \left(\psi_{B_2}^{\bar\Omega}\right)^{*}
    \psi_{B_1}^{\bar\Omega}
    }{
    S_{\mathbf 1\Omega}
    } =
    g_1 g_2\,.
\end{split}
\end{align}
Note that, for the diagonal case, \(\mathrm{Exp}(G)=\mathcal{I}\) and $\psi_a^j=S_{a j}$, and we assume that the dominant ground-state primary $|\Omega\rangle$ is unique.} 

Therefore, the Rényi entropy is given by
\begin{align}
    S_{n}(\mathcal{R}_\rho)
    =\frac{c_{\rm eff}}{6}\left(\frac{n+1}{n}\right)\log\!\left(\frac{L}{\epsilon}\right)
    +\log\!\big(p_\rho\big)+\log(g_1 g_2)+\cdots\,.\label{3.11m}
\end{align}
From this we can get the symmetry-resolved entanglement entropy (SREE) by taking $n\rightarrow 1$ limit. Our aim is not only to compute the SREE but also to relate the $p_\rho$'s to modular data and topological quantities, which we do now by explicitly identifying $p_{\rho}$ as the probability associated with each charge sector and subsequently deriving a closed form expression for it. 
 We start with the following definition of the probability associated with each charge sector, 
\begin{align}
    p_{\rho}=\frac{Z_{B_1|B_2}(q;\rho)}{Z_{B_1|B_2}(q)}\,.
 \end{align}
The annulus partition function in the open string channel is given by\footnote{Note that, in our further computations, we use the notation $Z(q, \mathcal{R}_{\rho})$ in place of ${Z_{B_1|B_2}(q;{\rho})}$.},
\begin{align}
    \begin{split}
        Z_{B_1|B_2}(q)=\sum_{\rho}n_{\rho B_1}^{\,\,B_2}\chi_{\rho}(q)
    = \sum_{\rho,i}n_{\rho B_1}^{\,\,B_2} S_{\rho i}\,\chi_i(\tilde q)\,
    \overset{\tilde q\to 0}{\sim}
    \sum_{\rho}n_{\rho B_1}^{\,\,B_2} S_{\rho \Omega}\,\chi_\Omega(\tilde q)+\cdots\,.
    \end{split}
\end{align}
Therefore,
\begin{align}
    p_{\rho}= \frac{n_{\rho B_1}^{\,\,B_2}\chi_{\rho}(q)}{\sum_{\sigma}n_{\sigma B_1}^{\,\,B_2}\chi_{\sigma}(q)} \overset{\tilde q\to 0}{\sim} \frac{n_{\rho B_1}^{\,\,B_2}S_{\rho \Omega}\chi_{\Omega}(\tilde q)}{\sum _{\sigma}n_{\sigma B_1}^{\,\,B_2}S_{\sigma \Omega}\chi_{\Omega}(\tilde q)}\xrightarrow[]{\textrm{div. by}\, S_{\mathbf{1}\Omega}} \frac{n_{\rho B_1}^{\,\,B_2 }\,d_{\rho}}{\sum_{\sigma}n_{\sigma B_1}^{\,\,B_2 }\,d_{\sigma}}\,.\label{3.14j}
\end{align}
{Again note that, $\Omega$ here labels the ground state and $d_{\rho}'s$ are computed with respect of it.} Also, $n_{\rho B_1}^{\,\,B_2}$ denotes the boundary fusion coefficient defined via the action of a topological defect line on a boundary state \footnote{We propose this interpretation of the formula without invoking the SymTFT construction, relying instead on a purely $1+1$D RCFT analysis, in contrast to \cite{Choi:2024wfm}. }:
\begin{align}
   \mathcal{L}_{\rho}\,\|B_1\rangle\!\rangle
   = n_{\rho B_1}^{\,\,B_2}\,\|B_2\rangle\!\rangle
   + \sum |\!|CS\rangle\!\rangle ,
   \label{5.31k}
\end{align}
where,  $\sum|\!|CS\rangle\!\rangle$ is the sum of other Cardy states {(or more generally the NIM-rep states)} with \textit{positive} coefficients and this positivity condition uniquely fixes the functional form of {$n_{\rho B_1}^{\,\,B_2}$}\footnote{We emphasize this point because the right-hand side can always be expanded as
\[
n_{\rho B_1}^{\,\,B_2}\,\|B_2\rangle\!\rangle \;+\; \sum \|IS\rangle\!\rangle,
\]
where $\|IS\rangle\!\rangle$ denote Ishibashi states. At the level of such an
Ishibashi-state decomposition, the coefficients
$n_{\rho B_1}^{\,\,B_2}$ need not be unique, and multiple
choices may solve the same equation. Our interpretation, however, uses the
additional input that every state on the right-hand side, other than
$\|B_2\rangle\!\rangle$ must be expressible in terms of Cardy states with a positive coefficient. With this
requirement, the fusion coefficient is fixed uniquely.
}. 
{
However, note that in the case of non-diagonal theories, the probabilities associated with the charge sectors can not be trivially extrapolated. We provide detailed comments on the issues as we discuss symmetry resolution in those cases in a later section.}
 \textcolor{black}{
\textcolor{black}{Now, using the state–operator correspondence together with the Cardy construction, one finds the explicit action (for diagonal RCFT) \cite{Petkova:2000ip},
\begin{align}
   \mathcal{L}_{\rho}\|B_i\rangle\!\rangle
   = \sum_{j=1}^{{\mathscr{N}}}\frac{S_{ij}}{\sqrt{S_{\mathbf{1}j}}}\,
     \mathcal{L}_{\rho}|j\rangle\!\rangle
   = \sum_{j=1}^{{\mathscr{N}}}\frac{S_{ij}}{\sqrt{S_{\mathbf{1}j}}}\,
     \frac{S_{\rho j}}{S_{\mathbf{1}j}}\,|j\rangle\!\rangle\,.
   \label{5.32h}
\end{align}}}

\smallskip
\noindent
{\color{black}{Here the action of a topological defect line, specifically the Verlinde line, which commutes with the chiral algebra of an RCFT, on an Ishibashi state is given by}}
\textcolor{black}{\begin{align}
    \mathcal L_\rho|\mu\rangle\!\rangle
    =
    \frac{S_{\rho\mu}}{S_{\mathbf 1\mu}}
    |\mu\rangle\!\rangle .
\end{align}}
This does not mean that \(|\mu\rangle\!\rangle\) is a primary state; it is an infinite coherent descendant state associated with the primary module \(\mathcal H_\mu\).
{For non-diagonal RCFTs, this simple diagonal action is no longer the correct
description in general.  The relevant boundary and annulus data are instead
encoded by NIM-rep coefficients, rather than by the ordinary Verlinde
coefficients alone~\cite{Behrend:1999bn,Petkova:2000ip}, as discussed in
Section~(\ref{sec4}).} 
Since $\mathcal{L}_{\rho}$ is a symmetry generator, it commutes with all Virasoro generators. 
Thus, even though Ishibashi states are not primary, the action of $\mathcal{L}_{\rho}$ on those states mimics that of primary fields, unlike Cardy states. From \eqref{5.32h}, one can then recombine terms to identify the coefficients $n_{\rho B_1}^{B_2}$. 
Open–closed duality further imposes the consistency condition
\begin{align}
    d_{B_1}d_{B_2}
    = \sum_{\rho\in \texttt{Irr}(\mathcal{C})}
      n_{\rho B_1}^{\,\,\,B_2}\, d_{\rho}\,.
    \label{3.20g}
\end{align} 
The derivation goes as follows (we take real, symmetric modular data).
\textcolor{black}{We again start with the annulus partition function in the open-string channel,
\begin{align}
    Z_{B_1|B_2}(q)
    = \sum_{\rho}n_{\rho B_1}^{\,\,B_2}\chi_{\rho}(q)
    = \sum_{\rho,i}n_{\rho B_1}^{\,\,B_2} S_{\rho i}\,\chi_i(\tilde q)\,
    \overset{\tilde q\to 0}{\sim}
    \sum_{\rho}n_{\rho B_1}^{\,\,B_2} S_{\rho \Omega}\,\chi_\Omega(\tilde q)+\cdots\,.
\end{align}
In the closed channel, the corresponding partition function reads
\begin{align}
\begin{split}
    Z_{B_1|B_2}(\tilde q)
    &= \langle\!\langle B_1\|\tilde q^{\color{black}{\frac{1}{2}(\,L_{0}+\bar{L}_0-\frac{c}{12})}}\|B_2\rangle\!\rangle
     = \sum_{i,j}\frac{S_{B_2 i}}{\sqrt{S_{\mathbf{1}i}}}\frac{S_{B_1 j}}{\sqrt{S_{\mathbf{1}j}}}\delta_{ij}\chi_j(\tilde q)\\
    &\overset{\tilde q\to 0}{\sim}
      \frac{S_{B_2 \Omega}S_{B_1 \Omega}}{S_{\mathbf{1}\Omega}}\,\chi_\Omega(\tilde q)+\cdots .
\end{split}
\end{align}
Matching the leading terms from both channels yields
\begin{align}
    \frac{1}{S_{\mathbf{1}\Omega}}\sum_{\rho}n_{\rho B_1}^{\,\,B_2}S_{\rho \Omega}
    = d_{B_1}d_{B_2}
    \quad\Longrightarrow\quad
    d_{B_1}d_{B_2}
    = \sum_{\rho}n_{\rho B_1}^{\,\,B_2}d_{\rho}\,
    \label{3.24j}
\end{align}
as claimed in \eqref{3.20g}. A similar derivation holds for more general modular data with complex entries. So have finally, $$
    p_{\rho}= \frac{n_{\rho B_1}^{\,\,B_2 }\,d_{\rho}}{d_{B_1}d_{B_2}}\,.
$$}
Now we \textit{explicitly} compute the formula \eqref{3.11m} with the above definition of $p_{\rho}$ for different boundary conditions. \textit{At this point, we emphasize once again that nowhere in the above derivation have we directly relied on any SymTFT-based construction}.  

{We again note that, for all cases considered here, \(d_\rho=S_{\rho\Omega}/S_{\mathbf{1}\Omega} \geq 1, \) denotes the ground-state/Perron--Frobenius quantum dimension. The label \(\Omega\) is the dominant ground-state label in the modular transformation matrix; in a non-unitary RCFT, this need not be the vacuum. This gives the positive weights in the sector probabilities, whereas vacuum-row quantum dimensions (which are not physical in case of theories where the vacuum state is not the ground state) can be negative or imaginary.}\\\\
\noindent\textbf{Weakly symmetric boundary condition}\\\\
For a weakly symmetric Cardy boundary state $n=N$, correspondingly the SREE in \eqref{3.11m} takes the form
\begin{align}
    {S}(\mathcal R_\rho)
    &= \frac{c_{\rm eff}}{3}\log\!\left(\frac{L}{\epsilon}\right)
    + \log(g_1 g_2)
    + \log\!\left(\frac{d_{\rho}\,N_{\rho B_1}^{B_2}}{d_{B_1}d_{B_2}}\right) + \cdots,
    \label{3.15k}
\end{align}
{The \(d_\rho,d_{B_1},d_{B_2}\) in \eqref{3.15k} are computed with respect to the ground-state, not from the vacuum, in non-unitary theories.} Here {$\log(g_1 g_2)$ is the boundary entropy} and, if a block contains several irreducible elements (this scenario may appear when the full resolution is not possible), a sum over the block index $\rho$ is implied. For this case, 
\begin{align}
   d_{B_1}d_{B_2}= d(B_1\otimes B_2)=\sum_{\rho} {\color{black}{N_{\rho B_1}^{B_2}\,}} d_{{\rho}},
\end{align}
{Here \({\color{black}{N_{\rho B_1}^{B_2}\,}}\) is the multiplicity of the line \(\mathcal L_\rho\) in the boundary fusion channel \(B_1\otimes B_2\). Channels with \({\color{black}{N_{\rho B_1}^{B_2}\,}}=0\) are absent from the corresponding block sum; in the multiplicity-free case, every present channel has \({\color{black}{N_{\rho B_1}^{B_2}\,}}=1\).} {For example, with a case where all the lines appear only once in the block sum i.e, ${\color{black}{N_{\rho B_1}^{B_2}\,}}=1,\,\forall \rho$, so that \eqref{3.15k} reduces to
\begin{align}
    {S}(\mathcal R_\rho)
    = \frac{c_{\rm eff}}{3}\log\!\left(\frac{L}{\epsilon}\right)
    + \log(g_1 g_2)
    + \log\!\left(\frac{d_{\rho}}{\sum\limits_{\sigma\in\texttt{Irr}(\mathcal{C})}d_{\sigma}}\right)+\cdots.
    \label{3.17j}
\end{align}}

\textcolor{black}{We like to emphasize that (\ref{3.15k}) can be computed only using the information of annulus multiplicities, quantum dimensions of the TDLs, and the information about the boundaries of the underlying RCFT. Once the open-channel TDL action is known, no separate reconstruction of the model-dependent boundary-Tube projectors is required, as done in \cite{Choi:2024wfm}.} \\

\noindent\textbf{Strongly (and possibly cloaking) boundary condition}\\
For a strongly symmetric boundary state, the topological defect line acts simply by its quantum dimension:
\begin{align}
    \mathcal{L}_{\rho}\,\|B\rangle\!\rangle = d_{\rho}\,\|B\rangle\!\rangle 
    \quad\Longrightarrow\quad 
    n_{\rho B}^{\;B}=d_{\rho}\,.
\end{align}
{Note that $d_{\rho}$ is the same Perron-Frobenius quantum dimension that we discussed earlier.}
Now, substituting this relation into \eqref{3.11m}  the SREE is given by \footnote{For cloaking states the formula in \eqref{3.24k} matches with \cite{Sierra_2024,Choi:2024wfm}. }
\begin{align}
     S(\mathcal R_\rho)
     = \frac{c_{\rm eff}}{3}
     \log\!\left(\frac{L}{\epsilon}\right)
     + \log(g_1 g_2)
     + \log\!\left(\frac{d_{\rho}^2}{\sum\limits_{\sigma \in \texttt{Irr}(\mathcal{C})}d_{\sigma}^2}\right)+\cdots\,.
     \label{3.24k}
\end{align}
\noindent The formula \eqref{3.24k} is expected to apply also to the cloaking boundary state \cite{Brehm:2022JPA, Brehm:2024arXiv} introduced in \eqref{2.8}. 
{\color{black}{In many diagonal unitary RCFTs, the cloaking boundary state coincides with the identity Ishibashi state $|\mathbb{I}\rangle\!\rangle$ (upto a normalization as discussed in (\ref{2.8}))  \cite{Brehm:2022JPA, Brehm:2024arXiv} corresponding to the vacuum, for which the defect acts by its usual quantum dimension}}
so that the Eq.~\eqref{3.24k} follows directly for this case as well, with the boundary entropy depending on the choice of the normalization, as will be demonstrated later. 
We stress, however, that although \eqref{3.24k} holds in these (cloaking) examples, the derivations (e.g 
(\ref{3.24j})) employed for the Cardy boundary states do not carry over in a straightforward way to cloaking boundaries (as cloaking states are not physical Cardy states and one should really emphasize that it is not a boundary condition, but may be thought of as a boundary condition with a topological point operator on it.)
\section{ Computing SREE in Unitary Rational CFT}\label{sec4}
In this section, we investigate symmetry resolution in both unitary diagonal and non-diagonal RCFTs. For diagonal models, we primarily focus on the Tambara-Yamagami (TY) symmetry resolution in the Ising and tetracritical Ising CFTs. For non-diagonal RCFTs, we examine two representative cases: the $E_6$ theory and its $\tfrac{1}{2}E_6$ fusion ring, and the $\mathbb{Z}_3$ symmetry resolution in the three-state Potts model.


\subsection{Symmetry resolution with respect to Tambara-Yamagami (TY) Fusion Category for Ising CFT}
The simplest unitary Virasoro minimal model, namely the critical Ising model, is the prototype example for two-dimensional rational CFT. It has three primary fields, $\{\mathbb{I}, \varepsilon, \sigma\}$, and their non-trivial fusions are
\begin{eqnarray}
\varepsilon\times\varepsilon = \mathbb{I}, \quad \varepsilon\times\sigma = \sigma,\quad \sigma\times\sigma = \mathbb{I} + \varepsilon\,.
\end{eqnarray}

This is a diagonal rational CFT with central charge $c=\frac{1}{2}$. So, it has three topological defects labeled by the primaries, namely, $\{\mathcal{L}_{\mathbb{I}}, \mathcal{L}_{\mathbb{\varepsilon}}, \mathcal{L}_{\mathbb{\sigma}} \}=\mathcal{C}$ and their action on the primaries is summarized in Table~(\ref{tabIsing}).
\begin{table}[H]
\centering
\begin{tabular}{cccc}
\toprule
Defect line & $\mathbb{I}$ & $\varepsilon$ & $\sigma$ \\
\midrule
$\mathcal{L}_{\mathbb{I}}$ & $1$ & $1$ & $1$ \\
$\mathcal{L}_{\mathbb{\varepsilon}}$ & $1$ & $1$ & $-1$\\
$\mathcal{L}_{\mathbb{\sigma}}$ & $\sqrt{2}$ & $-\sqrt{2}$ & $0$ \\
\bottomrule 
\end{tabular}
\caption{Action of the TY category on the primaries\,.} 
\label{tabIsing}
\end{table} 
\noindent
We want to study symmetry-resolved entanglement entropy for this TY fusion category symmetry. In order to do that, we need a strongly symmetric or weakly symmetric entangled boundary condition according to \cite{Choi:2023PRD, Choi:2024wfm, Heymann:2024vvf, Das:2024qdx}. But it turns out that for this case, there is no such boundary at the entangled surfaces. However, one can use the cloaking boundary condition \cite{Brehm:2022JPA, Brehm:2024arXiv} for the purpose of symmetry resolution.  
\noindent
\textcolor{black}{For the critical Ising model, it turns out that the Ishibashi state gives the cloaking boundary state that corresponds to the identity primary (upto a normalization) \cite{Brehm:2024arXiv}
\begin{eqnarray}
|\!|\gamma_\mathcal{J}\rangle\!\rangle =  \delta_{\mathcal{J}}\,|\mathbb{I}\rangle\!\rangle\,.   \label{cloakIsing} 
\end{eqnarray}
Now, we can compute the annulus partition function in the closed string channel when we sandwich the cloaking state \eqref{cloakIsing} at the boundary. 
\begin{align}
\begin{split}
Z(\tilde{q}) &= \langle\!\langle\gamma_\mathcal{J}|\!|\tilde{q}^{{\color{black}{\frac{1}{2}(\,L_{0}+\bar{L}_0-\frac{c}{12})}}}|\!|\gamma_\mathcal{J}\rangle\!\rangle\,,\\&
=| \delta_{\mathcal{J}}|^2\,\chi_1(\tilde{q})\,. 
\end{split}
\end{align}
After a modular transformation, we find, 
\begin{align}
\begin{split}
Z(\tilde{q}) &{=}| \delta_{\mathcal{J}}|^2 \Big(\frac{1}{2}\chi_1(q) + \frac{1}{2}\chi_2(q)+\frac{1}{\sqrt{2}}\chi_3(q)\Big)\,.\label{1.22}
\end{split}
\end{align}}
\noindent
\textcolor{black}{At this point, note that the non-integer character coefficients do not signal a failure of the modular transformation or of open–closed duality.
They show that the cloaking state is not a Cardy boundary state and that the transformed expression is not the trace
over a standard open BCFT Hilbert space with integer annulus multiplicities. Remarkably, as we will show below, if (\ref{1.22}) is used to compute the symmetry-resolved partition function via the action of the topological defect lines (TDLs), the resulting resolved entropy reproduces the correct formula as described in \cite{Choi:2024tri} upto the boundary entropy term, which will depend on the exact choice on the normalization factor $ \delta_{\mathcal{J}}\,.$}

\textcolor{black}{The TY fusion category has three elements, and there are three representations (each irreducible block is represented by $\mathcal{R}_\rho,\,\rho=1,2,3$). Hence, following the discussion presented in Sec.~(\ref{sec3}), the reduced density matrix can be decomposed into three blocks. From the Table~(\ref{tabIsing}), we can see the action of $\mathcal{C}$ on each primary is different; therefore, we can write the symmetry-resolved partition functions as,
\begin{align}
\begin{split}
Z(q^n, \mathcal{R}_1) = {\frac{|\delta_{\mathcal{J}}|^2}{2}\chi_1(q^n)}\,,\quad  Z(q^n, \mathcal{R}_2) = {\frac{|\delta_{\mathcal{J}}|^2}{2}\chi_2(q^n)}\,, 
\quad Z(q^n, \mathcal{R}_3) = {\frac{|\delta_{\mathcal{J}}|^2}{\sqrt{2}}\chi_3(q^n)}\,.
\end{split}
\end{align}
Now, using (\ref{eq3}), we can compute the symmetry-resolved $n$th R\'enyi entropy. 
As discussed in (\ref{3.11b}), in the high-energy limit, the leading contribution comes from the vacuum. Therefore,
\begin{eqnarray}
\frac{Z(q^n, \mathcal{R}_1)}{Z^n(q, \mathcal{R}_1)} \overset{\tilde{q} \to 0}{\sim} \left(\frac{|\delta_{\mathcal{J}}|^2}{4}\right)^{1-n} \widetilde{q}^{\,(n-\frac{1}{n})\frac{c}{24}}
\end{eqnarray}
 and SREE for this block becomes
\begin{eqnarray}
S_n(\mathcal{R}_1) = \frac{c}{6}\left(\frac{n+1}{n}\right)\log\left(\frac{L}{\epsilon}\right) + \log \frac{|\delta_{\mathcal{J}}|^2}{4}+ \ldots
\end{eqnarray} 
where the constant term 
\begin{eqnarray}
\log \frac{|\delta_{\mathcal{J}}|^2}{4} = \log\left(\frac{d_{\mathcal{R}_1}^{\,2}}{\sum\limits_{\mathcal{R}_i}d_{\mathcal{R}_i}^{\,2}}\right) + \log(\langle\!\langle\gamma_\mathcal{J}|\!|\Omega\rangle\langle\Omega|\gamma_\mathcal{J}\rangle\!\rangle)\,.
\end{eqnarray}
where, $d_{\mathcal{R}_1}=1$, $\sum\limits_{\mathcal{R}_i} d_{\mathcal{R}_i}^2 = 4$, and $\log(\langle\!\langle\gamma_\mathcal{J}|\!|\Omega\rangle\!\langle\Omega|\!|\gamma_\mathcal{J}\rangle\!\rangle)=|\delta_{\mathcal{J}}|^2$ corresponds to the Affleck-Ludwig boundary entropy \cite{Affleck:1991tk}. Similar result can be obtained for $S_n(\mathcal{R}_2)$, and
\begin{eqnarray}
S_n(\mathcal{R}_3) = \frac{c}{6}\left(\frac{n+1}{n}\right)\log\left(\frac{L}{\epsilon}\right) +  \log \frac{|\delta_{\mathcal{J}}|^2}{2}+ \ldots\,,
\end{eqnarray}
where the constant term
\begin{eqnarray}
\log \frac{|\delta_{\mathcal{J}}|^2}{2} = \log\left(\frac{d^{\, 2}_{\mathcal{R}_3}}{\sum\limits_{\mathcal{R}_i}d^{\, 2}_{\mathcal{R}_i}}\right) + \log(\langle\!\langle\gamma_\mathcal{J}|\!|\Omega\rangle\langle\Omega|\!|\gamma_\mathcal{J}\rangle\!\rangle)\,.
\end{eqnarray}
Here $d_{\mathcal{R}_3}=\sqrt{2}$ is the quantum dimension of the defect line $\mathcal{L}_\sigma$. We observe that equation \eqref{3.24k} is satisfied in this example when using the specified boundary states, known as the cloaking states. Through a detailed examination of this explicit case, it becomes clear that symmetry resolution in the Ising CFT with respect to the TY fusion category cannot be achieved using the standard procedure, as no weakly symmetric state exists in this case. However, by employing the cloaking state, we demonstrate that such a resolution becomes feasible. To the best of our knowledge, this work presents the first TY-resolution of the Ising CFT in the literature based solely on RCFT technique.}\par
\textcolor{black}{Before we end, we note that the probabilities associated with individual charge sectors remain unchanged irrespective of the choice of the normalization $\delta_J$, but the precise value of the boundary entropy depends on it. In summary, to fix the boundary entropy without any ambiguity, one needs to independently compute the normalization constant by following \cite{Brehm:2022JPA,Brehm:2024arXiv}.
In summary, although the complete RCFT boundary realization of this cloaking state remains to be understood, the known action of RCFT topological defect lines on the cloaking state is sufficient for symmetry resolution and computing normalized sector weights. We leave its detailed BCFT construction for future work.}

\subsection{Symmetry resolution for $\mathcal{M}(6,5)$}
We now analyze the symmetry resolution of the diagonal $\mathcal{M}(6,5)$ theory. Interestingly, in this case, there are two natural choices of sub-fusion categories, which we treat separately. For the larger sub-fusion category we find that no weakly symmetric boundary state exists, but we can still perform the resolution using an appropriate cloaking boundary state. In contrast, for the smaller sub-fusion category, a weakly symmetric boundary state is available, and the symmetry resolution proceeds accordingly.
\noindent
\\\\
\noindent\textbf{Tetracritical Ising model.} 
The diagonal minimal model corresponding to the modular data of $\mathcal{M}(6,5)$, the tetracritical Ising model \cite{Chang:2018iay, Choi:2023PRD}, is another example of a diagonal RCFT. The model $\mathcal{M}(6,5)$ has central charge $c=\tfrac{4}{5}$ and admits ten simple Verlinde lines, which we label as,
\begin{align}
\begin{split}
 & \textcolor{black}{{I}\equiv\mathcal{L}_{\Phi_{1,1}},\; C\equiv\mathcal{L}_{\Phi_{1,5}},\; M\equiv\mathcal{L}_{\Phi_{1,3}},\; W\equiv\mathcal{L}_{\Phi_{2,5}},\; MW\equiv\mathcal{L}_{\Phi_{2,3}},\; }\\&\textcolor{black}{CW\equiv\mathcal{L}_{\Phi_{2,1}},\;  N\equiv\mathcal{L}_{\Phi_{1,2}},\; CN\equiv\mathcal{L}_{\Phi_{1,4}},\; WN\equiv\mathcal{L}_{\Phi_{2,4}},\; CWN\equiv\mathcal{L}_{\Phi_{2,2}}\,.}
    \end{split}
\end{align}
with identification $\Phi_{r,s}\sim \Phi_{5-r,\, 6-s}$.
Among these lines the subset $\{I,\,C,\,N,\,CN,\,M\}$ closes into a TY-type sub-fusion with the nontrivial fusion rules\footnote{In contrast, for the Potts CFT, one of the simple TDLs is $N$, satisfying $N \times N = 1 + \eta + \bar{\eta}$, which is a $\mathbb{Z}_3$-TY line. In the diagonal theory, however, we instead encounter the composite line $M = \eta + \bar{\eta}$ with $N \times N = I + M$, which is not strictly of TY type.}
\begin{align}
    C^2=I, \qquad
    M^2=I+M+C, \qquad
    N^2=I+M.
\end{align}
We study symmetry resolution with respect to these five simple TDLs. The full fusion category contains ten primaries,
\begin{align}
  {\color{black}{  \mathbb{I}\equiv\Phi_{1,1},\; Y\equiv\Phi_{1,5},\;Z\equiv\Phi_{1,3},\;   X\equiv\Phi_{2,5},\; \sigma\equiv\Phi_{2,3},\; \varepsilon\equiv\Phi_{2,1},\;\Phi_{4,4},\; \Phi_{2,2},\; \Phi_{4,2},\; \Phi_{3,2}\,.}}
\end{align}
To identify the irreducible charge blocks, one first determines the action of the TDLs on these primaries. This action is displayed in the Table~(\ref{tabm}).

\begin{table}[H]
\centering
\begin{tabular}{lcccccccccc}
\toprule
Defect line & $\mathbb{I}$ & $Y$ & $Z$ & $X$ & $\sigma$  & $\varepsilon$ & $\Phi_{4,4}$ & $\Phi_{2,2}$ & $\Phi_{4,2}$ & $\Phi_{3,2}$ \\
\midrule
$I$   & $1$ & $1$ & $1$ & $1$ & $1$ & $1$ & $1$ & $1$ & $1$ & $1$  \\
$C$   & $1$ & $1$ & $1$ & $1$ & $1$  & $1$  & $-1$ & $-1$ & $-1$ & $-1$ \\
$M$   & $2$ & $2$ & $-1$ & $2$ & $-1$ & $2$ & $0$ & $0$ & $0$ & $0$ \\
$N$   & $\sqrt{3}$ & $-\sqrt{3}$ & $0$ & $\sqrt{3}$ & $0$ & $-\sqrt{3}$ & $-1$ & $1$ & $1$ & $-1$ \\
$CN$  & $\sqrt{3}$ & $-\sqrt{3}$ & $0$ & $\sqrt{3}$ & $0$ & $-\sqrt{3}$ & $1$ & $-1$ & $-1$ & $1$ \\
\bottomrule 
\end{tabular}
\caption{Action of the TDLs on the $\mathcal{M}(6,5)$ primaries of the tetracritical Ising model.}
\label{tabm}
\end{table}
\noindent
From Table~(\ref{tabm}), one sees that there are five irreducible blocks, each containing two primaries. In the absence of symmetric (weakly symmetric) boundary states, we choose the cloaking boundary, which we take to be the identity Ishibashi state,
\(
\|\gamma_\mathcal{J}\rangle\!\rangle=\|\mathbb{I}\rangle\!\rangle.
\)
\noindent
The annulus partition function in the closed-string channel is
\begin{align}
\begin{split}
    Z(\tilde q)
    &=\langle\!\langle\gamma_\mathcal{J}\|\tilde{q}^{\color{black}{\frac{1}{2}(\,L_{0}+\bar{L}_0-\frac{c}{12})}}\|\gamma_\mathcal{J}\rangle\!\rangle
    =|\delta_{\mathcal{J}}|^2\chi_{\mathbb{I}}(\tilde q)\,, \\
    &{=}|\delta_{\mathcal{J}}|^2\Bigg(\sqrt{\frac{1}{15}\!\left(\frac{5}{8}-\frac{\sqrt{5}}{8}\right)}\chi_{\mathbb{I}}(q)
      +\sqrt{\frac{1}{15}\!\left(\frac{5}{8}-\frac{\sqrt{5}}{8}\right)}\chi_{Y}(q)
      +\sqrt{\frac{1}{15}\!\left(\frac{\sqrt{5}}{8}+\frac{5}{8}\right)}\chi_{\varepsilon}(q)\\
    &\quad+\sqrt{\frac{1}{15}\!\left(\frac{\sqrt{5}}{8}+\frac{5}{8}\right)}\chi_{X}(q)
      +2\sqrt{\frac{1}{15}\!\left(\frac{5}{8}-\frac{\sqrt{5}}{8}\right)}\chi_{Z}(q)
      +2\sqrt{\frac{1}{15}\!\left(\frac{\sqrt{5}}{8}+\frac{5}{8}\right)}\chi_{\sigma}(q)\\
    &\quad+\sqrt{\frac{1}{5}\!\left(\frac{5}{8}-\frac{\sqrt{5}}{8}\right)}\chi_{4,4}(q)
      +\sqrt{\frac{1}{5}\!\left(\frac{5}{8}-\frac{\sqrt{5}}{8}\right)}\chi_{4,2}(q)\\
    &\quad+\sqrt{\frac{1}{5}\!\left(\frac{\sqrt{5}}{8}+\frac{5}{8}\right)}\chi_{2,2}(q)
      +\sqrt{\frac{1}{5}\!\left(\frac{\sqrt{5}}{8}+\frac{5}{8}\right)}\chi_{3,2}(q)\,\Bigg).
\end{split}
\end{align}
\noindent
The leading contribution to the block partition functions in the $\tilde q\to 0$ limit are given by
\begin{align}
\begin{split}
  &Z(q^n,\mathcal{R}_1)\sim \tfrac{|\delta_{\mathcal{J}}|^2}{12}\,\chi_{\mathbb{I}}(\tilde q^{1/n}),\qquad
  Z(q^n,\mathcal{R}_2)\sim \tfrac{|\delta_{\mathcal{J}}|^2}{12}\,\chi_{\mathbb{I}}(\tilde q^{1/n}),\\
  &Z(q^n,\mathcal{R}_3)\sim \tfrac{|\delta_{\mathcal{J}}|^2}{3}\,\chi_{\mathbb{I}}(\tilde q^{1/n}),\qquad
  Z(q^n,\mathcal{R}_4)\sim \tfrac{|\delta_{\mathcal{J}}|^2}{4}\,\chi_{\mathbb{I}}(\tilde q^{1/n}),\qquad
  Z(q^n,\mathcal{R}_5)\sim \tfrac{|\delta_{\mathcal{J}}|^2}{4}\,\chi_{\mathbb{I}}(\tilde q^{1/n})\,.
\end{split}
\end{align}
\noindent
Therefore, the SREE for the $\rho$-th irreducible block is
\begin{align}
\begin{split}
    S_{n}(\mathcal{R}_\rho)
    &=\frac{1}{1-n}\log\!\left(\frac{Z(q^n,\mathcal{R}_\rho)}{Z^n(q,\mathcal{R}_\rho)}\right),
    \quad\text{with}\quad
    \frac{Z(q^n,\mathcal{R}_\rho)}{Z^n(q,\mathcal{R}_\rho)}
    \xrightarrow{\tilde q\to 0}
    \left(p_\rho|\delta_{\mathcal{J}}|^2\right)^{\,1-n}\,\tilde q^{\big(n-\frac{1}{n}\big)\frac{c_{\rm eff}}{24}},
\end{split}
\end{align}
where the coefficients are
\(
p_\rho=\tfrac{1}{12},\tfrac{1}{12},\tfrac{1}{3},\tfrac{1}{4},\tfrac{1}{4}
\)
for the five irreducible blocks, respectively. In the cloaking (identity Ishibashi) boundary choice, this yields the compact form as \eqref{3.24k},
\begin{align}
   {S}_{n}(\mathcal R_\rho)
   = \frac{c_{\rm eff}}{6}\left(\frac{n+1}{n}\right)\log\!\left(\frac{L}{\epsilon}\right)
   + \log(g^2)
   + \log\!\left(\frac{d_{\rho}^2}{\sum\limits_{\sigma} d_{\sigma}^2}\right)+\cdots, \,\textrm{with,}\,\, g^2=|\delta_{\mathcal{J}}|^2
   \label{5.25k}
\end{align}
with the quantum dimensions for the five lines given by
\[
d_{I}=1,\quad d_{M}=2,\quad d_{C}=1,\quad d_{N}=\sqrt{3},\quad d_{CN}=\sqrt{3}\,.
\]
Hence
\begin{align}
\sum_{\rho} d_{\rho}^2 = 1^2 + 2^2 + 1^2 + (\sqrt{3})^2 + (\sqrt{3})^2 = 12\,,
\end{align}
which reproduces the numerical prefactors $p_\rho=\tfrac{1}{12},\tfrac{1}{12},\tfrac{1}{3},\tfrac{1}{4},\tfrac{1}{4}$ quoted above for the five irreducible blocks. 
\\\\
\noindent\textbf{Resolution with weakly symmetric boundary state:}\\
As observed above,  symmetric boundary states do not exist within the sub-fusion sector under consideration. However, one can identify a \emph{weakly symmetric} boundary state by resolving the minimal model $\mathcal{M}(6,5)$ with respect to the smaller subset of Verlinde lines $\{I,\,C,\,M\}$. From the fusion algebra restricted to this subset, it is immediate that the boundary state $\|M\rangle\!\rangle$ satisfies the weak symmetry condition. Consequently, the annulus partition function becomes
\begin{align}
    Z(\tilde q)
    = \langle\!\langle M\|\;\tilde q^{\color{black}{\frac{1}{2}(\,L_{0}+\bar{L}_0-\frac{c}{12})}}\;\|M\rangle\!\rangle
    = \chi_{\mathbb{I}}(q) + \chi_{Y}(q) + \chi_{Z}(q)\,.
\end{align}
The action of the topological defect lines shows that there are three irreducible blocks. Their annulus block partition functions in the $\tilde q\to 0$ limit behave as
\begin{align}
\begin{split}
    Z(q^n,\mathcal{R}_1) &\sim \tfrac{1}{2}\sqrt{\tfrac{1}{30}\!\left(5-\sqrt{5}\right)}\;\chi_{\mathbb{I}}(\tilde q^{1/n}),\,\,\,\,\,\,
    Z(q^n,\mathcal{R}_2)\sim \tfrac{1}{2}\sqrt{\tfrac{1}{30}\!\left(5-\sqrt{5}\right)}\;\chi_{\mathbb{I}}(\tilde q^{1/n})\,,\\&
   \hspace{2 cm} Z(q^n,\mathcal{R}_3) \sim \sqrt{\tfrac{1}{30}\!\left(5-\sqrt{5}\right)}\;\chi_{\mathbb{I}}(\tilde q^{1/n})\,.
\end{split}
\end{align}
The boundary entropy for this choice is
\begin{align}
    \log(g^2)
    = \log\big|\langle\!\langle M\|\,\mathbb{I}\rangle\!\rangle\big|^2
    = \log\!\sqrt{\tfrac{2}{15}\!\left(5-\sqrt{5}\right)}.
\end{align}
With the weakly symmetric boundary $\|M\rangle\!\rangle$ the SREE for the block $\rho$ again reads \eqref{3.15k},
\begin{align}
\begin{split}
     S_{n}(\mathcal R_\rho)
    &= \frac{c_{\rm eff}}{6}\left(\frac{n+1}{n}\right)\log\!\left(\frac{L}{\epsilon}\right)
    + \log(g^2)
    + \log\!\left(\frac{d_{\rho}\,N_{\rho B_1}^{B_2}}{d(B_1\otimes B_2)}\right)+\cdots,
\end{split}
\end{align}
where, for the lines under consideration,
\begin{align}
    d_{I}=1,\quad d_{C}=1,\quad d_{M}=2,
\end{align}
and the relevant fusion coefficients are
\begin{align}
    N_{\mathbb{I},Z}^{\,Z}=1,\qquad {\color{black}{N_{Y,Y}^{\,Z}=0}},\qquad N_{Z,Z}^{\,Z}=1.
\end{align}
Therefore, the SREE for the $\rho$-th irreducible block is
\begin{align}
\begin{split}
    S_{n}(\mathcal{R}_\rho)
    &=\frac{1}{1-n}\log\!\left(\frac{Z(q^n,\mathcal{R}_\rho)}{Z^n(q,\mathcal{R}_\rho)}\right),
    \quad\text{with}\quad
    \frac{Z(q^n,\mathcal{R}_\rho)}{Z^n(q,\mathcal{R}_\rho)}
    \xrightarrow{\tilde q\to 0}
    p_\rho^{\,1-n}\,\tilde q^{\big(n-\frac{1}{n}\big)\frac{c_{\rm eff}}{24}},
\end{split}
\end{align}
where the coefficients are
\(
p_\rho=\tfrac{1}{6},\tfrac{1}{6},\tfrac{1}{3}
\)
for the three irreducible blocks respectively.

\noindent
\noindent

\noindent
\subsection{Symmetry resolution in non-diagonal RCFTs}
Before turning to non-unitary RCFTs, we discuss two non-diagonal examples. The $\frac{1}{2}E_6$ example illustrates that the closed-channel defect eigenvalues and the NIM-rep spectrum need not determine the action of the symmetry generators on the complete open-channel Hilbert space. The three-state Potts model provides a contrasting example in which the $\mathbb{Z}_3$ action is known and the usual RCFT symmetry resolution \cite{Haghighat:2023sax} can be performed. \\

\subsubsection*{$E_6$ and its  $\tfrac{1}{2}E_6$ symmetry resolution?}
We consider $SU(2)_{10}$ {WZW model} of $E_6$-type \cite{DiFrancesco:1997nk}. The torus partition function is given by,
\begin{align}
    Z(\tau)=|\chi_0+\chi_{3}|^2+|\chi_{\frac{3}{2}}+\chi_{\frac{7}{2}}|^2+|\chi_{2}+\chi_{5}|^2
\end{align}
where, $\chi_{j}(\tau)$ is spin $j$ $SU(2)$ affine character. The modular S-matrix for $SU(2)_{10}$ is a $11\times 11$ matrix which can be constructed by the following,
\begin{align}
    S_{jj'}=\sqrt{\frac{2}{k+2}}\sin\left(\frac{(2j+1)(2j'+1)\pi}{k+2}\right),\,\textrm{with,}\,\,k=10,\,\,\textrm{and,}\, j,j'=0,\frac{1}{2},\cdots, 5\,.
\end{align}
The $E_6$ fusion ring consists of six simple TDLs: $\mathcal{L}_1,\mathcal{L}_2,\cdots,\mathcal{L}_6$, with the following fusion rules.
\begin{align}
\begin{split}
&{\color{black}{\mathcal{L}_2\times \mathcal{L}_2=\mathcal{L}_1+\mathcal{L}_3}},\qquad
{\color{black}{\mathcal{L}_3\times\mathcal{L}_2=\mathcal{L}_2+\mathcal{L}_4+\mathcal{L}_6}},\qquad
\mathcal{L}_3 \times \mathcal{L}_3
=
\mathcal{L}_1 + 2\,\mathcal{L}_3 + \mathcal{L}_5,
 \\
&{\color{black}{\mathcal{L}_4\times\mathcal{L}_2=\mathcal{L}_3+\mathcal{L}_5}},\qquad
\mathcal{L}_4 \times \mathcal{L}_3
=
\mathcal{L}_2 + \mathcal{L}_4 + \mathcal{L}_6,
\qquad
\mathcal{L}_4 \times \mathcal{L}_4
=
\mathcal{L}_1 + \mathcal{L}_3,\\
&{\color{black}{\mathcal{L}_5\times\mathcal{L}_2=\mathcal{L}_4}},\qquad\mathcal{L}_5 \times \mathcal{L}_3
=
\mathcal{L}_3,\qquad
\mathcal{L}_5 \times \mathcal{L}_4
=
\mathcal{L}_2,
\qquad
\mathcal{L}_5 \times \mathcal{L}_5
=
\mathcal{L}_1,
\\
& {\color{black}{\mathcal{L}_6\times\mathcal{L}_2=\mathcal{L}_3}},\qquad
\mathcal{L}_6 \times \mathcal{L}_3
=
\mathcal{L}_2 + \mathcal{L}_4,
\qquad
\mathcal{L}_6 \times \mathcal{L}_4
=
\mathcal{L}_3,\qquad
\mathcal{L}_6 \times \mathcal{L}_5
=
\mathcal{L}_6,
\\
&\mathcal{L}_6 \times \mathcal{L}_6
=
\mathcal{L}_1 + \mathcal{L}_5.
\end{split}
\end{align}
Consequently, the $\frac{1}{2}E_6$ fusion ring consists of three simple TDLs: $\mathcal{L}_1=I,\,\mathcal{L}_{5}=X,\,\mathcal{L}_{3}=Y$. The fusion of the TDLs can be derived from the matrix ($\psi$) constructed from the eigenvector of the adjacency matrix of $E_6$, which can be derived from the Dynkin diagram.
\begin{equation}
\mathcal{A}(E_6)=\left(
\begin{array}{cccccc}
 0 & 1 & 0 & 0 & 0 & 0 \\
 1 & 0 & 1 & 0 & 0 & 0 \\
 0 & 1 & 0 & 1 & 0 & 1 \\
 0 & 0 & 1 & 0 & 1 & 0 \\
 0 & 0 & 0 & 1 & 0 & 0 \\
 0 & 0 & 1 & 0 & 0 & 0 \\
\end{array}
\right)
  \end{equation}
  {The \(\psi\)-matrix is the orthonormal eigenvector matrix of \(\mathcal A(E_6)\). Its row index \(a=1,\ldots,6\) labels graph-boundary nodes, while its column index \(j\in {\rm Exp}(E_6)=\{0,\frac32,2,3,\frac72,5\}\) labels the Ishibashi sectors appearing in the modular invariant. Equivalently,
\begin{equation}
    \sum_{b=1}^{6}\mathcal A(E_6)_{ab}\psi_b^{\,j}
    =
    2\cos\!\left(\frac{(2j+1)\pi}{12}\right)\psi_a^{\,j},
    \qquad
    \sum_{a=1}^{6}\psi_a^{\,j}\psi_a^{\,j'}=\delta_{j j'}.
    \label{E6PsiDefinition}
\end{equation}}
In the ordering of graph nodes and exponents above, this matrix is
\begin{align}
    \psi_{a}^{j}=\left(
\begin{array}{cccccc}
 \alpha & \frac{1}{2} & \beta & \beta & \frac{1}{2} & \alpha \\
 \beta & \frac{1}{2} & \alpha & -\alpha & -\frac{1}{2} & -\beta \\
 \gamma & 0 & -\delta & -\delta & 0 & \gamma \\
 \beta & -\frac{1}{2} & \alpha & -\alpha & \frac{1}{2} & -\beta \\
 \alpha & -\frac{1}{2} & \beta & \beta & -\frac{1}{2} & \alpha \\
 \delta & 0 & -\gamma & \gamma & 0 & -\delta
\end{array}
\right),
    \label{E6PsiMatrix}
\end{align}
where
\[
    \alpha=\frac{1}{2}\sqrt{\frac{3-\sqrt3}{6}},\qquad
    \beta=\frac{1}{2}\sqrt{\frac{3+\sqrt3}{6}},\qquad
    \gamma=\frac{1}{2}\sqrt{\frac{3+\sqrt3}{3}},\qquad
    \delta=\frac{1}{2}\sqrt{\frac{3-\sqrt3}{3}}.
\]
The graph fusion coefficients are obtained from the modified Verlinde formula \cite{DiFrancesco:1997nk, Behrend:1998mu,Behrend:1999bn,  Petkova:2000ip},
\begin{align}
   \widehat{ \mathcal{N}}_{ab}^c=\sum _{j\in {\rm Exp}(E_6)}\frac{\psi_{a}^{j}\,\, \psi_{b}^j\,\, (\psi_c^j)^*}{\psi_1^j}\,.
\end{align}
Here \(\widehat{\mathcal N}_{ab}^{\,c}\) is the graph-fusion coefficient of the TDLs in the A-D-E modular classification. The same \(\psi\)-matrix gives the NIM-rep coefficients
\begin{equation}
    n_{ia}^{\,b}
    =
    \sum_{j\in {\rm Exp}(E_6)}
\frac{\psi_a^{\,j}\left(\psi_b^{\,j}\right)^*S_{ij}}{S_{\mathbf{1}j}},
    \label{E6NIMfrompsi}
\end{equation}
where the \(SU(2)_{10}\) label \(i=0,\ldots,10\) runs over the full  spectrum. {{Equivalently, the corresponding \(E_6\) NIM-rep data are the non-negative integer matrices \(n_i\), \(i=0,\ldots,10\), acting on the six graph-boundary labels and generated by
\begin{equation}
    n_0=\mathbbm 1_6,\qquad n_1=\mathcal A(E_6),\qquad
    n_{i+1}=\mathcal A(E_6)n_i-n_{i-1}\quad (i=1,\ldots,9),
    \qquad n_{11}=0,
    \label{E6NIMrecursion}
\end{equation}
which is the \(SU(2)_{10}\) fusion recursion represented on the \(E_6\) graph. With rows and columns ordered as the graph nodes \((1,\ldots,6)\), the full matrices are
\begin{center}
\resizebox{0.76\textwidth}{!}{$
\begin{array}{@{}c@{\quad}c@{\quad}c@{}}
n_0=\left(\begin{smallmatrix}
1&0&0&0&0&0\\
0&1&0&0&0&0\\
0&0&1&0&0&0\\
0&0&0&1&0&0\\
0&0&0&0&1&0\\
0&0&0&0&0&1
\end{smallmatrix}\right)
&
n_1=\left(\begin{smallmatrix}
0&1&0&0&0&0\\
1&0&1&0&0&0\\
0&1&0&1&0&1\\
0&0&1&0&1&0\\
0&0&0&1&0&0\\
0&0&1&0&0&0
\end{smallmatrix}\right)
&
n_2=\left(\begin{smallmatrix}
0&0&1&0&0&0\\
0&1&0&1&0&1\\
1&0&2&0&1&0\\
0&1&0&1&0&1\\
0&0&1&0&0&0\\
0&1&0&1&0&0
\end{smallmatrix}\right)
\\[1.3em]
n_3=\left(\begin{smallmatrix}
0&0&0&1&0&1\\
0&0&2&0&1&0\\
0&2&0&2&0&1\\
1&0&2&0&0&0\\
0&1&0&0&0&1\\
1&0&1&0&1&0
\end{smallmatrix}\right)
&
n_4=\left(\begin{smallmatrix}
0&0&1&0&1&0\\
0&1&0&2&0&1\\
1&0&3&0&1&0\\
0&2&0&1&0&1\\
1&0&1&0&0&0\\
0&1&0&1&0&1
\end{smallmatrix}\right)
&
n_5=\left(\begin{smallmatrix}
0&1&0&1&0&0\\
1&0&2&0&1&0\\
0&2&0&2&0&2\\
1&0&2&0&1&0\\
0&1&0&1&0&0\\
0&0&2&0&0&0
\end{smallmatrix}\right)
\\[1.3em]
n_6=\left(\begin{smallmatrix}
1&0&1&0&0&0\\
0&2&0&1&0&1\\
1&0&3&0&1&0\\
0&1&0&2&0&1\\
0&0&1&0&1&0\\
0&1&0&1&0&1
\end{smallmatrix}\right)
&
n_7=\left(\begin{smallmatrix}
0&1&0&0&0&1\\
1&0&2&0&0&0\\
0&2&0&2&0&1\\
0&0&2&0&1&0\\
0&0&0&1&0&1\\
1&0&1&0&1&0
\end{smallmatrix}\right)
&
n_8=\left(\begin{smallmatrix}
0&0&1&0&0&0\\
0&1&0&1&0&1\\
1&0&2&0&1&0\\
0&1&0&1&0&1\\
0&0&1&0&0&0\\
0&1&0&1&0&0
\end{smallmatrix}\right)
\\[1.3em]
n_9=\left(\begin{smallmatrix}
0&0&0&1&0&0\\
0&0&1&0&1&0\\
0&1&0&1&0&1\\
1&0&1&0&0&0\\
0&1&0&0&0&0\\
0&0&1&0&0&0
\end{smallmatrix}\right)
&
n_{10}=\left(\begin{smallmatrix}
0&0&0&0&1&0\\
0&0&0&1&0&0\\
0&0&1&0&0&0\\
0&1&0&0&0&0\\
1&0&0&0&0&0\\
0&0&0&0&0&1
\end{smallmatrix}\right)
&
\end{array}
$}
\end{center}
The annulus multiplicity between graph boundaries \(a,b\) is \(n_{ia}^{\,b}=(n_i)_{ab}\). In particular, the \(\tfrac12E_6\) object \(Y=\mathcal L_3\) is the third node, so \(n_{iY}^{\,Y}=(n_i)_{33}\).}}

The exponents select the subset of primaries that are preserved in the $E_6$ modular invariant (more precisely, it selects the primaries that are present in the diagonal block of the multiplicity matrix) partition function, these label the Ishibashi states. Now the fusion of $\frac{1}{2}E_6$ TDLs are given by \cite{hagge2007nonbraidedfusioncategoriesrank, Chang:2018iay},
\begin{align}
    X^2=I,\quad Y^2=I+X+2Y,\quad XY=YX=Y\,.
\end{align}
{Strictly speaking, as we discussed earlier, for this non-diagonal modular invariant the boundary labels are graph-boundary labels rather than Cardy labels in the diagonal Verlinde sense.} The graph boundary $|\!|Y\rangle\!\rangle$ is weakly symmetric with respect to the subfusion $\{I,\,X,\,Y\}$. Now the closed string partition function with this choice of boundary is given by\footnote{$\mathcal{I}$ denotes the full spectrum, which is, for our case, $SU(2)_{10}$.},
\begin{align}
\begin{split}
    Z_{\textrm{closed}}(\tilde q)&=\langle\!\langle Y|\!|\tilde q^{\color{black}{\frac{1}{2}(\,L_{0}+\bar{L}_0-\frac{c}{12})}}|\!|Y\rangle \!\rangle\,,\\ &
    =\sum_{j\in \textrm{Exp}(E_6)}\frac{(\psi_{3}^j)^2}{S_{\mathbf{1}j}}\chi_{j}(\tilde q)=\sum_{\substack{j\in \textrm{Exp}(E_6)\\ i\in \mathcal{I}}}\frac{(\psi_{3}^j)^2}{S_{\mathbf{1}j}}S_{ji} \chi_{i}(q):=\sum_{\substack{ i\in \mathcal{I}}}n_{ia}^a \chi_{i}(q)\,,\\ &
    =\chi_0(q)+2 \chi_1(q)+3 \chi_2(q)+3 \chi_3(q)+2 \chi_4(q)+\chi_{5}(q)=Z_{\textrm{open}}(q) \label{4.31c}
    \end{split}
\end{align}
 {For the $Y|Y$ annulus the corresponding $E_6$ NIM-rep entry is
\begin{equation}
    n_{iY}^{\,Y}=(1,0,2,0,3,0,3,0,2,0,1).
\end{equation}
Equivalently, with \(\chi_j\equiv\chi^{SU(2)_{10}}_{2j}\), this is the open-channel spectrum in \eqref{4.31c}.} {From \eqref{4.31c}, we see that the open-string spectrum associated with the boundary state \(|\!|Y\rangle\!\rangle\) contains SU(2)$_{10}$ primaries that are not present in the \(E_6\) spectrum. Thus, the corresponding open-channel Hilbert space is not contained entirely within the physical spectrum of the \(E_6\) theory. Consequently, the intrinsic \(\tfrac12 E_6\) fusion ring does not act diagonally by scalar eigenvalues on all open-channel primaries.}{ Indeed, the usual scalar action is defined only on the exponents of \(G\):
\begin{align}
\mathcal L_a \phi_j
=
\frac{\psi_a^j}{\psi_1^j}\phi_j,
\qquad
\mathcal L_i \phi_j
=
\frac{S_{ij}}{S_{\mathbf{1}j}}\phi_j,
\qquad
j\in \mathrm{Exp}(E_6),\; i\in\mathcal I . \label{4.38}
\end{align}
For primaries outside \(\mathrm{Exp}(E_6)\), however, the action is not captured
by these scalar eigenvalues. Instead, one generally obtains a matrix action,
\begin{align}
\mathcal L_a \phi_i
=
\sum_j (D_a)_{i}{}^{j}\phi_j,
\qquad
i\notin \mathrm{Exp}(E_6), \label{4.39}
\end{align}
which depends on the embedding into the full SU(2)$_{10}$ theory. Therefore a symmetry resolution based only on the intrinsic \(\tfrac12 E_6\) fusion data is not canonical. One may still decompose the open spectrum using the full defect algebra, but this decomposition is no longer the usual diagonal resolution by scalar projectors intrinsic to the \(E_6\) theory. In this sense, for the extra open-channel primaries, the \(\tfrac12 E_6\) data alone do not define a symmetry resolution in the standard sense.}

\textcolor{black}{However, note that the appearance of these additional primaries does not point towards any inconsistency. In a non-diagonal RCFT, the open-channel spectrum need not coincide with the set of closed-channel Ishibashi exponents. The difficulty is instead that the $\tfrac{1}{2}E_6$ defect eigenvalues in \eqref{4.38} specify the TDL action only on the primaries of $E_6$ appearing in the open-channel partition function \eqref{4.31c}. They do not determine the matrices $D_a$ in \eqref{4.39}, which describe the action on the remaining primaries and their multiplicity spaces. Consequently, the modular invariant, the $\tfrac{1}{2}E_6$ fusion rules, and the $E_6$ NIM-rep determine the $Y|Y$ annulus spectrum, but they are not sufficient to construct the projectors that would resolve the complete open-channel Hilbert space according to the action of the $\tfrac{1}{2}E_6$ TDLs. We therefore do not assign $\tfrac{1}{2}E_6$ sector probabilities or sector-resolved entropies in
this example. In particular, we do not interpret a decomposition of the boundary fusion rule
\[
Y^2=I+X+2Y
\]
as a conventional symmetry resolution of the full RCFT annulus Hilbert space. This conclusion does not imply that a half-$E_6$ resolution cannot exist. The TQFT based prescription of \cite{Choi:2024wfm} can formally define the corresponding sector projectors. To evaluate those projectors as RCFT operators, however, one must additionally know their action on all the
$SU(2)_{10}$ primaries and multiplicity spaces appearing in the $Y|Y$ annulus. This requires theory-dependent boundary-junction data, including the relevant boundary $\widetilde F$-symbols \cite{Choi:2024wfm, Choi:2024tri}, which are not determined by the modular invariance, fusion rules, and NIM-rep alone, and their computation will require a separate study and is beyond the scope of the present work. }

\textcolor{black}{Thus, the $\tfrac{1}{2}E_6$ example identifies a precise limitation of the direct RCFT construction. The conventional symmetry resolution requires the action of the symmetry-generating TDLs on the complete open-channel Hilbert space. When this action is not known, the resolution cannot be completed from the available RCFT data. Determining the missing action and comparing it with the formal TQFT projectors is an interesting problem that we leave for future work. This limitation does not affect the diagonal examples considered in this paper or the non-diagonal three-state Potts example discussed next, for which the required open-channel symmetry action is known.}
\noindent
\subsubsection*{Three-state Potts and its $\mathbb{Z}_3$ resolution}
In three state Potts model, the $\mathbb{Z}_3$ fusion ring consists of three simple TDLs: $I, \,\eta,\,\eta^2$. The fusion of these TDLs is given by,
\begin{align}
    \eta\times \eta=\eta^2,\,\eta\times \eta^2=I,\,\eta^2\times \eta^2=\eta\,.\label{4.44y}
\end{align}
As we observe, there is no 
$\mathbb{Z}_3$‑symmetric boundary with respect to the fusion ring \eqref{4.44y}. By definition, symmetric boundaries in the group‑like case require that the state obtained after the action of defect lines coincides with the original state.\footnote{The action of a defect on boundaries coincides with the fusion ring structure of the primaries.} However, one can enlarge the fusion ring to what known as the Tambara–Yamagami $\mathbb{Z}_3$, which includes an additional simple object $N$. This extended fusion ring admits the following fusion rules \cite{Haghighat:2023sax}:
\begin{align}
    N\times N=I+\eta+\eta^2,\,N\times \eta=N,\,N\times \bar \eta=N\,.
\end{align}
{As we see, $N$ can be treated as a \textit{strongly symmetric boundary} with respect to $\mathbb{Z}_3$ fusion ring,} and the annulus open string partition function is given by \cite{Affleck:1998nq, Behrend:1998mu, Behrend:1999bn}, 
\begin{align}
\begin{split}
    Z(\tilde q)&=\langle\!\langle N|\!|\tilde q^{\color{black}{\frac{1}{2}(\,L_{0}+\bar{L}_0-\frac{c}{12})}}|\!|N\rangle\!\rangle\,,\\ &
    =Z(q)=\chi_1(q)+\chi_{Y}(q)+2\chi_{Z}(q)\,.
    \end{split}
\end{align}
As we see, unlike the $E_6$ case, the open string spectrum is solely in the Potts spectrum. Now, we study symmetry resolution with respect to the three ($I,\,\eta,\,\eta^2$) invertible simple TDLs.
\begin{table}[h!]
\centering
\setlength{\tabcolsep}{12pt}
\renewcommand{\arraystretch}{1.2}
\begin{tabular}{c|cccc}
  & $I$ & $Y$ & $Z$ & $Z^*$ \\
\hline
$I$ & 1 & 1 &1 &1 \\
$\eta$ & $1$ & $1$ & $\omega$ & $\omega^2$ \\
$\eta^2$ & $1$ & $1$ & $\omega^2$ & $\omega$  \\
\end{tabular}
\caption{Action of TDLs on the open string spectrum primaries \cite{Sinha:2023hum, Gu:2023yhm}.}
\label{tab3}
\end{table}
From Table~\eqref{tab3}, we see that there are three irreducible blocks corresponding to the three simple TDLs: $(I,Y),\,Z,\,Z^*$.
The leading contributions to the block partition functions in the $\tilde q\to 0$ limit are
\begin{align}
\begin{split}
  &Z(q^n,\mathcal{R}_1)\sim \sqrt{\tfrac{1}{30} (5-\sqrt{5})}\,\chi_{\mathbb{I}}(\tilde q^{1/n}),\quad
  Z(q^n,\mathcal{R}_2)\sim \sqrt{\tfrac{1}{30} (5-\sqrt{5})}\,\chi_{\mathbb{I}}(\tilde q^{1/n}),\\& \hspace{3 cm} Z(q^n,\mathcal{R}_3)\sim \sqrt{\tfrac{1}{30} (5-\sqrt{5})}\chi_{\mathbb{I}}(\tilde q^{1/n})\,.
\end{split}
\end{align}
Therefore, the symmetry resolved Rényi entropy for the $\rho$-th irreducible block is 
\begin{align}
\begin{split}
    S_{n}(\mathcal{R}_\rho)
    &=\frac{1}{1-n}\log\!\left(\frac{Z(q^n,\mathcal{R}_\rho)}{Z^n(q,\mathcal{R}_\rho)}\right),
    \quad\text{with}\quad
    \frac{Z(q^n,\mathcal{R}_\rho)}{Z^n(q,\mathcal{R}_\rho)}
    \xrightarrow{\tilde q\to 0}
    p_\rho^{\,1-n}\,\tilde q^{\big(n-\frac{1}{n}\big)\frac{c_{\rm eff}}{24}},
\end{split}
\end{align}
where the coefficients are
\(
p_\rho=\tfrac{1}{3}, \forall\rho
\).
Then it yields the following for SREE  \eqref{3.17j},
\begin{align}
\begin{split}
   {S}(\mathcal R_\rho)
  & = \frac{c}{3}\log\!\left(\frac{L}{\epsilon}\right)
   + \log(g^2)
   + \log\!\left(\frac{d_{\rho}}{\sum d_{\rho}}\right) + \ldots, \,\textrm{with,}\,\, g^2=3 \sqrt{\frac{1}{30}(5-\sqrt{5})},\,d_{\rho}=1\\ &
   \to  \frac{c}{3}\log\!\left(\frac{L}{\epsilon}\right)
   + \log(g^2)
   + \log\!\left(\frac{1}{|G|}\right) + \ldots, \,|G|=\textrm{Cardinality of $\mathbb{Z}_3\,.$}
   \end{split}
\end{align}
The quantum dimensions for the three defect lines are $d_\rho=1$. In \cite{Capizzi:2021kys}, the authors have obtained the same result using the form factor bootstrap method, {utilizing the correlators of twist-field operators to compute the charged moment}.
\section{SREE in Non-Unitary Rational CFT}\label{sec5}
As we already discussed in the introduction, symmetry resolution has mainly been developed in unitary settings, i.e, for unitary fusion categories, modular data, and RCFTs. The non-unitary case remains largely unexplored. By bulk–boundary correspondence, unitary TQFTs admit unitary boundary RCFT  \cite{Witten:1988hf, Moore:1988qv, Moore:1989vd, Dijkgraaf:1989pz, Verlinde:1988sn, Fuchs:2009iz}, and extending this logic beyond unitarity is a natural question \cite{Gang:2023rei}. Recent work shows that topological twists of certain $3d$ $\mathcal{N}=4$ rank-zero (S-fold) SCFTs yield non-unitary TQFTs with modular data generalizing the Haagerup–Izumi constructions \cite{Gang:2023ggt,Gang:2022kpe}.
 Furthermore, modular data of non-unitary rational CFTs have been constructed using bulk-boundary correspondence and used the method of \cite{Bantay:2005vk, Bantay:2007zz}\footnote{For other interesting applications of this method please refer to \cite{Govindarajan:2025jlq, Govindarajan:2025rgh}.} to compute the characters\footnote{One other method to compute characters of a rational CFT is known as MMS approach, by analyzing modular invariant linear differential equation (MLDE) \cite{Mathur:1988na, Mathur:1988gt, Naculich:1988xv}. For more details, readers are further referred to \cite{Hampapura:2015cea, Das:2021uvd, Chandra:2018pjq}.  }.
We begin our analysis by examining the simplest non-unitary RCFT, known as the Lee-Yang minimal model. 


\textcolor{black}{Before proceeding further, we would like to make a few comments regarding the definition of the entanglement entropy for a non-unitary theory. For a normalized diagonalizable non-Hermitian reduced density operator for a given symmetry sector,
\[
\widehat\rho_{A,\rho}|r_a\rangle=\omega_a|r_a\rangle,
\qquad
\langle l_a|\widehat\rho_{A,\rho}
=\omega_a\langle l_a|,
\qquad
\langle l_a|r_b\rangle=\delta_{ab}\,.
\]
Then, writing the right and left eigenvectors as the columns of $R$ and $L$ gives
$L^\dagger R=\mathbf 1$ and
\[
\widehat\rho_{A,\rho}
=R\,{\rm diag}(\omega_a)L^\dagger .
\]
Now, \cite{Tu:2021xje} defines the following,
\[
|\widehat\rho_{A,\rho}|_{\rm sp}
=R\,{\rm diag}(|\omega_a|)L^\dagger ,
\]
and hence
\begin{equation}
S^{\rm gen}_{n,\rho}
=\frac{1}{1-n}
\log\!\sum_a\omega_a|\omega_a|^{\,n-1}\,. \label{eqq}
\end{equation}
Here $|\widehat\rho|_{\rm sp}$ denotes the modulus of its eigenvalues in a biorthogonal basis. It is not generally
$\sqrt{\widehat\rho^\dagger\widehat\rho}$. This definition is included only to distinguish the possible non-Hermitian prescriptions and does not enter our calculations. We also refer the readers to \cite{Fossati:2023zyz} for the symmetry resolution for non-Hermitian CFT. All our previous results follow from the annulus replica ratio in (\ref{eq3}) and are therefore unchanged. Moreover, for the ground state primary, the conformal dimension $\Delta_{\Omega}\leq 0$ changes the dominant modular contribution and yields $c_{\rm eff}=c-24\Delta_\Omega\,,$ but does not by itself imply negative eigenvalues or justify the prescription in (\ref{eqq}), although a negative norm squared ground state may appear for some non-unitary RCFT.}

\subsection{Lee-Yang Minimal Model $\mathcal{M}(5, 2)$: a simplest example}
The Lee-Yang minimal model is the simplest non-unitary two-primary rational CFT with central charge $c=-\frac{22}{5}$ {{and conformal dimension $h=-\frac{1}{5}$}}. This model possesses the Fibonacci fusion category symmetry, an example of non-invertible symmetry. The non-trivial fusion of the primary is given by 
\begin{eqnarray}
\Phi_2\times \Phi_2 = \Phi_1 + \Phi_2
\end{eqnarray}
where $\Phi_1$ corresponds to the vacuum state and $\Phi_2$ corresponds to a non-vacuum state, in this case, the non-vacuum state is the ground state $|\Omega\rangle$. {In unitary conformal field theories, the vacuum state coincides with the ground state, whereas in non‑unitary theories the vacuum need not be the ground state}. It is a diagonal RCFT, and its modular S-matrix and characters are given by
\begin{eqnarray}
&&S = \frac{2}{\sqrt{5}}\begin{pmatrix}
-\sin(\frac{2\pi}{5}) & \sin(\frac{4\pi}{5}) \\
\sin(\frac{4\pi}{5}) & \sin(\frac{2\pi}{5})\,
\end{pmatrix},\\
&&\chi_1(\tau) = q^{\frac{11}{60}}\left(1+q^2+q^3+\ldots\right)\,,\\
&&\chi_2(\tau) = q^{-\frac{1}{60}}\left(1+q+q^2+\ldots\right)\,. 
\end{eqnarray} 
\noindent
We now turn to the study of symmetry‑resolved entanglement entropy in the Fibonacci fusion category for this non‑unitary minimal model, employing the techniques of Boundary Conformal Field Theory (BCFT) discussed earlier. The weakly symmetric Cardy boundary state, together with the annulus partition function in the open‑string channel, are given by:
\begin{eqnarray}
&&|\Phi_2\rangle = -i\left(1-\frac{2}{\sqrt{5}}\right)^{\frac{1}{4}} |1\rangle\!\rangle + \frac{\sqrt{5+3 \sqrt{5}}  }{\sqrt[4]{10
   \left(5+\sqrt{5}\right)}} |2\rangle\!\rangle, \\
&&Z(q) = \chi_1(q) + \chi_2(q)\,.   
\end{eqnarray} 
The Fibonacci fusion category $\mathcal{C}=\{\mathcal{L}_{\Phi_1}, \mathcal{L}_{\Phi_2}\}$ has two elements; there are two irreps. Therefore, the reduced density matrix has two diagonal blocks. The action of $\mathcal{C}$ on the primaries is given in the  Table~(\ref{Tabb}).
\begin{table}[H]
\centering
\begin{tabular}{ccc}
\toprule
Defect line & $\Phi_1$ & $\Phi_2$ \\
\midrule
$\mathcal{L}_{\Phi_1}$ & $1$ & $1$ \\
$\mathcal{L}_{\Phi_2}$ & $\frac{1-\sqrt{5}}{2}$ & $\frac{1+\sqrt{5}}{2}$\\
\bottomrule 
\end{tabular}
\caption{Action of the Fibonacci category on the primaries.} \label{Tabb}
\end{table} 
\noindent
This action distinguishes two sectors. Therefore, the partition function is fully resolved, and in each block, there is a contribution from a single character. So, the symmetry-resolved partition functions are
\begin{align}
\begin{split}
    &Z(q^n, \mathcal{R}_1) = {\chi_1(q^n)}, \,\,\,\,\,\,\,\,\,
Z(q^n, \mathcal{R}_2) = {\chi_2(q^n)}\,.
\end{split}  
\end{align}

We can now proceed to compute the symmetry‑resolved entanglement entropy (SREE). Since this is a non‑unitary rational conformal field theory, the leading contribution arises from the ground state rather than from the vacuum. In this example, {$|\Omega\rangle=|\Phi_2\rangle\!\rangle$} is a ground state. Therefore,
\begin{eqnarray}
\frac{Z(q^n, \mathcal{R}_1)}{Z^n(q, \mathcal{R}_1)} \overset{\tilde{q} \to 0}{\sim} \left(\sqrt{\frac{5-\sqrt{5}}{10}}\right)^{1-n} \widetilde{q}^{\,(n-\frac{1}{n})\frac{c_{\rm eff}}{24}}
\end{eqnarray}
 and we get the SREE for the block $\mathcal{R}_1$ as
\begin{eqnarray}
S(\mathcal{R}_1) = \frac{c_{\rm eff}}{3}\log\left(\frac{L}{\epsilon}\right) + \frac{1}{2}\log \frac{5-\sqrt{5}}{10}+ \ldots
\end{eqnarray} 
where the constant term \footnote{For non-unitary theory, bra and ket are not complex conjugate of  each other in general. But, in our case it does and our results matches with those in \cite{Fukusumi:2025xrj}. We thank Yoshiki Fukusumi for pointing out {this} connection to us.}
\begin{eqnarray}
\frac{1}{2}\log \frac{5-\sqrt{5}}{10} = \log\left(\frac{d_{\mathcal{R}_1}}{\sum\limits_{\mathcal{R}_i}d_{\mathcal{R}_i}}\right) + \log(\langle\!\langle\Phi_2|\!|\Omega\rangle\!\langle\Omega|\!|\Phi_2\rangle\!\rangle)\,.
\end{eqnarray}
Here $d_{\mathcal{R}_1}=1$, $\sum\limits_{\mathcal{R}_i} d_{\mathcal{R}_i} = \frac{3+ \sqrt{5}}{2}$, and $\log(\langle\!\langle\Phi_2|\!|\Omega\rangle\!\langle\Omega|\!|\Phi_2\rangle\!\rangle)=\log\left(\frac{5+3\sqrt{5}}{\sqrt{10(5+\sqrt{5})}}\right)$ corresponds to the Affleck-Ludwig boundary entropy \cite{Affleck:1991tk}. Similarly, we have
\begin{eqnarray}
S(\mathcal{R}_2) = \frac{c_{\rm eff}}{3}\log\left(\frac{L}{\epsilon}\right) + \frac{1}{2}\log \frac{5+\sqrt{5}}{10}+ \ldots
\end{eqnarray}
and the constant term
\begin{eqnarray}
\frac{1}{2}\log \frac{5+\sqrt{5}}{10} = \log\left(\frac{d_{\mathcal{R}_2}}{\sum\limits_{\mathcal{R}_i}d_{\mathcal{R}_i}}\right) + \log(\langle\!\langle\Phi_2|\!|\Omega\rangle\!\langle\Omega|\!|\Phi_2\rangle\!\rangle)
\end{eqnarray}
where $d_{\mathcal{R}_2}=\frac{1+\sqrt{5}}{2}$ is the quantum dimension of $\mathcal{L}_{\Phi_2}\,.$ {Again, to remind the readers, these quantum dimensions are the ground-state/Perron--Frobenius quantum dimensions.}

\subsection{The Haagerup Rational CFT $\mathcal{R}_{\kappa=3}$}

We now turn to one of the central topics of our paper, the symmetry resolution of RCFTs with generalized Haagerup fusion category. We begin with the simplest example, $\mathcal{R}_{k=3}$, realized as the tensor product of the Virasoro minimal models $\mathcal{M}(2,5)$ and $\mathcal{M}(3,5)$ . The modular S-matrix is given by \cite{Gang:2023ggt},
\begin{align}
    \begin{split}
        S=\left(
\begin{array}{cccccccc}
 -a_0 & a_0 & a_3 & -a_3 & a_2 & -a_2 & a_2 & -a_2 \\
 a_0 & a_0 & a_3 & a_3 & -a_2 & -a_2 & -a_2 & -a_2 \\
 a_3 & a_3 & a_0 & a_0 & a_2 & a_2 & a_2 & a_2 \\
 -a_3 & a_3 & a_0 & -a_0 & -a_2 & a_2 & -a_2 & a_2 \\
 a_2 & -a_2 & a_2 & -a_2 & 2 a_2 \cos \left(\frac{\pi }{5}\right) & 2 a_2 \cos \left(\frac{2 \pi }{5}\right) & 2 a_2 \cos \left(\frac{3 \pi }{5}\right) & 2 a_2 \cos \left(\frac{4 \pi }{5}\right) \\
 -a_2 & -a_2 & a_2 & a_2 & 2 a_2 \cos \left(\frac{2 \pi }{5}\right) & 2 a_2 \cos \left(\frac{4 \pi }{5}\right) & 2 a_2 \cos \left(\frac{6 \pi }{5}\right) & 2 a_2 \cos \left(\frac{8 \pi }{5}\right) \\
 a_2 & -a_2 & a_2 & -a_2 & 2 a_2 \cos \left(\frac{3 \pi }{5}\right) & 2 a_2 \cos \left(\frac{6 \pi }{5}\right) & 2 a_2 \cos \left(\frac{9 \pi }{5}\right) & 2 a_2 \cos \left(\frac{12 \pi }{5}\right) \\
 -a_2 & -a_2 & a_2 & a_2 & 2 a_2 \cos \left(\frac{4 \pi }{5}\right) & 2 a_2 \cos \left(\frac{8 \pi }{5}\right) & 2 a_2 \cos \left(\frac{12 \pi }{5}\right) & 2 a_2 \cos \left(\frac{16 \pi }{5}\right) \\
\end{array}
\right)
    \end{split}
\end{align}
where, $a_0=\frac{1}{\sqrt{8 (k+2)}}+\frac{1}{\sqrt{8 (k-2)}}, a_1=\frac{1}{\sqrt{2 (k-2)}}, a_2=\frac{1}{\sqrt{2 (k+2)}},\,a_3=\frac{1}{\sqrt{8 (k-2)}}-\frac{1}{\sqrt{8 (k+2)}} $, with $k=3$. The full fusion can be immediately found by the Verlinde formula. However, by carefully inspecting the full fusion one can find that the following primaries $\Phi_1, \Phi_4,\,\Phi_5,\,\Phi_7$ form a sub-fusion. The sub-fusion is given by,
\begin{align}
    \begin{split}
&\Phi_4\times\Phi_4=\Phi_1+\Phi_4+\Phi_5+\Phi_7,\,\\ &\Phi_5\times \Phi_4= \Phi_4+\Phi_7,\,\\ &
    \Phi_5\times \Phi_5=\Phi_1+\Phi_5,\\ &
    \Phi_7\times \Phi_4=\Phi_4+\Phi_5,\\ &
    \Phi_7\times \Phi_5=\Phi_4,\\ &
    \Phi_7\times \Phi_7=\Phi_1+\Phi_7\,.    
    \end{split}
\end{align}
The sub-fusion category has four elements and, accordingly, four irreducible representations \(\{\mathcal{R}_i\}_{i=1}^4\)\,. Consequently, the reduced density matrix of subsystem \(A\), \(\rho_A\), decomposes into four distinct (mutually orthogonal) blocks, each carrying a unique character. However, the attainable resolution depends on whether the defect lines act distinctly on the primaries: primaries with identical defect action remain in the same block.
To proceed further, we investigate the action of (topological) defects on the primary fields. 
We carry out our analysis using the diagonal RCFT associated with the $\mathcal{R}_{k=3}$ modular data, where primaries and topological defects are in one-to-one correspondence.
Equivalently, each defect \(\mathcal{L}_a\) (labeled by a primary \(a\)) acts diagonally on primaries:
\begin{align}
\mathcal{L}_a  \Phi_i \;=\; \lambda_a(i)\,\Phi_i,
\qquad 
\lambda_a(i) \;=\; \frac{S_{a i}}{S_{\mathbf{1} i}},
\end{align}
where \(S\) is the modular \(S\)-matrix. Table~(\ref{tab:defect-vs-primary}) summarizes the action of defects on the primaries. 

\begin{table}[h!]
\centering
\setlength{\tabcolsep}{12pt}
\renewcommand{\arraystretch}{1.2}
\begin{tabular}{c|cccc}
  & $\Phi_{1}$ & $\Phi_{4}$ & $\Phi_{5}$ & $\Phi_{7}$ \\
\hline
$\mathcal{L}_{1}$ & 1 & 1 &1 &1 \\
$\mathcal{L}_{4}$ & $\frac{1}{2} \left(3-\sqrt{5}\right)$ & $\frac{1}{2} \left(\sqrt{5}+3\right)$ & $-1$ & $-1$ \\
$\mathcal{L}_{5}$ & $\frac{1}{2} \left(1-\sqrt{5}\right)$ & $\frac{1}{2} \left(\sqrt{5}+1\right)$ & $\frac{1}{2} \left(\sqrt{5}+1\right)$ & $\frac{1}{2} \left(1-\sqrt{5}\right)$ \\
$\mathcal{L}_{7}$ & $\frac{1}{2} \left(1-\sqrt{5}\right)$ & $\frac{1}{2} \left(\sqrt{5}+1\right)$ & $\frac{1}{2} \left(1-\sqrt{5}\right)$ & $\frac{1}{2} \left(\sqrt{5}+1\right)$ \\
\end{tabular}
\caption{Eigenvalues $\lambda_{a}(i)$ for the action of defect $\mathcal{L}_a$ on primary $\Phi_i$ in a diagonal RCFT, with 
$\displaystyle \lambda_{a}(i)=\frac{S_{a i}}{S_{\mathbf{1} i}}$.}
\label{tab:defect-vs-primary}
\end{table}
\noindent
In this non-unitary RCFT, the state of lowest conformal weight (the “ground state”) is {\(|\Omega\rangle=|\Phi_{3}\rangle\!\rangle\)}, not the vacuum. Consequently, the quantum dimensions ({Perron--Frobenius quantum dimensions}) of the topological defect lines
\(\mathcal{L}_{a}\) are evaluated relative to \(|\Phi_{3}\rangle\!\rangle\) via the modular \(S\)-matrix:
\begin{align}
  d_{a}
  \;=\;
  \frac{S_{a3}}{S_{\mathbf{1}3}}\,.
\end{align}
For the defect lines of interest, we obtain
\begin{align}
  d_{1} &= 1\,, &
  d_{4} &= \tfrac{1}{2}\!\left(\sqrt{5}+3\right)\,, &
  d_{5} &= \tfrac{1}{2}\!\left(\sqrt{5}+1\right)\,, &
  d_{7} &= \tfrac{1}{2}\!\left(\sqrt{5}+1\right)\,.
\end{align}

To compute the annulus partition function, we choose the (weakly) symmetric boundary state
\(|\!|\, \Phi_{4}\rangle\!\rangle\). The closed–string channel amplitude is then
\begin{align}
  Z(\tilde q)
  \;=\;
 \langle\! \langle \Phi_{4} \,|\!|\, \tilde q^{\color{black}{\frac{1}{2}(\,L_{0}+\bar{L}_0-\frac{c}{12})}} \,|\!|\, \Phi_{4}\rangle\!\rangle\,,\label{5.19d}
\end{align}
where \(\tilde q\) is the modular parameter in the closed channel. The partition function in \eqref{5.19d} can be computed by expanding the boundary Cardy states in terms of Ishibashi states as,
\begin{align}
    |\!|{B}_i\rangle\!\rangle=\sum_{i}\frac{S_{ij}}{\sqrt{S_{\mathbf{1}j}}}|j\rangle\!\rangle, \,\textrm{with}, \,{B_i}=\Phi_4,
\end{align}
where $|j\rangle\!\rangle$'s are the Ishibashi states and using the following property of Ishibashi states: $\langle\!\langle i|\tilde q^{\color{black}{\frac{1}{2}(L_0+\bar{L}_0-\frac{c}{12})}}|j\rangle\!\rangle=\chi_{i}(\tilde{q})\delta_{ij}$. In the open string channel, the partition function takes the following form (\ref{eq4}),
\begin{align}
\begin{split}Z(q)&
=\chi_1(q)+\chi_4(q)+\chi_5(q)+\chi_{7}(q)\,.
    \end{split}
\end{align}
As shown in the Table~(\ref{tab:defect-vs-primary}), the action of defect lines $\mathcal{L}_a$ is clearly different on the primaries, hence we have four irreducible blocks. The leading contribution to the partition function at $\tilde q \to 0$ limit is given by,
\begin{align}
    \begin{split}
     &   Z(q^n,\mathcal{R}_1)\sim \frac{5-\sqrt{5}}{10\sqrt{2}}\chi_{3}(\tilde q^{\frac{1}{n}}),\,\,\,\,  Z(q^n,\mathcal{R}_2)\sim \frac{1}{2} \sqrt{\frac{3}{5}+\frac{1}{\sqrt{5}}}\,\chi_{3}(\tilde q^{\frac{1}{n}}), \,\, \,\\ & Z(q^n,\mathcal{R}_3)\sim \frac{1}{\sqrt{10}}\,\chi_{3}(\tilde q^{\frac{1}{n}}),\, \qquad Z(q^n,\mathcal{R}_4)\sim \frac{1}{\sqrt{10}}\chi_{3}(\tilde q^{\frac{1}{n}})\,.
     \end{split}
\end{align}
Therefore, the SREE for the first irreducible block is given by,
\begin{align}
    \begin{split}
        S(\mathcal{R}_1)=\frac{c_{\textrm{eff}}}{3}\log\left(\frac{L}{\epsilon}\right)+\log\left(\frac{5-\sqrt{5}}{10\sqrt{2}}\right)+\cdots
    \end{split}
\end{align}
Therefore the consistency condition from \eqref{3.15k} is the following,
\begin{align}
    \begin{split}
        \log\left(\frac{5-\sqrt{5}}{10\sqrt{2}}\right)\overset{?}{=} \log(g^2)+\log\left(\frac{d_{\rho} \,N_{\rho B_1}^{B_2}}{d(B_1\otimes B_2)}\right)\,.\label{5.27g}
    \end{split}
\end{align}
Now computing the RHS independently, we find,
\begin{align}
    g^2=|\langle\!\langle \Phi_4|\!|\Omega\rangle|^2=\sqrt{\frac{2}{5}}+\frac{1}{\sqrt{2}}, \,\,
  \log\left(\frac{d_{\rho} \,N_{\rho B_1}^{B_2}}{d(B_1\otimes B_2)}\right)=\log\left(\frac{7-3 \sqrt{5}}{2}\right)\,.\label{5.29g}
\end{align}
Putting \eqref{5.29g} in \eqref{5.27g}, one can show that the equality in \eqref{3.15k} indeed holds. Note that we choose the boundary as weakly symmetric, for which we notice that $d(B_1\otimes B_2)=\sum_{\rho}d_{\rho}$, and the expression for the entanglement entropy reduces to (\ref{3.17j})\,.

Next, we turn to the example of $\mathcal{R}_{k=4}$ \cite{Gang:2023ggt}, where several subtleties arise. In particular, the relevant sub-fusion ring exhibits non-trivial multiplicities, and we find that the standard methods do not yield a complete symmetry resolution. We make a detailed comment on that.
\subsection{The Haagerup Rational CFT $\mathcal{R}_{\kappa=4}$}

Haagerup $\mathcal{R}_{\kappa=4}$ is also a non-unitary rational conformal field theory with ten primaries with central charge $c=-11$ and conformal dimensions are $h_i=\{\frac{3}{2},\, -\frac{1}{2},\, 0,\, -\frac{3}{8},\, -\frac{11}{24},\, -\frac{1}{3},\, -\frac{1}{8},\, \frac{1}{6},\, \frac{13}{24}\}$ \cite{Gang:2023ggt}, which can also be realized as non-unitary super-Virasoro minimal model $S\mathcal{M}(2,12)$. The modular S-matrix by which the characters transform under $\tau\to -\frac{1}{\tau}$ is given by \cite{Gang:2023ggt},

\begin{eqnarray} 
\scalebox{0.80}{$
S=\begin{pmatrix}
a_0 & a_0 & a_3 & a_3 & -a_1 & a_2 & -a_2 & a_2 & -a_2 & a_2 \\
a_0 & a_0 & a_3 & a_3 & a_1 & -a_2 & -a_2 & -a_2 & -a_2 & -a_2 \\
a_3 & a_3 & a_0 & a_0 & a_1 & a_2 & a_2 & a_2 & a_2 & a_2 \\
a_3 & a_3 & a_0 & a_0 & -a_1 & -a_2 & a_2 & -a_2 & a_2 & -a_2 \\
-a_1 & a_1 & a_1 & -a_1 & 2a_1\cos\left(\frac{\pi}{2}\right) & 0 & 0 & 0 & 0 & 0 \\
a_2 & -a_2 & a_2 & -a_2 & 0 & 2a_2\cos\left(\frac{\pi}{6}\right) & 2a_2\cos\left(\frac{2\pi}{6}\right) & 2a_2\cos\left(\frac{3\pi}{6}\right) & 2a_2\cos\left(\frac{4\pi}{6}\right) & 2a_2\cos\left(\frac{5\pi}{6}\right) \\
-a_2 & -a_2 & a_2 & a_2 & 0 & 2a_2\cos\left(\frac{2\pi}{6}\right) & 2a_2\cos\left(\frac{4\pi}{6}\right) & 2a_2\cos\left(\frac{6\pi}{6}\right) & 2a_2\cos\left(\frac{8\pi}{6}\right) & 2a_2\cos\left(\frac{10\pi}{6}\right) \\
a_2 & -a_2 & a_2 & -a_2 & 0 & 2a_2\cos\left(\frac{3\pi}{6}\right) & 2a_2\cos\left(\frac{6\pi}{6}\right) & 2a_2\cos\left(\frac{9\pi}{6}\right) & 2a_2\cos\left(\frac{12\pi}{6}\right) & 2a_2\cos\left(\frac{15\pi}{6}\right) \\
-a_2 & -a_2 & a_2 & a_2 & 0 & 2a_2\cos\left(\frac{4\pi}{6}\right) & 2a_2\cos\left(\frac{8\pi}{6}\right) & 2a_2\cos\left(\frac{12\pi}{6}\right) & 2a_2\cos\left(\frac{16\pi}{6}\right) & 2a_2\cos\left(\frac{20\pi}{6}\right) \\
a_2 & -a_2 & a_2 & -a_2 & 0 & 2a_2\cos\left(\frac{5\pi}{6}\right) & 2a_2\cos\left(\frac{10\pi}{6}\right) & 2a_2\cos\left(\frac{15\pi}{6}\right) & 2a_2\cos\left(\frac{20\pi}{6}\right) & 2a_2\cos\left(\frac{25\pi}{6}\right)
\end{pmatrix}
$}
\end{eqnarray}
where, $a_0  = \frac{1}{\sqrt{8(\kappa - 2)}} + \frac{1}{\sqrt{8(\kappa + 2)}},\,  a_1  = \frac{1}{\sqrt{2(\kappa - 2)}},\,  a_2  = \frac{1}{\sqrt{2(\kappa + 2)}},\,  a_3  = \frac{1}{\sqrt{8(\kappa - 2)}} - \frac{1}{\sqrt{8(\kappa + 2)}},\, \text{with}\, \kappa=4$. Therefore, we can compute the fusion rule of the primaries, and we have noticed that there is a sub-fusion algebra generated by $\mathcal{C}=\{\mathcal{L}_{\Phi_1}, \mathcal{L}_{\Phi_2}, \mathcal{L}_{\Phi_3}, \mathcal{L}_{\Phi_4}, \mathcal{L}_{\Phi_7}, \mathcal{L}_{\Phi_9}\}$.

\begin{minipage}[t]{0.48\textwidth} 
\begin{align*}
\Phi_2 \times \Phi_2 &= \Phi_1\,, \\
\Phi_3 \times \Phi_2 &= \Phi_4\,, \\
\Phi_3 \times \Phi_3 &= \Phi_1 + \Phi_3 + \Phi_4 + \Phi_7 + \Phi_9 \,,\\
\Phi_4 \times \Phi_2 &= \Phi_3 \,,\\
\Phi_4 \times \Phi_3 &= \Phi_2 + \Phi_3 + \Phi_4 + \Phi_7 + \Phi_9 \,,\\
\Phi_4 \times \Phi_4 &= \Phi_1 + \Phi_3 + \Phi_4 + \Phi_7 + \Phi_9\,, \\
\Phi_5 \times \Phi_2 &= \Phi_5\,, \\
\end{align*}
\end{minipage}%
\hfill 
\begin{minipage}[t]{0.48\textwidth} 
\begin{align}
\begin{array}{rcl}
\Phi_7 \times \Phi_2 &=& \Phi_9 \,,\\
\Phi_7 \times \Phi_3 &=& \Phi_3 + \Phi_4 + \Phi_7\,, \\
\Phi_7 \times \Phi_4 &= &\Phi_3 + \Phi_4 + \Phi_9\,, \\
\Phi_7 \times \Phi_7 &= &\Phi_1 + \Phi_3 + \Phi_9\,, \\ 
\Phi_9 \times \Phi_2 &=& \Phi_7\,, \\
\Phi_9 \times \Phi_3 &=& \Phi_3 + \Phi_4 + \Phi_9\,, \\
\Phi_9 \times \Phi_4 &=& \Phi_3 + \Phi_4 + \Phi_7\,, \\
\Phi_9 \times \Phi_7 &=& \Phi_2 + \Phi_4 + \Phi_7\,, \\
\Phi_9 \times \Phi_9 &=& \Phi_1 + \Phi_3 + \Phi_9\,. \\
\label{fr}
\end{array} 
\end{align}
\end{minipage}

 We also notice that the boundary state $|\!|\, \Phi_{5}\rangle\!\rangle$ is weakly symmetric under this fusion category $\mathcal{C}$.
\begin{align}
\begin{split}
&\mathcal{L}_{\Phi_1} |\!|\, \Phi_{5}\rangle\!\rangle = |\!|\, \Phi_{5}\rangle\!\rangle\,, \\&
\mathcal{L}_{\Phi_2}|\!|\, \Phi_{5}\rangle\!\rangle = |\!|\, \Phi_{5}\rangle\!\rangle\,, \\&
\mathcal{L}_{\Phi_3} |\!|\, \Phi_{5}\rangle\!\rangle = |\!|\, \Phi_{5}\rangle\!\rangle \oplus |\!|\, \Phi_{5}\rangle\!\rangle\oplus |\!|\, \Phi_{6}\rangle\!\rangle \oplus |\!|\, \Phi_{8}\rangle\!\rangle \oplus |\!|\, \Phi_{10}\rangle\!\rangle\,,\\&
\mathcal{L}_{\Phi_4} |\!|\, \Phi_{5}\rangle\!\rangle = |\!|\, \Phi_{5}\rangle\!\rangle \oplus |\!|\, \Phi_{5}\rangle\!\rangle\oplus |\!|\, \Phi_{6}\rangle\!\rangle \oplus |\!|\, \Phi_{8}\rangle\!\rangle \oplus |\!|\, \Phi_{10}\rangle\!\rangle\,,\\&
\mathcal{L}_{\Phi_7} |\!|\, \Phi_{5}\rangle\!\rangle = |\!|\, \Phi_{5}\rangle\!\rangle \oplus |\!|\, \Phi_{5}\rangle\!\rangle\oplus |\!|\, \Phi_{6}\rangle\!\rangle \oplus |\!|\, \Phi_{8}\rangle\!\rangle \oplus |\!|\, \Phi_{10}\rangle\!\rangle\,,\\&
\mathcal{L}_{\Phi_9} |\!|\, \Phi_{5}\rangle\!\rangle = |\!|\, \Phi_{5}\rangle\!\rangle \oplus |\!|\, \Phi_{5}\rangle\!\rangle\oplus |\!|\, \Phi_{6}\rangle\!\rangle \oplus |\!|\, \Phi_{8}\rangle\!\rangle \oplus |\!|\, \Phi_{10}\rangle\!\rangle\,.
\end{split}
\end{align}

The untwisted partition function in the closed string channel and open string channel are given respectively by,
\begin{align}
\begin{split}
Z(q^n)  &=  \frac{3}{3+\sqrt{3}}\, \chi_1(\tilde{q}^\frac{1}{n}) + \frac{3}{3+\sqrt{3}}\, \chi_2(\tilde{q}^\frac{1}{n}) + \frac{3}{3-\sqrt{3}}\, \chi_3(\tilde{q}^\frac{1}{n}) + \frac{3}{3-\sqrt{3}}\, \chi_4(\tilde{q}^\frac{1}{n})\,, \\
&=  \chi_1(q^n) + \chi_2(q^n) + 2\chi_3(q^n) + 2\chi_4(q^n) + \chi_7(q^n) + \chi_9(q^n)\,.
\end{split}
\end{align}
\textit{Note that the chosen weakly symmetric boundary state, $|\!|{B}\rangle\!\rangle= |\!|\, \Phi_{5}\rangle\!\rangle$ does not belong to the sub-fusion}. However, from the fusion table, one can immediately see that ${B}\otimes {B}$ again generates all the elements of the sub-fusion and is given by,{\begin{align}
    \begin{split}
|\!|\Phi_5\rangle\!\rangle\otimes  |\!|\Phi_5\rangle\!\rangle= |\!|\Phi_1\rangle\!\rangle+ |\!|\Phi_2\rangle\!\rangle+ 2|\!|\Phi_3\rangle\!\rangle+ 2|\!|\Phi_4\rangle\!\rangle+ |\!|\Phi_7\rangle\!\rangle+|\!|\Phi_9\rangle\!\rangle
    \end{split}
\end{align}
As we see, this provides a non-trivial example of fusion with non-trivial multiplicities. It is important to note that the corresponding sub-fusion
does not itself contain multiplicities. However, what enters in the SREE expression is whether the boundary fusion involves non-trivial
multiplicities. In the present case, this is precisely what happens.}
 From Table~\eqref{tab6}, it is clear that the actions of the six defect lines are the same for the following blocks: $\{\Phi_1,\,\Phi_2\},\,\{\Phi_3,\,\Phi_4\},\,\{\Phi_7,\Phi_9\}$. Therefore, there will be three blocks (reducible) to be resolved.
\begin{table}[H]
\begin{tabular}{ccccccccccc}
\toprule
Defect line/Bulk primary & $\Phi_1$& $\Phi_2$& $\Phi_3$& $\Phi_4$& $\Phi_5$& $\Phi_6$& $\Phi_7$& $\Phi_8$& $\Phi_9$& $\Phi_{10}$\\
\midrule
$\mathcal{L}_{\Phi_1}$ & $1$ & $1$ &$1$ &$1$ &$1$ &$1$ &$1$ &$1$ &$1$ &$1$ \\
$\mathcal{L}_{\Phi_2}$ & $1$ & $1$ &$1$ &$1$ &$-1$ &$-1$ &$1$ &$-1$ &$1$ &$-1$ \\
$\mathcal{L}_{\Phi_3}$ & $2-\sqrt{3}$ & $2-\sqrt{3}$ &$2+\sqrt{3}$ &$2+\sqrt{3}$ &$-1$ &$1$ &$-1$ &$1$ &$-1$ &$1$ \\
$\mathcal{L}_{\Phi_4}$ & $2-\sqrt{3}$ & $2-\sqrt{3}$ &$2+\sqrt{3}$ &$2+\sqrt{3}$ &$1$ &$-1$ &$-1$ &$-1$ &$-1$ &$-1$ \\
$\mathcal{L}_{\Phi_7}$ & $1-\sqrt{3}$ & $1-\sqrt{3}$ &$1+\sqrt{3}$ &$1+\sqrt{3}$ &$0$ &$1$ &$1$ &$-2$ &$1$ &$1$ \\
$\mathcal{L}_{\Phi_9}$ & $1-\sqrt{3}$ & $1-\sqrt{3}$ &$1+\sqrt{3}$ &$1+\sqrt{3}$ &$0$ &$-1$ &$1$ &$2$ &$1$ &$-1$ \\
\bottomrule
\end{tabular}
\caption{Action of the defect lines on primaries.}
\label{tab6}
\end{table}
\noindent
The leading contribution to the partition function at $\tilde q \to 0$ limit is given by,
 \begin{align}
     \begin{split}
         Z(q^n,\mathcal{R}_1)\overset{\tilde q\to 0}{\sim} \frac{1}{6} \left(3-\sqrt{3}\right)\,\chi_3(\tilde q^{\frac{1}{n}}),\,Z(q^n,\mathcal{R}_2)\overset{\tilde q\to 0}{\sim} \left(1+\frac{1}{\sqrt{3}}\right)\,\chi_3(\tilde q^{\frac{1}{n}}),\,Z(q^n,\mathcal{R}_3)=\frac{1}{\sqrt{3}}\,\chi_3(\tilde q^{\frac{1}{n}})\,.
     \end{split}
 \end{align}
The \textit{partially} resolved entanglement entropy for each block is given by,
\begin{align}
    \begin{split}
       & S(\mathcal{R}_1)=\frac{c_{\textrm{eff}}}{3}\log\left(\frac{L}{\epsilon}\right)+\log\left(\frac{1}{6} \left(3-\sqrt{3}\right)\right)+\cdots\,,\\ &
       S(\mathcal{R}_2)=\frac{c_{\textrm{eff}}}{3}\log\left(\frac{L}{\epsilon}\right)+\log\left(1+\frac{1}{\sqrt{3}}\right)+\cdots\,,\\&
         S(\mathcal{R}_3)=\frac{c_{\textrm{eff}}}{3}\log\left(\frac{L}{\epsilon}\right)+\log\left(\frac{1}{\sqrt{3}}\right)+\cdots\,.
    \end{split}
\end{align}
Now we ask the same question as before by making a comparison with \eqref{3.15k},
\begin{align}
    \begin{split}
     &  \textrm{block-1}:\,\,  \log\left(\frac{1}{6} \left(3-\sqrt{3}\right)\right)\overset{?}{=} \log(g^2)+\log\left(\frac{d_{\rho} \,N_{\rho B_1}^{B_2}}{d(B_1\otimes B_2)}\right),\,\rho\in \mathcal{R}_1\,,\\ &
     \textrm{block-2}:\,\, \log\left(1+\frac{1}{\sqrt{3}}\right)\overset{?}{=} \log(g^2)+\log\left(\frac{d_{\rho} \,N_{\rho B_1}^{B_2}}{d(B_1\otimes B_2)}\right),\,\rho\in \mathcal{R}_2\,,\\ &
     \textrm{block-3}:\log\left(\frac{1}{\sqrt{3}}\right)\overset{?}{=} \log(g^2)+\log\left(\frac{d_{\rho} \,N_{\rho B_1}^{B_2}}{d(B_1\otimes B_2)}\right),\,\rho\in \mathcal{R}_3\,.\label{5.36j} 
    \end{split}
\end{align}
Again, independent computation shows,
\begin{align}
\begin{split}
  &g^2=|\langle\!\langle \Phi_5|\!|\Omega\rangle|^2= \frac{1}{2} \left(\sqrt{3}+3\right),\,\\&\frac{\sum_{i\in \rho}d_{i} \,N_{i B_1}^{B_2}}{d(B_1\otimes B_2)}=\left\{\frac{2}{\left(\sqrt{3}+3\right)^2},\frac{2}{3},\,\left(\frac{1}{\sqrt{3}}-\frac{1}{3}\right)\right\}, \quad \forall i\in \rho=1,2,3\quad 
 (\textrm{respectively})\,.\label{5.37j}
  \end{split}
\end{align}
\noindent Again, ground state $|\Omega\rangle=|\Phi_3\rangle\!\rangle\,.$ Substituting \eqref{5.37j} into \eqref{5.36j} confirms that the equality of \eqref{3.15k} remains valid. However, \eqref{3.17j} does not apply here: the multiplicity-free assumption is violated, the fusion exhibits nontrivial multiplicities, and also the resulting blocks are not made up from single characters. Apart from that, one may notice that the \textit{defect lines discussed above do not fully resolve the symmetry sector}. Can we further refine the decomposition into smaller irreducible blocks? The answer is \emph{to some extent}-yes, thanks to \cite{Heymann:2024vvf}. \\\\
\noindent\textbf{Prescription.} \noindent\textbf{Refinement via an auxiliary defect.} Introduce an additional defect line associated with the boundary state $ |\!|\, \Phi_{5}\rangle\!\rangle$, denoted by $\mathcal{L}_{\Phi_5}$, and examine its action on the primaries of the chosen subcategory. If this defect acts differently on the primaries within a single block $\mathcal{R}_i$, that is, if the images/eigenvalues under $\mathcal{L}_{\Phi_5}$ are not identical, then $\mathcal{R}_i$ splits into two irreducible blocks; otherwise, no further refinement occurs. 
\noindent Applying this prescription to our case, the action of $\mathcal{L}_{\Phi_5}$ separates the primaries in the first two blocks, thereby splitting each into two irreducible blocks, while it fails to further decompose the third block. Consequently, \textit{we obtain five blocks in total—four irreducible and one remaining reducible}. A straightforward calculation, in the same spirit as above, once again confirms \eqref{3.15k}.

\subsection{The Haagerup Rational CFT $\mathcal{R}_{\kappa=5}$} 

Finally, we consider another example of a generalized Haagerup fusion category, namely $\mathcal{R}_{\kappa=5}$ \cite{Gang:2023ggt}. 
Below, we discuss the fusion of primaries in the fusion category. We again find that a sub-fusion exists and is given by,

\begin{minipage}[t]{0.48\textwidth} 
\begin{align*}
\Phi_4\times\Phi_4& = \Phi_1+\Phi_4+\Phi_5+\Phi_7+\Phi_9+\Phi_{11}\,,\\\Phi_4\times\Phi_5& = \Phi_4+\mathbf{2}\Phi_5+\Phi_7+\Phi_9+\Phi_{11}\,,\\
\Phi_4\times\Phi_7& = \Phi_4+\Phi_5+\Phi_9+\Phi_{11}\,,\\
\Phi_4\times\Phi_9 &= \Phi_4+\Phi_5+\Phi_7+\Phi_{11}\,,\\
\Phi_4\times\Phi_{11}&= \Phi_4+\Phi_5+\Phi_7+\Phi_9\,,\\
\Phi_5\times\Phi_5& = \Phi_1+\mathbf{2}\Phi_4+\mathbf{2}\Phi_5+\Phi_7+\Phi_9+\Phi_{11}\,,\\
\Phi_5\times\Phi_7 &= \Phi_4+\Phi_5+\Phi_7+\Phi_9+\Phi_{11}\,,\\
\Phi_5\times\Phi_9 &= \Phi_4+\Phi_5+\Phi_7+\Phi_9+\Phi_{11}\,,\\
\Phi_5\times\Phi_{11}&= \Phi_4+\Phi_5+\Phi_7+\Phi_9+\Phi_{11}\,,\\
\end{align*}
\hfill
\end{minipage}%
\begin{minipage}[t]{0.48\textwidth} 
\begin{align*}
\Phi_7\times\Phi_7& = \Phi_1+\Phi_5+\Phi_7+\Phi_9\,,\\
\Phi_7\times\Phi_9 &= \Phi_4+\Phi_5+\Phi_7\,,\\
\Phi_7\times\Phi_{11}&= \Phi_4+\Phi_5+\Phi_{11}\,,\\
\Phi_9\times\Phi_9& = \Phi_1+\Phi_5+\Phi_9+\Phi_{11}\,,\\
\Phi_9\times\Phi_{11}&= \Phi_4+\Phi_5+\Phi_9\,,\\
\Phi_{11}\times\Phi_{11}&= \Phi_1+\Phi_5+\Phi_7+\Phi_{11}\,.
\end{align*}
\end{minipage}%

\begin{table}[H]
\begin{tabular}{ccccccc}
\toprule
Defect line/Bulk primary & $\Phi_1$& $\Phi_4$& $\Phi_5$& $\Phi_7$& $\Phi_9$& $\Phi_{11}$\\
\midrule
$\mathcal{L}_{\Phi_1}$ & $1$ & $1$ &$1$ &$1$ &$1$ &$1$  \\
$\mathcal{L}_{\Phi_4}$ & $\frac{5-\sqrt{21}}{2}$ & $\frac{5+\sqrt{21}}{2}$ &$1$ &$-1$ &$-1$ &$-1$\\
$\mathcal{L}_{\Phi_5}$ & $\frac{7-\sqrt{21}}{2}$ & $\frac{7+\sqrt{21}}{2}$ &$-1$ &$0$ &$0$ &$0$\\
$\mathcal{L}_{\Phi_7}$ & $\frac{3-\sqrt{21}}{2}$ & $\frac{3+\sqrt{21}}{2}$ &$0$ &$2\cos(\frac{\pi}{7})$ &$2\sin(\frac{\pi}{14})$ &$-2\sin(\frac{3\pi}{14})$ \\
$\mathcal{L}_{\Phi_9}$ & $\frac{3-\sqrt{21}}{2}$ & $\frac{3+\sqrt{21}}{2}$ &$0$ &$2\sin(\frac{\pi}{14})$ &$-2\sin(\frac{3\pi}{14})$ &$2\cos(\frac{\pi}{7})$ \\
$\mathcal{L}_{\Phi_{11}}$ & $\frac{3-\sqrt{21}}{2}$ & $\frac{3+\sqrt{21}}{2}$ &$0$ &$-2\sin(\frac{3\pi}{14})$ &$2\cos(\frac{\pi}{7})$ &$2\sin(\frac{\pi}{14})$ \\
\bottomrule
\end{tabular}
\caption{Action of the defect lines on primaries.}
\label{Hk5}
\end{table}

The sub-fusion category contains six elements, $\mathcal{C}=\{\mathcal{L}_{\Phi_1}, \mathcal{L}_{\Phi_4}, \mathcal{L}_{\Phi_5}, \mathcal{L}_{\Phi_7}, \mathcal{L}_{\Phi_9}, \mathcal{L}_{\Phi_{11}}\}$, so it has six irreducible representations, $\mathcal{R}_i$. From Table~(\ref{Hk5}), we observe that the action of $\mathcal{C}$ on the primaries is distinct. Consequently, we conclude that a single character contributes to each representation (or block) in the reduced density matrix $\rho_A$. In this setting, we take $|\!|\Phi_4\rangle\!\rangle$ as a weakly symmetric boundary state. The corresponding partition function, together with the symmetry‑resolved partition functions in the open‑string channel, are given by:
\begin{eqnarray}
Z(q) = \chi_1(q)+\chi_4(q)+\chi_5(q)+\chi_7(q)+\chi_9(q) + \chi_{11}(q)
\end{eqnarray}
and, 
\begin{align}
\begin{split}
&Z(q^n, \mathcal{R}_1) = {\chi_1(q^n)} \,,
\,\,\,\,\,\,Z(q^n, \mathcal{R}_2) ={\chi_4(q^n)}\,, 
\,\,\,\,\,\,\,Z(q^n, \mathcal{R}_3) ={\chi_5(q^n)}\,, \\&
Z(q^n, \mathcal{R}_4) = {\chi_7(q^n)}\,, 
\,\,\,\,\,\,\,Z(q^n, \mathcal{R}_5) = {\chi_9(q^n)}\,, 
\,\,\,\,\,\,\,Z(q^n, \mathcal{R}_6) = {\chi_{11}(q^n)}\,. 
\end{split}
\end{align}   
This is a non-unitary rational CFT; therefore, the leading contribution comes from the ground state rather than the vacuum in the high energy limit $\tilde{q} \to 0\,.$ In this example, $|\Phi_3\rangle \!\rangle$ is a ground state. 
 Finally, we get the following expressions for the SREE for the different blocks.
\begin{eqnarray}
S_n(\mathcal{R}_1) = \frac{c_{\rm eff}}{6}\left(\frac{n+1}{n}\right)\log\left(\frac{L}{\epsilon}\right) + \log \frac{1}{2}\left(\frac{1}{\sqrt{6}}-\frac{1}{\sqrt{14}}\right)+ \ldots\,.
\end{eqnarray} 
where the constant term 
\begin{eqnarray}
\log \frac{1}{2}\left(\frac{1}{\sqrt{6}}-\frac{1}{\sqrt{14}}\right) = \log\left(\frac{d_{\mathcal{R}_1}}{\sum\limits_{\mathcal{R}_i}d_{\mathcal{R}_i}}\right) + \log(\langle\!\langle\Phi_4|\!|\Omega\rangle\langle\Omega|\!|\Phi_4\rangle\!\rangle)\,.
\end{eqnarray}
Here $d_{\mathcal{R}_1}=1$, $\sum\limits_{\mathcal{R}_i} d_{\mathcal{R}_i} = \frac{23+5\sqrt{21}}{2}$, and $\log(\langle\!\langle\Phi_4|\!|\Omega\rangle\langle\Omega|\!|\Phi_4\rangle\!\rangle)=\frac{1}{2}\log\left(\frac{55}{42}+2\sqrt{\frac{3}{7}}\right)$ corresponds to the Affleck-Ludwig boundary entropy \cite{Affleck:1991tk}. Similarly, we have
\begin{eqnarray}
S_n(\mathcal{R}_2) = \frac{c_{\rm eff}}{6}\left(\frac{n+1}{n}\right)\log\left(\frac{L}{\epsilon}\right) + \log \frac{1}{2}\left(\sqrt{\frac{5+\sqrt{21}}{21}}\right)+ \ldots\,.
\end{eqnarray}
and the constant term
\begin{eqnarray}
\log \frac{1}{2}\left(\sqrt{\frac{5+\sqrt{21}}{21}}\right) = \log\left(\frac{d_{\mathcal{R}_2}}{\sum\limits_{\mathcal{R}_i}d_{\mathcal{R}_i}}\right) + \log(\langle\!\langle\Phi_4|\!|\Omega\rangle\langle\Omega|\!|\Phi_4\rangle\!\rangle)
\end{eqnarray}
where $d_{\mathcal{R}_2}=\frac{5+\sqrt{21}}{2}$ is the quantum dimension of $\mathcal{L}_{\Phi_4}\,.$ {Again, these are Perron--Frobenius quantum dimensions computed with respect to the ground state.}
\section{Summary and Conclusion}\label{Sec6}

In this work, we have investigated the symmetry-resolved entanglement entropy (SREE) in $(1+1)$-dimensional critical systems possessing \emph{categorical} symmetries. Our interest has been in a broad class of rational conformal field theories (RCFTs), including examples that are (1) non-unitary, (2) non-diagonal, and (3) exhibit multiplicities in their fusion rules, for which symmetry resolution is substantially more intricate than in the conventional group-like or unitary diagonal cases.
The general motivation for our study comes from the ongoing effort to understand how entanglement structures encode internal and categorical symmetry data in quantum field theories. 
However, \textit{explicit} computations of SREE in non-unitary and non-diagonal theories, especially those with non-invertible symmetries, have remained limited. This work aims to fill that gap.

\medskip
\noindent\textbf{Summary of Results
} \\
Our analysis and main results may be summarized as follows:
\begin{itemize}

\item \textbf{\textit{General Framework for Categorical Symmetry Resolution.}}  
We first discussed the role of boundary conditions in symmetry resolution and introduced a general notion of computing SREE. We provide a general formula for the SREE that uniformly applies to symmetric (weakly or strongly) and cloaking boundary conditions, as well as to fusion rings with multiplicities, without invoking any SymTFT construction. Our derivation relies solely on a purely two-dimensional RCFT analysis. This formalism applies not only to weakly and strongly symmetric boundary conditions, but also to the more subtle \emph{cloaking} boundary conditions, and it naturally accommodates fusion categories with multiplicities. \textcolor{black}{The formula applies whenever the compatible defect action on the annulus Hilbert space is specified}.

\item \textbf{\textit{Replica Computations, Universal Scaling, and Boundary entropy.}}
We carried out explicit replica computations of the von~Neumann entropy associated with each symmetry sector, identifying both the universal leading contributions and the subleading corrections associated with boundary entropy. For several unitary and non-unitary models, we evaluated the SREE via first principles and made a connection with the generalized Affleck--Ludwig boundary entropy formula. We began with the Tambara--Yamagami (TY) fusion category structure in the Ising CFT, where the absence of weakly/strongly symmetric boundary conditions necessitates cloaking boundary states. We also analyze the SREE for the tetracritical Ising model. Interestingly, in this case, we found that there were two natural choices of sub-fusion categories. For the larger sub-fusion category, we find that no weakly symmetric boundary state exists, but we can still perform the resolution using an appropriate cloaking boundary state. In contrast, for the smaller sub-fusion category, a weakly symmetric boundary state was available. We subsequently turned to non-diagonal RCFTs. \textcolor{black}{In the $SU(2)_{10}$ theory with the $E_6$ modular invariant, the graph
boundary $\|Y\rangle\!\rangle$ is weakly symmetric under the $\tfrac{1}{2}E_6$ fusion algebra. However, the $Y|Y$ annulus contains primaries that are absent from the closed-channel $E_6$ Ishibashi exponent set. The available defect eigenvalues, therefore, do not determine the TDL action on the complete open-channel Hilbert space. Consequently, the conventional RCFT symmetry resolution cannot be completed from the modular invariant and NIM-rep data
alone. This is a limitation of the available RCFT data, rather than evidence that no sector decomposition exists. A TQFT based resolution may still be defined once the required boundary and junction data are supplied. We leave this for future investigations.} By contrast, in the three-state Potts model, the relevant $\mathbb Z_3$ action is known, and the RCFT symmetry resolution can be performed using an appropriate weakly symmetric boundary condition. \textcolor{black}{More generally,
whenever this action is available, our method provides an explicit and readily computable RCFT based symmetry resolution. Clearly identifying both this operational scope and the additional information that may be required for certain cases is an important outcome of our analysis.}

\item \textbf{\textit{Non-Unitary Symmetry Resolution and Generalized Haagerup RCFTs.}}
We then studied the symmetry resolution in non-unitary settings. As a starting point, we computed the SREE for the Lee--Yang minimal model. Following this, we addressed generalized Haagerup RCFTs, whose modular data encode non-invertible symmetries and which may not necessarily admit a well-defined VOA realization. This makes the implementation of boundary-based SREE calculations particularly subtle. We reviewed the modular and fusion data of the Haagerup family $\mathcal{R}_{k}$, discussed the structure of simple objects and fusion channels, and emphasized the non-invertible symmetry interpretation. By carefully identifying admissible Cardy boundary states, we explicitly computed the symmetry-resolved entanglement entropies for $\mathcal{R}_{k=3,4,5}$. Our findings show that with an appropriate choice of weakly symmetric boundary conditions, \emph{full symmetry-sector resolution} is achievable for $\mathcal{R}_{3}$ and $\mathcal{R}_{5}$. However, for $\mathcal{R}_{4}$, a complete resolution is \emph{not} possible when restricting to defect subfusion sectors alone. We demonstrated that introducing boundary defects allows one to improve and, in some cases, complete the resolution.

\end{itemize}

\medskip

\noindent
  \textbf{\textit{Outlook and Open Problem.}} 
\noindent
There are several interesting future directions possible. We list a few of them below.
It is highly compelling to consider extending these computations beyond the realm of rational conformal field theories (RCFTs). \textit{This naturally raises several fundamental questions, for instance, whether the existence of a vertex operator algebra (VOA) is a necessary condition for the existence of a corresponding symmetry topological field theory (SymTFT)?} A particularly promising direction is to refine and generalize the existing framework so as to address the challenges of symmetry resolution in non-diagonal RCFTs, where modular invariants and fusion structures become considerably more intricate.

In addition to this, our current analysis has primarily focused on line operators, which correspond to 0-form defects. A natural next step would be to explore the broader landscape of higher-form defects, such as condensation or domain-wall defects, which play a crucial role in the categorical and geometric understanding of symmetry breaking and topological phase transitions. Another significant avenue of investigation lies in the explicit computation of the anomaly inflow action. Determining this action would allow one to identify and characterize the electric and magnetic defect lines, \textit{along with their associated bulk linkings}. Such an analysis could shed light on the interplay between boundary symmetries and bulk topological data, ultimately contributing to a deeper understanding of the correspondence between conformal and topological field theories.

\normalsize







\appendix

\section*{Acknowledgments}
We would like to thank Brandon Rayhaun for detailed comments on an initial version of the manuscript.
A.B and J.S would also like to thank Arpit Das and Thomas Quella for discussions on related topics and comments on some parts of the manuscript. J.S is indebted to Suresh Govindarajan, Chethan N. Gowdigere, and Akhila S. for useful discussions. S.G would like to thank Jayashish Das for the discussion on the related topic.  S.P (PMRF ID: 1703278) is supported by the Prime Minister’s Research Fellowship of the Government of India. S.G is partially supported by the Prime Minister’s Research Fellowship of the Government of India and by the ANRF grant ANRF/ARG/2025/001338/PS. S.G would like to thank Institut des Hautes Études Scientifiques (IHES), France, for their kind hospitality during the course of the work. A.B thanks the participants of the BIRS-CMI workshop ``Quantum Gravity and Information Theory: Modern Developments" (\url{https://www.birs.ca/events/2025/5-day-workshops/25w5386}) held at CMI, Chennai, during 9-14th November 2025, where part of the work was presented by J.S. J.S would like to thank the organizers of the BIRS-CMI workshop 2025 for giving warm hospitality. J.S is supported by the project CRG/2023/001120 and OTH/SERB/13451 (50:50 Postdoctoral scheme of IIT Gandhinagar).  A.B is supported by the Core Research Grant (CRG/2023/001120) from the Anusandhan National Research Foundation, India. A.B also
acknowledges the associateship program of the Indian Academy of Science, Bengaluru. We acknowledge the use of ChatGPT 5.6 (Sol) to improve the scientific language and clarity of the manuscript.

\section{Another non-unitary RCFT: $\mathcal{M}(7, 2)$} \label{App}

Very recently, the Rational Conformal Field Theory $\mathcal{M}(7, 2)$ has attracted growing interest in the context of RG flows and non‑invertible symmetries \cite{Ambrosino:2025yug, Katsevich:2025ojk, Blakeney:2025ext, Fukusumi:2025xrj}. The model $\mathcal{M}(7, 2)$ is a diagonal, non‑unitary Virasoro minimal model with central charge $c=-\frac{68}{7}$. It has three Virasoro primaries and their non‑trivial fusion rules given by
\begin{eqnarray}
\Phi_2\times\Phi_2 &=& \Phi_1 + \Phi_3\,, \\
\Phi_2\times\Phi_3 &=& \Phi_2 + \Phi_3\,, \\
\Phi_3\times\Phi_3 &=& \Phi_1 + \Phi_2 + \Phi_3\,.
\end{eqnarray}
\noindent
Correspondingly, there are three Verlinde lines (defect lines) namely, $\mathcal{L}_{\Phi_1}, \mathcal{L}_{\Phi_2}$, and $\mathcal{L}_{\Phi_3}$ with quantum dimensions(Perron--Frobenius quantum dimensions) $1, \frac{\cos{\frac{3\pi}{14}}}{\sin{\frac{\pi}{7}}}\sim 1.802, \frac{1}{2\sin{\frac{\pi}{14}}}\sim 2.247$ respectively and three Cardy boundary states ${|\!|\Phi_1\rangle\!\rangle, |\!|\Phi_2\rangle\!\rangle, |\!|\Phi_3\rangle\!\rangle}\,.$ The action of the defect lines on the primaries is given in the Table~(\ref{tabM72}). 
\begin{table}[H]
\centering
\begin{tabular}{cccc}
\toprule
Defect line & $\Phi_1$ & $\Phi_2$ & $\Phi_3$ \\
\midrule
$\mathcal{L}_{\Phi_1}$ & $1$ & $1$ & $1$ \\
$\mathcal{L}_{\Phi_2}$ & $-2\sin{\frac{3\pi}{14}}$ & $2\sin{\frac{\pi}{14}}$ & $\frac{\cos{\frac{3\pi}{14}}}{\sin{\frac{\pi}{7}}}$\\
$\mathcal{L}_{\Phi_3}$ & $\frac{\sin{\frac{\pi}{7}}}{\cos{\frac{3\pi}{14}}}$ & $-\frac{1}{2\sin{\frac{3\pi}{14}}}$ & $\frac{1}{2\sin{\frac{\pi}{14}}}$ \\
\bottomrule 
\end{tabular}
\caption{Action of the $\textrm{psu}(2)_5$ category on the primaries of $\mathcal{M}(7, 2)$.} 
\label{tabM72}
\end{table} 
\noindent
It turns out that the Cardy boundary state $|\!|\Phi_3\rangle\!\rangle$ is {weakly symmetric} under the above fusion category, known as $\textrm{psu}(2)_5$ (is a self-dual, multiplicity-free fusion category). This Cardy boundary state in terms of Ishibashi states is given by
\begin{eqnarray}
|\!|\Phi_3\rangle\!\rangle = \frac{\sqrt{2}}{7^{\frac{1}{4}}} \frac{\sin{\frac{\pi}{7}}}{\sqrt{\cos{\frac{3\pi}{14}}}} |\Phi_1\rangle\!\rangle - \frac{i\sqrt{2}}{7^{\frac{1}{4}}} \frac{\cos{\frac{3\pi}{14}}}{\sqrt{\cos{\frac{\pi}{14}}}} |\Phi_2\rangle\!\rangle + \frac{\sqrt{2}}{7^{\frac{1}{4}}} \frac{\cos{\frac{\pi}{14}}}{\sqrt{\sin{\frac{\pi}{7}}}} |\Phi_3\rangle\!\rangle\,.    
\end{eqnarray}

The annulus partition function in the closed and open-string channels with boundary state $|\!|\Phi_3\rangle\!\rangle$ is given by
\begin{eqnarray}
 Z(q) &=& \langle\!\langle\Phi_3|\!|\tilde{q}^{\color{black}{\frac{1}{2}(\,L_{0}+\bar{L}_0-\frac{c}{12})}}|\!|\Phi_3\rangle\!\rangle\,, \\
 &=& \chi_1(q) + \chi_2(q)+ \chi_3(q)\,.
\end{eqnarray}

From the action of the defect lines on the primaries listed in Table~(\ref{tabM72}), we can write the resolved-partition function in the three sectors, namely,
\begin{align}
\begin{split}
&Z( q^n, \mathcal{R}_1) = {\chi_1(q^n)}\,,\, 
Z(q^n, \mathcal{R}_2) = {\chi_2(q^n)}\,,\,\,Z( q^n, \mathcal{R}_3)={\chi_3(q^n)}\,.
\end{split}
\end{align}
The $\mathcal{M}(7, 2)$ is a non-unitary RCFT; therefore, the leading contribution in the high energy limit comes from the ground state rather than the vacuum. In this example, $|\Omega\rangle=|\Phi_3\rangle\!\rangle$ is a ground state. Therefore, the symmetry-resolved $n$th R\'enyi entropy is
\begin{eqnarray}
S_n(\mathcal{R}_1) = \frac{c_{\rm eff}}{6}\left(\frac{n+1}{n}\right)\log\left(\frac{L}{\epsilon}\right) + \log \frac{2\sin{\frac{\pi}{7}}}{\sqrt{7}}+ \ldots
\end{eqnarray} 
where the constant term 
\begin{eqnarray}
\log \frac{2\sin{\frac{\pi}{7}}}{\sqrt{7}} = \log\left(\frac{d_{\mathcal{R}_1}}{\sum\limits_{\mathcal{R}_i}d_{\mathcal{R}_i}}\right) + \log(\langle\!\langle\Phi_3|\!|\Omega\rangle\langle\Omega|\!|\Phi_3\rangle\!\rangle)\,.
\end{eqnarray}
where, $d_{\mathcal{R}_1}=1$, $\sum\limits_{\mathcal{R}_i} d_{\mathcal{R}_i} = 1+ \frac{\cos{\frac{3\pi}{14}}}{\sin{\frac{\pi}{7}}}+\frac{1}{2\sin{\frac{\pi}{14}}}\sim 5.049$, and $\log(\langle\!\langle\Phi_3|\!|\Omega\rangle\langle\Omega|\Phi_3\rangle\!\rangle)=\log\left(\frac{2}{\sqrt{7}}\frac{\cos^2{\frac{\pi}{14}}}{\sin{\frac{\pi}{7}}}\right)\sim 0.504$ corresponds to the Affleck-Ludwig boundary entropy \cite{Affleck:1991tk}. Similarly, for the other two sectors
\begin{eqnarray}
S_n(\mathcal{R}_2) &=& \frac{c_{\rm eff}}{6}\left(\frac{n+1}{n}\right)\log\left(\frac{L}{\epsilon}\right) + \log \frac{2\cos{\frac{3\pi}{14}}}{\sqrt{7}}+ \ldots \\
S_n(\mathcal{R}_3) &=& \frac{c_{\rm eff}}{6}\left(\frac{n+1}{n}\right)\log\left(\frac{L}{\epsilon}\right) + \log \frac{2\cos{\frac{\pi}{14}}}{\sqrt{7}}+ \ldots
\end{eqnarray} 
and the constant terms are given by respectively
\begin{eqnarray}
\log \frac{2\cos{\frac{3\pi}{14}}}{\sqrt{7}} &=& \log\left(\frac{d_{\mathcal{R}_2}}{\sum\limits_{\mathcal{R}_i}d_{\mathcal{R}_i}}\right) + \log(\langle\!\langle\Phi_3|\!|\Omega\rangle\langle\Omega|\!|\Phi_3\rangle\!\rangle)\,,\\
\log \frac{2\cos{\frac{\pi}{14}}}{\sqrt{7}} &=& \log\left(\frac{d_{\mathcal{R}_3}}{\sum\limits_{\mathcal{R}_i}d_{\mathcal{R}_i}}\right) + \log(\langle\!\langle\Phi_3|\!|\Omega\rangle\langle\Omega|\!|\Phi_3\rangle\!\rangle) 
\end{eqnarray}
where, $d_{\mathcal{R}_2}=\frac{\cos{\frac{3\pi}{14}}}{\sin{\frac{\pi}{7}}}\sim 1.802$, $d_{\mathcal{R}_3}=\frac{1}{2\sin{\frac{\pi}{14}}}\sim 2.247\,.$ 


\bibliography{refN}

@book{Wigner:1959,
  author    = {Wigner, E. P.},
  title     = {Group Theory and Its Application to the Quantum Mechanics of Atomic Spectra},
  publisher = {Academic Press},
  address   = {New York},
  year      = {1959}
}

@article{Cardy:1986gw,
  author       = {Cardy, J. L.},
  title        = {Effect of Boundary Conditions on the Operator Content of Two-Dimensional Conformally Invariant Theories},
  journal      = {Nucl. Phys. B},
  volume       = {275},
  pages        = {200--218},
  year         = {1986},
  doi          = {10.1016/0550-3213(86)90596-1}
}

@article{Verlinde:1988sn,
  author       = {Verlinde, E. P.},
  title        = {Fusion Rules and Modular Transformations in 2D Conformal Field Theory},
  journal      = {Nucl. Phys. B},
  volume       = {300},
  pages        = {360--376},
  year         = {1988},
  doi          = {10.1016/0550-3213(88)90603-7}
}

@article{Cardy:1989ir,
  author       = {Cardy, J. L.},
  title        = {{Boundary Conditions, Fusion Rules and the Verlinde Formula}},
  journal      = {Nucl. Phys. B},
  volume       = {324},
  pages        = {581--596},
  year         = {1989},
  doi          = {10.1016/0550-3213(89)90521-X}
}

@inproceedings{Cardy:1989vyr,
  author       = {Cardy, J.},
  title        = {Boundary Conditions in Conformal Field Theory},
  booktitle    = {Adv. Stud. Pure Math.},
  volume       = {19},
  pages        = {127--148},
  year         = {1989}
}

@misc{Moore:1989vd,
  author       = {Moore, G. W. and Seiberg, N.},
  title        = {{Lectures on RCFT}},
  howpublished = {RU-89-32},
  year         = {1989}
}

@article{Affleck:1991tk,
  author       = {Affleck, I. and Ludwig, A. W. W.},
  title        = {Universal noninteger `ground state degeneracy' in critical quantum systems},
  journal      = {Phys. Rev. Lett.},
  volume       = {67},
  pages        = {161--164},
  year         = {1991},
  doi          = {10.1103/PhysRevLett.67.161}
}

@article{Cardy:1991tv,
  author       = {Cardy, J. L. and Lewellen, D. C.},
  title        = {Bulk and boundary operators in conformal field theory},
  journal      = {Phys. Lett. B},
  volume       = {259},
  pages        = {274--278},
  year         = {1991},
  doi          = {10.1016/0370-2693(91)90828-E}
}

@article{Petkova:2000ip,
  author       = {Petkova, V. B. and Zuber, J. B.},
  title        = {Generalized twisted partition functions},
  journal      = {Phys. Lett. B},
  volume       = {504},
  pages        = {157--164},
  year         = {2001},
  doi          = {10.1016/S0370-2693(01)00276-3},
  eprint       = {hep-th/0011021},
  archivePrefix= {arXiv}
}

@misc{Cardy:2004hm,
  author       = {Cardy, J. L.},
  title        = {Boundary conformal field theory},
  year         = {2004},
  eprint       = {hep-th/0411189},
  archivePrefix= {arXiv}
}

@article{Bantay:2005vk,
    author = "Bantay, P. and Gannon, T.",
    title = "{Conformal characters and the modular representation}",
    eprint = "hep-th/0512011",
    archivePrefix = "arXiv",
    doi = "10.1088/1126-6708/2006/02/005",
    journal = "JHEP",
    volume = "02",
    pages = "005",
    year = "2006"
}

@article{Bantay:2007zz,
    author = "Bantay, Peter and Gannon, Terry",
    title = "{Vector-valued modular functions for the modular group and the hypergeometric equation}",
    doi = "10.4310/CNTP.2007.v1.n4.a2",
    journal = "Commun. Num. Theor. Phys.",
    volume = "1",
    pages = "651--680",
    year = "2007"
}

@article{Evans:2010yr,
  author       = {Evans, D. E. and Gannon, T.},
  title        = {{The exoticness and realisability of twisted Haagerup-Izumi modular data}},
  journal      = {Commun. Math. Phys.},
  volume       = {307},
  pages        = {463--512},
  year         = {2011},
  doi          = {10.1007/s00220-011-1329-3},
  eprint       = {1006.1326},
  archivePrefix= {arXiv}
}

@article{Gannon:2025birs,
  author       = {Gannon, T.},
  title        = {{CFT} and Modular Forms, \href{www.birs.ca/events/2025/5-day-workshops/25w5318/videos/watch/202507211030-Gannon.html}{Workshop 25w5318: Recent Developments in Logarithmic Conformal Field Theory at BIRS}},
journal      = {July 21, 2025},
  volume       = {},
  pages        = {},
  year         = {},
  doi          = {},
  eprint       = {}
}

@article{Gaiotto:2014kfa,
  author       = {Gaiotto, D. and Kapustin, A. and Seiberg, N. and Willett, B.},
  title        = {Generalized Global Symmetries},
  journal      = {JHEP},
  volume       = {02},
  pages        = {172},
  year         = {2015},
  doi          = {10.1007/JHEP02(2015)172},
  eprint       = {1412.5148},
  archivePrefix= {arXiv}
}

@article{Huang:2021nvb,
  author       = {Huang, T. C. and Lin, Y. H. and Ohmori, K. and Tachikawa, Y. and Tezuka, M.},
  title        = {{Numerical Evidence for a Haagerup Conformal Field Theory}},
  journal      = {Phys. Rev. Lett.},
  volume       = {128},
  pages        = {231603},
  year         = {2022},
  doi          = {10.1103/PhysRevLett.128.231603},
  eprint       = {2110.03008},
  archivePrefix= {arXiv}
}

@article{Vanhove:2021zop,
  author       = {Vanhove, R. and others},
  title        = {{Critical Lattice Model for a Haagerup Conformal Field Theory}},
  journal      = {Phys. Rev. Lett.},
  volume       = {128},
  pages        = {231602},
  year         = {2022},
  doi          = {10.1103/PhysRevLett.128.231602},
  eprint       = {2110.03532},
  archivePrefix= {arXiv}
}

@article{Jia:2024wnu,
author = {Jia, Zhian},
title = {{Quantum Cluster State Model with Haagerup Fusion Category Symmetry}},
journal = {Advanced Quantum Technologies},
volume = {},
number = {},
pages = {e00305},
year = {2025},
doi = {https://doi.org/10.1002/qute.202500305},
book = {https://advanced.onlinelibrary.wiley.com/doi/abs/10.1002/qute.202500305},
eprint = {2412.19657},
archivePrefix= {arXiv},
primaryClass= {math.QA}
}

@article{Choi:2022jqy,
  author       = {Choi, Y. and Lam, H. T. and Shao, S. H.},
  title        = {Noninvertible Global Symmetries in the Standard Model},
  journal      = {Phys. Rev. Lett.},
  volume       = {129},
  pages        = {161601},
  year         = {2022},
  doi          = {10.1103/PhysRevLett.129.161601},
  eprint       = {2205.05086},
  archivePrefix= {arXiv}
}

@article{Kusuki:2023bsp,
  author       = {Kusuki, Y. and others},
  title        = {Symmetry-resolved entanglement entropy, spectra \& boundary conformal field theory},
  journal      = {JHEP},
  volume       = {11},
  pages        = {216},
  year         = {2023},
  doi          = {10.1007/JHEP11(2023)216},
  eprint       = {2309.03287},
  archivePrefix= {arXiv}
}

@article{Islam:2015mom,
    author = "Islam, Rajibul and Ma, Ruichao and Preiss, Philipp M. and Tai, M. Eric and Lukin, Alexander and Rispoli, Matthew and Greiner, Markus",
    title = "{Measuring entanglement entropy through the interference of quantum many-body twins}",
    journal = {Nature},
    volume = {528},
    pages = {77-83},
    eprint = "1509.01160",
    archivePrefix = "arXiv",
    primaryClass = "cond-mat.quant-gas",
    doi = {10.1038/nature15750},
    year = {2015},
    URL = {https://doi.org/10.1038/nature15750}
}

@article{
doi:10.1126/science.aau0818,
author = {Alexander Lukin  and Matthew Rispoli  and Robert Schittko  and M. Eric Tai  and Adam M. Kaufman  and Soonwon Choi  and Vedika Khemani  and Julian Léonard  and Markus Greiner },
title = {Probing entanglement in a many-body–localized system},
journal = {Science},
volume = {364},
number = {6437},
pages = {256-260},
year = {2019},
doi = {10.1126/science.aau0818},
URL = {https://www.science.org/doi/abs/10.1126/science.aau0818},
eprint = {https://www.science.org/doi/pdf/10.1126/science.aau0818}}

@article{Gang:2022kpe,
    author = "Gang, Dongmin and Kim, Dongyeob",
    title = "{Generalized non-unitary Haagerup-Izumi modular data from 3D S-fold SCFTs}",
    eprint = "2211.13561",
    archivePrefix = "arXiv",
    primaryClass = "hep-th",
    doi = "10.1007/JHEP03(2023)185",
    journal = "JHEP",
    volume = "03",
    pages = "185",
    year = "2023"
}

@article{Gang:2023ggt,
  author       = {Gang, D. and Kim, D. and Lee, S.},
  title        = {{A non-unitary bulk-boundary correspondence: Non-unitary Haagerup RCFTs from S-fold SCFTs}},
  journal      = {SciPost Phys.},
  volume       = {17},
  pages        = {064},
  year         = {2024},
  doi          = {10.21468/SciPostPhys.17.2.064},
  eprint       = {2310.14877},
  archivePrefix= {arXiv}
}

@article{Gang:2023rei,
    author = "Gang, Dongmin and Kim, Heeyeon and Stubbs, Spencer",
    title = "{Three-Dimensional Topological Field Theories and Nonunitary Minimal Models}",
    eprint = "2310.09080",
    archivePrefix = "arXiv",
    primaryClass = "hep-th",
    doi = "10.1103/PhysRevLett.132.131601",
    journal = "Phys. Rev. Lett.",
    volume = "132",
    number = "13",
    pages = "131601",
    year = "2024"
}

@article{Choi:2023PRD,
  author       = {Choi, J. and Rayhaun, B. C. and Sanghavi, Y. and Shao, S.-H.},
  title        = {Remarks on boundaries, anomalies, and noninvertible symmetries},
  journal      = {Phys. Rev. D},
  volume       = {108},
  pages        = {125005},
  year         = {2023},
  doi          = {10.1103/PhysRevD.108.125005}
}

@article{Heymann:2024vvf,
  author       = {Heymann, J. and Quella, T.},
  title        = {Revisiting the symmetry-resolved entanglement for noninvertible symmetries in 1+1d conformal field theories},
  journal      = {Phys. Rev. D},
  volume       = {112},
  number       = {2},
  pages        = {025004},
  year         = {2025},
  doi          = {10.1103/PhysRevD.112.025004},
  eprint       = {2409.02315},
  archivePrefix= {arXiv},
  primaryClass = {hep-th}
}

@article{Banerjee:2024ldl,
  author       = {Banerjee, A. and others},
  title        = {Symmetry resolution in non-Lorentzian field theories},
  journal      = {JHEP},
  volume       = {06},
  pages        = {121},
  year         = {2024},
  doi          = {10.1007/JHEP06(2024)121},
  eprint       = {2404.02206},
  archivePrefix= {arXiv}
}

@article{Donnelly:2018ppr,
    author = "Donnelly, William and Wong, Gabriel",
    title = "{Entanglement branes, modular flow, and extended topological quantum field theory}",
    eprint = "1811.10785",
    archivePrefix = "arXiv",
    primaryClass = "hep-th",
    doi = "10.1007/JHEP10(2019)016",
    journal = "JHEP",
    volume = "10",
    pages = "016",
    year = "2019"
}

@article{Brehm:2022JPA,
  author       = {Brehm, E. M. and Runkel, I.},
  title        = {{Lattice models from CFT on surfaces with holes: I. Torus partition function via two lattice cells}},
  journal      = {J. Phys. A},
  volume       = {55},
  pages        = {235001},
  year         = {2022},
  doi          = {10.1088/1751-8121/ac6a91}
}

@article{Brehm:2024arXiv,
  author       = {Brehm, E. M. and Runkel, I.},
  title        = {{Lattice models from CFT on surfaces with holes II: Cloaking boundary conditions and loop models}},
  year         = {2024},
  eprint       = {2410.19938},
  archivePrefix= {arXiv},
  primaryClass = {hep-th}
}

@article{Choi:2024wfm,
  author       = {Choi, Y. and Rayhaun, B. C. and Zheng, Y.},
  title        = {{Noninvertible Symmetry-Resolved Affleck-Ludwig-Cardy Formula and Entanglement Entropy from the Boundary Tube Algebra}},
  journal      = {Phys. Rev. Lett.},
  volume       = {133},
  pages        = {251602},
  year         = {2024},
  doi          = {10.1103/PhysRevLett.133.251602},
  eprint       = {2409.02806},
  archivePrefix= {arXiv}
}

@article{Copetti:2024dcz,
  author       = {Copetti, C. and Cordova, L. and Komatsu, S.},
  title        = {S-matrix bootstrap and non-invertible symmetries},
  journal      = {JHEP},
  volume       = {03},
  pages        = {204},
  year         = {2025},
  doi          = {10.1007/JHEP03(2025)204},
  eprint       = {2408.13132},
  archivePrefix= {arXiv},
  primaryClass = {hep-th}
}

@article{Choi:2024tri,
    author = "Choi, Yichul and Rayhaun, Brandon C. and Zheng, Yunqin",
    title = "{Generalized Tube Algebras, Symmetry-Resolved Partition Functions, and Twisted Boundary States}",
    eprint = "2409.02159",
    archivePrefix = "arXiv",
    primaryClass = "hep-th",
    month = "9",
    year = "2024"
}

@article{Roy:2025hew,
    author = "Roy, Ananda and Lukyanov, Sergei L. and Saleur, Hubert",
    title = "{Boundary conditions for the entanglement cut in two-dimensional conformal field theories}",
    eprint = "2503.12674",
    archivePrefix = "arXiv",
    primaryClass = "quant-ph",
    doi = "10.1103/677y-pxs5",
    journal = "Phys. Rev. B",
    volume = "112",
    number = "12",
    pages = "L121406",
    year = "2025"
}

@article{Ambrosino:2025yug,
    author = "Ambrosino, Federico and Negro, Stefano",
    title = "{Minimal Model Renormalization Group Flows: Noninvertible Symmetries and Nonperturbative Description}",
    eprint = "2501.07511",
    archivePrefix = "arXiv",
    primaryClass = "hep-th",
    reportNumber = "DESY-25-003",
    doi = "10.1103/dg1s-5vp6",
    journal = "Phys. Rev. Lett.",
    volume = "135",
    number = "2",
    pages = "021602",
    year = "2025"
}

@article{Katsevich:2025ojk,
    author = "Katsevich, Andrei and Klebanov, Igor R. and Sun, Zimo and Tarnopolsky, Grigory",
    title = "{Towards a Quintic Ginzburg-Landau Description of the $(2,7)$ Minimal Model}",
    eprint = "2510.19085",
    archivePrefix = "arXiv",
    primaryClass = "hep-th",
    month = "10",
    year = "2025"
}

@article{Blakeney:2025ext,
    author = "Blakeney, Matthew and Corcoran, Luke and de Leeuw, Marius and Pozsgay, Balazs and Vernier, Eric",
    title = "{Temperley-Lieb integrable models and fusion categories}",
    eprint = "2510.19902",
    archivePrefix = "arXiv",
    primaryClass = "cond-mat.str-el",
    month = "10",
    year = "2025"
}

@article{Das:2024qdx,
    author = "Das, Arpit and Molina-Vilaplana, Javier and Saura-Bastida, Pablo",
    title = "{Generalized symmetry resolution of entanglement in conformal field theory for twisted and anyonic sectors}",
    eprint = "2409.02162",
    archivePrefix = "arXiv",
    primaryClass = "hep-th",
    doi = "10.1103/PhysRevD.110.125005",
    journal = "Phys. Rev. D",
    volume = "110",
    number = "12",
    pages = "125005",
    year = "2024"
}

@article{Govindarajan:2025rgh,
    author = "Govindarajan, Suresh and Santara, Jagannath",
    title = "{Two approaches to the holomorphic modular bootstrap}",
    eprint = "2503.23761",
    archivePrefix = "arXiv",
    primaryClass = "hep-th",
    doi = "10.1007/JHEP10(2025)181",
    journal = "JHEP",
    volume = "10",
    pages = "181",
    year = "2025"
}

@article{Govindarajan:2025jlq,
    author = "Govindarajan, Suresh and Sadanandan, Akhila and Santara, Jagannath",
    title = "{Quasi-Characters for three-character Rational Conformal Field Theories}",
    eprint = "2510.24248",
    archivePrefix = "arXiv",
    primaryClass = "hep-th",
    month = "10",
    year = "2025"
}

@article{Mathur:1988na,
    author = "Mathur, Samir D. and Mukhi, Sunil and Sen, Ashoke",
    title = "{On the Classification of Rational Conformal Field Theories}",
    reportNumber = "TIFR/TH/88-39",
    doi = "10.1016/0370-2693(88)91765-0",
    journal = "Phys. Lett. B",
    volume = "213",
    pages = "303--308",
    year = "1988"
}

@article{Mathur:1988gt,
    author = "Mathur, Samir D. and Mukhi, Sunil and Sen, Ashoke",
    title = "{Reconstruction of Conformal Field Theories From Modular Geometry on the Torus}",
    reportNumber = "TIFR/TH/88-50",
    doi = "10.1016/0550-3213(89)90615-9",
    journal = "Nucl. Phys. B",
    volume = "318",
    pages = "483--540",
    year = "1989"
}

@article{Naculich:1988xv,
    author = "Naculich, Stephen G.",
    title = "{DIFFERENTIAL EQUATIONS FOR RATIONAL CONFORMAL CHARACTERS}",
    reportNumber = "BRX-TH-257",
    doi = "10.1016/0550-3213(89)90150-8",
    journal = "Nucl. Phys. B",
    volume = "323",
    pages = "423--440",
    year = "1989"
}

@article{Hampapura:2015cea,
    author = "Hampapura, Harsha R. and Mukhi, Sunil",
    title = "{On 2d Conformal Field Theories with Two Characters}",
    eprint = "1510.04478",
    archivePrefix = "arXiv",
    primaryClass = "hep-th",
    doi = "10.1007/JHEP01(2016)005",
    journal = "JHEP",
    volume = "01",
    pages = "005",
    year = "2016"
}

@article{Chandra:2018pjq,
    author = "Chandra, A. Ramesh and Mukhi, Sunil",
    title = "{Towards a Classification of Two-Character Rational Conformal Field Theories}",
    eprint = "1810.09472",
    archivePrefix = "arXiv",
    primaryClass = "hep-th",
    doi = "10.1007/JHEP04(2019)153",
    journal = "JHEP",
    volume = "04",
    pages = "153",
    year = "2019"
}

@article{Das:2021uvd,
    author = "Das, Arpit and Gowdigere, Chethan N. and Santara, Jagannath",
    title = "{Classifying three-character RCFTs with Wronskian index equalling 0 or 2}",
    eprint = "2108.01060",
    archivePrefix = "arXiv",
    primaryClass = "hep-th",
    doi = "10.1007/JHEP11(2021)195",
    journal = "JHEP",
    volume = "11",
    pages = "195",
    year = "2021"
}

@article{Lin:2022dhv,
    author = "Lin, Ying-Hsuan and Okada, Masaki and Seifnashri, Sahand and Tachikawa, Yuji",
    title = "{Asymptotic density of states in 2d CFTs with non-invertible symmetries}",
    eprint = "2208.05495",
    archivePrefix = "arXiv",
    primaryClass = "hep-th",
    reportNumber = "YITP-SB-2022-29",
    doi = "10.1007/JHEP03(2023)094",
    journal = "JHEP",
    volume = "03",
    pages = "094",
    year = "2023"
}

@book{DiFrancesco:1997nk,
	address = {New York},
	author = {Di Francesco, P. and Mathieu, P. and Senechal, D.},
	doi = {10.1007/978-1-4612-2256-9},
	isbn = {9780387947853, 9781461274759},
	publisher = {Springer-Verlag},
	series = {Graduate Texts in Contemporary Physics},
	slaccitation = {%%CITATION = INSPIRE-454643;%%},
	title = {{Conformal Field Theory}},
	year = {1997}
}

@article{Chang:2018iay,
    author = "Chang, Chi-Ming and Lin, Ying-Hsuan and Shao, Shu-Heng and Wang, Yifan and Yin, Xi",
    title = "{Topological Defect Lines and Renormalization Group Flows in Two Dimensions}",
    eprint = "1802.04445",
    archivePrefix = "arXiv",
    primaryClass = "hep-th",
    reportNumber = "CALT-TH-2017-067, CALT-TH 2017-067, PUPT-2546",
    doi = "10.1007/JHEP01(2019)026",
    journal = "JHEP",
    volume = "01",
    pages = "026",
    year = "2019"
}

@article{Seiberg:2023cdc,
    author = "Seiberg, Nathan and Shao, Shu-Heng",
    title = "{Majorana chain and Ising model - (non-invertible) translations, anomalies, and emanant symmetries}",
    eprint = "2307.02534",
    archivePrefix = "arXiv",
    primaryClass = "cond-mat.str-el",
    reportNumber = "YITP-SB-2023-14",
    doi = "10.21468/SciPostPhys.16.3.064",
    journal = "SciPost Phys.",
    volume = "16",
    number = "3",
    pages = "064",
    year = "2024"
}

@article{Cordova:2022ieu,
    author = "Cordova, Clay and Ohmori, Kantaro",
    title = "{Noninvertible Chiral Symmetry and Exponential Hierarchies}",
    eprint = "2205.06243",
    archivePrefix = "arXiv",
    primaryClass = "hep-th",
    doi = "10.1103/PhysRevX.13.011034",
    journal = "Phys. Rev. X",
    volume = "13",
    number = "1",
    pages = "011034",
    year = "2023"
}

@inproceedings{Cordova:2022ruw,
    author = "Cordova, Clay and Dumitrescu, Thomas T. and Intriligator, Kenneth and Shao, Shu-Heng",
    title = "{Snowmass White Paper: Generalized Symmetries in Quantum Field Theory and Beyond}",
    booktitle = "{Snowmass 2021}",
    eprint = "2205.09545",
    archivePrefix = "arXiv",
    primaryClass = "hep-th",
    month = "5",
    year = "2022"
}

@inproceedings{Costa:2024wks,
    author = "Costa, Davi and others",
    title = "{Simons Lectures on Categorical Symmetries}",
    eprint = "2411.09082",
    archivePrefix = "arXiv",
    primaryClass = "math-ph",
    month = "11",
    year = "2024"
}

@article{Kitaev:2005dm,
    author = "Kitaev, Alexei and Preskill, John",
    title = "{Topological entanglement entropy}",
    eprint = "hep-th/0510092",
    archivePrefix = "arXiv",
    reportNumber = "CALT-68-2578",
    doi = "10.1103/PhysRevLett.96.110404",
    journal = "Phys. Rev. Lett.",
    volume = "96",
    pages = "110404",
    year = "2006"
}

@article{PhysRevLett.96.110405,
  title = {Detecting Topological Order in a Ground State Wave Function},
  author = {Levin, Michael and Wen, Xiao-Gang},
  journal = {Phys. Rev. Lett.},
  volume = {96},
  issue = {11},
  pages = {110405},
  numpages = {4},
  year = {2006},
  month = {Mar},
  publisher = {American Physical Society},
  doi = {10.1103/PhysRevLett.96.110405},
  url = {https://link.aps.org/doi/10.1103/PhysRevLett.96.110405}
}

@article{Benedetti:2024dku,
    author = "Benedetti, Valentin and Casini, Horacio and Kawahigashi, Yasuyuki and Longo, Roberto and Magan, Javier M.",
    title = "{Modular invariance as completeness}",
    eprint = "2408.04011",
    archivePrefix = "arXiv",
    primaryClass = "hep-th",
    doi = "10.1103/PhysRevD.110.125004",
    journal = "Phys. Rev. D",
    volume = "110",
    number = "12",
    pages = "125004",
    year = "2024"
}

@article{Magan:2021myk,
    author = "Magan, Javier M.",
    title = "{Proof of the universal density of charged states in QFT}",
    eprint = "2111.02418",
    archivePrefix = "arXiv",
    primaryClass = "hep-th",
    doi = "10.1007/JHEP12(2021)100",
    journal = "JHEP",
    volume = "12",
    pages = "100",
    year = "2021"
}

@article{Northe:2023khz,
    author = "Northe, Christian",
    title = "{Entanglement Resolution with Respect to Conformal Symmetry}",
    eprint = "2303.07724",
    archivePrefix = "arXiv",
    primaryClass = "hep-th",
    doi = "10.1103/PhysRevLett.131.151601",
    journal = "Phys. Rev. Lett.",
    volume = "131",
    number = "15",
    pages = "151601",
    year = "2023"
}

@article{Murciano:2020vgh,
    author = "Murciano, Sara and Di Giulio, Giuseppe and Calabrese, Pasquale",
    title = "{Entanglement and symmetry resolution in two dimensional free quantum field theories}",
    eprint = "2006.09069",
    archivePrefix = "arXiv",
    primaryClass = "hep-th",
    doi = "10.1007/JHEP08(2020)073",
    journal = "JHEP",
    volume = "08",
    pages = "073",
    year = "2020"
}

@article{DiGiulio:2022jjd,
    author = "Di Giulio, Giuseppe and Meyer, Ren\'e and Northe, Christian and Scheppach, Henri and Zhao, Suting",
    title = "{On the boundary conformal field theory approach to symmetry-resolved entanglement}",
    eprint = "2212.09767",
    archivePrefix = "arXiv",
    primaryClass = "hep-th",
    doi = "10.21468/SciPostPhysCore.6.3.049",
    journal = "SciPost Phys. Core",
    volume = "6",
    pages = "049",
    year = "2023"
}

@article{Bhardwaj:2023wzd,
    author = "Bhardwaj, Lakshya and Schafer-Nameki, Sakura",
    title = "{Generalized charges, part I: Invertible symmetries and higher representations}",
    eprint = "2304.02660",
    archivePrefix = "arXiv",
    primaryClass = "hep-th",
    doi = "10.21468/SciPostPhys.16.4.093",
    journal = "SciPost Phys.",
    volume = "16",
    number = "4",
    pages = "093",
    year = "2024"
}

@article{Bhardwaj:2023ayw,
    author = "Bhardwaj, Lakshya and Schafer-Nameki, Sakura",
    title = "{Generalized charges, part II: Non-invertible symmetries and the symmetry TFT}",
    eprint = "2305.17159",
    archivePrefix = "arXiv",
    primaryClass = "hep-th",
    doi = "10.21468/SciPostPhys.19.4.098",
    journal = "SciPost Phys.",
    volume = "19",
    number = "4",
    pages = "098",
    year = "2025"
}

@inproceedings{Iqbal:2024pee,
    author = "Iqbal, Nabil",
    title = "{Jena lectures on generalized global symmetries: principles and applications}",
    eprint = "2407.20815",
    archivePrefix = "arXiv",
    primaryClass = "hep-th",
    month = "7",
    year = "2024"
}

@article{Bartsch:2023wvv,
    author = "Bartsch, Thomas and Bullimore, Mathew and Grigoletto, Andrea",
    title = "{Representation theory for categorical symmetries}",
    eprint = "2305.17165",
    archivePrefix = "arXiv",
    primaryClass = "hep-th",
    month = "5",
    year = "2023"
}

@article{Bhardwaj:2017xup,
    author = "Bhardwaj, Lakshya and Tachikawa, Yuji",
    title = "{On finite symmetries and their gauging in two dimensions}",
    eprint = "1704.02330",
    archivePrefix = "arXiv",
    primaryClass = "hep-th",
    reportNumber = "IPMU-17-0049",
    doi = "10.1007/JHEP03(2018)189",
    journal = "JHEP",
    volume = "03",
    pages = "189",
    year = "2018"
}

@article{Gu:2023yhm,
    author = "Gu, Xia and Xie, Xianjin",
    title = "{Generalized Cardy conditions of topological defect lines}",
    eprint = "2310.15487",
    archivePrefix = "arXiv",
    primaryClass = "hep-th",
    month = "10",
    year = "2023"
}

@article{Casini:2019kex,
    author = "Casini, Horacio and Huerta, Marina and Mag\'an, Javier M. and Pontello, Diego",
    title = "{Entanglement entropy and superselection sectors. Part I. Global symmetries}",
    eprint = "1905.10487",
    archivePrefix = "arXiv",
    primaryClass = "hep-th",
    doi = "10.1007/JHEP02(2020)014",
    journal = "JHEP",
    volume = "02",
    pages = "014",
    year = "2020"
}

@article{AliAhmad:2025bnd,
    author = "Ali Ahmad, Shadi and Klinger, Marc S. and Wang, Yifan",
    title = "{The Many Faces of Non-invertible Symmetries}",
    eprint = "2509.18072",
    archivePrefix = "arXiv",
    primaryClass = "hep-th",
    month = "9",
    year = "2025"
}

@article{McGreevy:2022oyu,
    author = "McGreevy, John",
    title = "{Generalized Symmetries in Condensed Matter}",
    eprint = "2204.03045",
    archivePrefix = "arXiv",
    primaryClass = "cond-mat.str-el",
    doi = "10.1146/annurev-conmatphys-040721-021029",
    journal = "Ann. Rev. Condensed Matter Phys.",
    volume = "14",
    pages = "57--82",
    year = "2023"
}

@article{Castro-Alvaredo:2024azg,
    author = "Castro-Alvaredo, Olalla A. and Santamar\'\i{}a-Sanz, Luc\'\i{}a",
    title = "{Symmetry Resolved Measures in Quantum Field Theory: a Short Review}",
    eprint = "2403.06652",
    archivePrefix = "arXiv",
    primaryClass = "hep-th",
    month = "3",
    year = "2024"
}

@article{Bottini:2025hri,
    author = "Bottini, Lea E. and Schafer-Nameki, Sakura",
    title = "{Construction of a Gapless Phase with Haagerup Symmetry}",
    eprint = "2410.19040",
    archivePrefix = "arXiv",
    primaryClass = "hep-th",
    doi = "10.1103/PhysRevLett.134.191602",
    journal = "Phys. Rev. Lett.",
    volume = "134",
    number = "19",
    pages = "191602",
    year = "2025"
}

@article{Haagerup1994,
  author = {Haagerup, Uffe},
  title = {Principal graphs of subfactors in the index range $4< [M:N] <
  3+ \sqrt{2}$},
  booktitle = {Subfactors (Kyuzeso, 1993)},
  journal = {\em Subfactors (Proceedings of Taniguchi Symposium,
  Kyuzeso, 1993)},
  volume = "1",
  pages = {1--38},
  year = {1994},
  }

@article{Schafer-Nameki:2023jdn,
    author = "Schafer-Nameki, Sakura",
    title = "{ICTP lectures on (non-)invertible generalized symmetries}",
    eprint = "2305.18296",
    archivePrefix = "arXiv",
    primaryClass = "hep-th",
    doi = "10.1016/j.physrep.2024.01.007",
    journal = "Phys. Rept.",
    volume = "1063",
    pages = "1--55",
    year = "2024"
}

@inproceedings{Shao:2023gho,
    author = "Shao, Shu-Heng",
    title = "{What's Done Cannot Be Undone: TASI Lectures on Non-Invertible Symmetries}",
    booktitle = "{Theoretical Advanced Study Institute in Elementary Particle Physics 2023}: {Aspects of Symmetry}",
    eprint = "2308.00747",
    archivePrefix = "arXiv",
    primaryClass = "hep-th",
    reportNumber = "YITP-SB-2023-19",
    month = "8",
    year = "2023"
}

@article{Ohmori:2014eia,
    author = "Ohmori, Kantaro and Tachikawa, Yuji",
    title = "{Physics at the entangling surface}",
    eprint = "1406.4167",
    archivePrefix = "arXiv",
    primaryClass = "hep-th",
    reportNumber = "IPMU-14-0131, UT-14-28, IPMU-14-0131, UT-14-28",
    doi = "10.1088/1742-5468/2015/04/P04010",
    journal = "J. Stat. Mech.",
    volume = "1504",
    pages = "P04010",
    year = "2015"
}

@article{Cardy_2016,
doi = {10.1088/1742-5468/2016/12/123103},
url = {https://doi.org/10.1088/1742-5468/2016/12/123103},
year = {2016},
month = {dec},
publisher = {IOP Publishing and SISSA},
volume = {2016},
number = {12},
pages = {123103},
author = {Cardy, John and Tonni, Erik},
title = {{Entanglement Hamiltonians in two-dimensional conformal field theory}},
journal = {Journal of Statistical Mechanics: Theory and Experiment}
}

@article{Ardonne_2010,
   title={{Clebsch–Gordan and 6j-coefficients for rank 2 quantum groups}},
   volume={43},
   ISSN={1751-8121},
   url={http://dx.doi.org/10.1088/1751-8113/43/39/395205},
   DOI={10.1088/1751-8113/43/39/395205},
   number={39},
   journal={Journal of Physics A: Mathematical and Theoretical},
   publisher={IOP Publishing},
   author={Ardonne, Eddy and Slingerland, Joost},
   year={2010},
   month=aug, pages={395205} }

@article{vercleyen2023lowrankfusionrings,
      title={On Low Rank Fusion Rings}, 
      author={Gert Vercleyen and Joost Slingerland},
      year={2023},
      eprint={2205.15637},
      archivePrefix={arXiv},
      primaryClass={math-ph},
      url={https://arxiv.org/abs/2205.15637}, 
}

@article{Capizzi:2021kys,
    author = "Capizzi, Luca and Horv{\'a}th, D{\'a}vid X. and Calabrese, Pasquale and Castro-Alvaredo, Olalla A.",
    title = "{Entanglement of the 3-state Potts model via form factor bootstrap: total and symmetry resolved entropies}",
    eprint = "2108.10935",
    archivePrefix = "arXiv",
    primaryClass = "hep-th",
    doi = "10.1007/JHEP05(2022)113",
    journal = "JHEP",
    volume = "05",
    pages = "113",
    year = "2022"
}

@article{Affleck:1998nq,
    author = "Affleck, Ian and Oshikawa, Masaki and Saleur, Hubert",
    title = "{Boundary critical phenomena in the three state Potts model}",
    eprint = "cond-mat/9804117",
    archivePrefix = "arXiv",
    reportNumber = "NSF-ITP-98-019",
    doi = "10.1088/0305-4470/31/28/003",
    journal = "J. Phys. A",
    volume = "31",
    pages = "5827",
    year = "1998"
}

@article{Behrend:1998mu,
    author = "Behrend, Roger E. and Pearce, Paul A. and Zuber, Jean-Bernard",
    title = "{Integrable boundaries, conformal boundary conditions and A-D-E fusion rules}",
    eprint = "hep-th/9807142",
    archivePrefix = "arXiv",
    reportNumber = "SACLAY-SPH-T-98-076",
    doi = "10.1088/0305-4470/31/50/001",
    journal = "J. Phys. A",
    volume = "31",
    pages = "L763--L770",
    year = "1998"
}

@article{Behrend:1999bn,
    author = "Behrend, Roger E. and Pearce, Paul A. and Petkova, Valentina B. and Zuber, Jean-Bernard",
    title = "{Boundary conditions in rational conformal field theories}",
    eprint = "hep-th/9908036",
    archivePrefix = "arXiv",
    reportNumber = "SACLAY-SPH-T-99-085",
    doi = "10.1016/S0550-3213(99)00592-1",
    journal = "Nucl. Phys. B",
    volume = "570",
    pages = "525--589",
    year = "2000"
}

@misc{hagge2007nonbraidedfusioncategoriesrank,
      title={Some non-braided fusion categories of rank 3}, 
      author={Tobias J. Hagge and Seung-Moon Hong},
      year={2007},
      eprint={0704.0208},
      archivePrefix={arXiv},
      primaryClass={math.GT},
      url={https://arxiv.org/abs/0704.0208}, 
}

@article{Heidenreich:2021xpr,
    author = "Heidenreich, Ben and McNamara, Jacob and Montero, Miguel and Reece, Matthew and Rudelius, Tom and Valenzuela, Irene",
    title = "{Non-invertible global symmetries and completeness of the spectrum}",
    eprint = "2104.07036",
    archivePrefix = "arXiv",
    primaryClass = "hep-th",
    reportNumber = "ACFI-T21-03",
    doi = "10.1007/JHEP09(2021)203",
    journal = "JHEP",
    volume = "09",
    pages = "203",
    year = "2021"
}

@article{Kitaev:1997wr,
    author = "Kitaev, A. Yu.",
    title = "{Fault tolerant quantum computation by anyons}",
    eprint = "quant-ph/9707021",
    archivePrefix = "arXiv",
    doi = "10.1016/S0003-4916(02)00018-0",
    journal = "Annals Phys.",
    volume = "303",
    pages = "2--30",
    year = "2003"
}

@article{Dennis:2001nw,
    author = "Dennis, Eric and Kitaev, Alexei and Landahl, Andrew and Preskill, John",
    title = "{Topological quantum memory}",
    eprint = "quant-ph/0110143",
    archivePrefix = "arXiv",
    reportNumber = "CALT-68-2346, CALT-68-2346",
    doi = "10.1063/1.1499754",
    journal = "J. Math. Phys.",
    volume = "43",
    pages = "4452--4505",
    year = "2002"
}

@article{Feiguin:2006ydp,
    author = "Feiguin, Adrian and Trebst, Simon and Ludwig, Andreas W. W. and Troyer, Matthias and Kitaev, Alexei and Wang, Zhenghan and Freedman, Michael H.",
    title = "{Interacting anyons in topological quantum liquids: The golden chain}",
    eprint = "cond-mat/0612341",
    archivePrefix = "arXiv",
    doi = "10.1103/PhysRevLett.98.160409",
    journal = "Phys. Rev. Lett.",
    volume = "98",
    pages = "160409",
    year = "2007"
}

@article{Aharonov:2005nkd,
    author = "Aharonov, Dorit and Kitaev, Alexei and Preskill, John",
    title = "{Fault-Tolerant Quantum Computation with Long-Range Correlated Noise}",
    eprint = "quant-ph/0510231",
    archivePrefix = "arXiv",
    reportNumber = "CALT-68-2576",
    doi = "10.1103/PhysRevLett.96.050504",
    journal = "Phys. Rev. Lett.",
    volume = "96",
    number = "5",
    pages = "050504",
    year = "2006"
}

@article{Preskill:1999he,
    author = "Preskill, John",
    title = "{Quantum information and physics: some future directions}",
    eprint = "quant-ph/9904022",
    archivePrefix = "arXiv",
    reportNumber = "CALT-68-2219, CALT-68-2219",
    doi = "10.1080/09500340008244031",
    journal = "J. Mod. Opt.",
    volume = "47",
    pages = "127--137",
    year = "2000"
}

@article{PRXQuantum.5.037001,
  title = {Quantum Computing for High-Energy Physics: State of the Art and Challenges},
  author = {Di Meglio, Alberto and Jansen, Karl and Tavernelli, Ivano and Alexandrou, Constantia and Arunachalam, Srinivasan and Bauer, Christian W. and Borras, Kerstin and Carrazza, Stefano and Crippa, Arianna and Croft, Vincent and de Putter, Roland and Delgado, Andrea and Dunjko, Vedran and Egger, Daniel J. and Fern\'andez-Combarro, Elias and Fuchs, Elina and Funcke, Lena and Gonz\'alez-Cuadra, Daniel and Grossi, Michele and Halimeh, Jad C. and Holmes, Zo\"e and K\"uhn, Stefan and Lacroix, Denis and Lewis, Randy and Lucchesi, Donatella and Martinez, Miriam Lucio and Meloni, Federico and Mezzacapo, Antonio and Montangero, Simone and Nagano, Lento and Pascuzzi, Vincent R. and Radescu, Voica and Ortega, Enrique Rico and Roggero, Alessandro and Schuhmacher, Julian and Seixas, Joao and Silvi, Pietro and Spentzouris, Panagiotis and Tacchino, Francesco and Temme, Kristan and Terashi, Koji and Tura, Jordi and T\"uys\"uz, Cenk and Vallecorsa, Sofia and Wiese, Uwe-Jens and Yoo, Shinjae and Zhang, Jinglei},
  journal = {PRX Quantum},
  volume = {5},
  issue = {3},
  pages = {037001},
  numpages = {49},
  year = {2024},
  month = {Aug},
  publisher = {American Physical Society},
  doi = {10.1103/PRXQuantum.5.037001},
  url = {https://link.aps.org/doi/10.1103/PRXQuantum.5.037001}
}

@article{Benini:2025lav,
    author = "Benini, Francesco and Calabrese, Pasquale and Fossati, Michele and Singh, Amartya Harsh and Venuti, Marco",
    title = "{Entanglement Asymmetry for Higher and Noninvertible Symmetries}",
    eprint = "2509.16311",
    archivePrefix = "arXiv",
    primaryClass = "hep-th",
    reportNumber = "SISSA 10/2025/FISI",
    month = "9",
    year = "2025"
}

@article{Witten:1988hf,
    author = "Witten, Edward",
    editor = "Mitra, Asoke N.",
    title = "{Quantum Field Theory and the Jones Polynomial}",
    reportNumber = "IASSNS-HEP-88-33",
    doi = "10.1007/BF01217730",
    journal = "Commun. Math. Phys.",
    volume = "121",
    pages = "351--399",
    year = "1989"
}

@article{Moore:1988qv,
    author = "Moore, Gregory W. and Seiberg, Nathan",
    title = "{Classical and Quantum Conformal Field Theory}",
    reportNumber = "IASSNS-HEP-88-39",
    doi = "10.1007/BF01238857",
    journal = "Commun. Math. Phys.",
    volume = "123",
    pages = "177",
    year = "1989"
}

@article{Dijkgraaf:1989pz,
    author = "Dijkgraaf, Robbert and Witten, Edward",
    title = "{Topological Gauge Theories and Group Cohomology}",
    reportNumber = "THU-89-9, IASSNS-HEP-89-33",
    doi = "10.1007/BF02096988",
    journal = "Commun. Math. Phys.",
    volume = "129",
    pages = "393",
    year = "1990"
}

@article{Fuchs:2009iz,
    author = "Fuchs, Jurgen and Runkel, Ingo and Schweigert, Christoph",
    title = "{Twenty-five years of two-dimensional rational conformal field theory}",
    eprint = "0910.3145",
    archivePrefix = "arXiv",
    primaryClass = "hep-th",
    reportNumber = "ZMP-HH-09-21, HAMBURGER-BEITRAGE-ZUR-MATHEMATIK-349, HAMBURGER-BEITRAEGE-ZUR-MATHEMATIK-NR.-394",
    doi = "10.1063/1.3277118",
    journal = "J. Math. Phys.",
    volume = "51",
    pages = "015210",
    year = "2010"
}

@book{Recknagel:2013uja,
    author = "Recknagel, Andreas and Schomerus, Volker",
    title = "{Boundary Conformal Field Theory and the Worldsheet Approach to D-Branes}",
    doi = "10.1017/CBO9780511806476",
    isbn = "978-0-521-83223-6, 978-0-521-83223-6, 978-1-107-49612-5",
    publisher = "Cambridge University Press",
    series = "Cambridge Monographs on Mathematical Physics",
    month = "11",
    year = "2013"
}

@article{Northe:2024tnm,
    author = "Northe, Christian",
    title = "{Young Researchers School 2024 Maynooth: lectures on CFT, BCFT and DCFT}",
    eprint = "2411.03381",
    archivePrefix = "arXiv",
    primaryClass = "hep-th",
    doi = "10.1088/1751-8121/ada1b4",
    journal = "J. Phys. A",
    volume = "58",
    number = "10",
    pages = "103001",
    year = "2025"
}

@article{Ishibashi:1988kg,
    author = "Ishibashi, Nobuyuki",
    title = "{The Boundary and Crosscap States in Conformal Field Theories}",
    reportNumber = "UT-530-TOKYO",
    doi = "10.1142/S0217732389000320",
    journal = "Mod. Phys. Lett. A",
    volume = "4",
    pages = "251",
    year = "1989"
}

@article{german3,
   title={Equipartition of the entanglement entropy},
   volume={98},
   ISSN={2469-9969},
   url={http://dx.doi.org/10.1103/PhysRevB.98.041106},
   DOI={10.1103/physrevb.98.041106},
   number={4},
   journal={Phys. Rev. B},
   publisher={American Physical Society (APS)},
   author={Xavier, J. C. and Alcaraz, F. C. and Sierra, G.},
   eprint = "1804.06357",
   year={2018}
}

@article{Bonsignori:2020laa,
    author = "Bonsignori, Riccarda and Calabrese, Pasquale",
    title = "{Boundary effects on symmetry resolved entanglement}",
    eprint = "2009.08508",
    archivePrefix = "arXiv",
    primaryClass = "cond-mat.stat-mech",
    doi = "10.1088/1751-8121/abcc3a",
    journal = "J. Phys. A",
    volume = "54",
    number = "1",
    pages = "015005",
    year = "2021"
}

@article{GarciaEtxebarria:2022vzq,
    author = "Garc{\'\i}a Etxebarria, I{\~n}aki",
    title = "{Branes and Non-Invertible Symmetries}",
    eprint = "2208.07508",
    archivePrefix = "arXiv",
    primaryClass = "hep-th",
    doi = "10.1002/prop.202200154",
    journal = "Fortsch. Phys.",
    volume = "70",
    number = "11",
    pages = "2200154",
    year = "2022"
}

@article{Haghighat:2023sax,
    author = "Haghighat, Babak and Sun, Youran",
    title = "{Topological Defect Lines in bosonized Parafermionic CFTs}",
    eprint = "2306.16555",
    archivePrefix = "arXiv",
    primaryClass = "hep-th",
    doi = "10.4310/ATMP.241031012317",
    journal = "Adv. Theor. Math. Phys.",
    volume = "28",
    pages = "1987--2023",
    year = "2024"
}

@article{Sinha:2023hum,
    author = "Sinha, Madhav and Yan, Fei and Grans-Samuelsson, Linnea and Roy, Ananda and Saleur, Hubert",
    title = "{Lattice realizations of topological defects in the critical (1+1)-d three-state Potts model}",
    eprint = "2310.19703",
    archivePrefix = "arXiv",
    primaryClass = "hep-th",
    doi = "10.1007/JHEP07(2024)225",
    journal = "JHEP",
    volume = "07",
    pages = "225",
    year = "2024"
}

@article{Calabrese:2021wvi,
    author = "Calabrese, Pasquale and Dubail, J{\'e}r{\^o}me and Murciano, Sara",
    title = "{Symmetry-resolved entanglement entropy in Wess-Zumino-Witten models}",
    eprint = "2106.15946",
    archivePrefix = "arXiv",
    primaryClass = "hep-th",
    doi = "10.1007/JHEP10(2021)067",
    journal = "JHEP",
    volume = "10",
    pages = "067",
    year = "2021"
}

@article{Hung:2019bnq,
    author = "Hung, Ling Yan and Wong, Gabriel",
    title = "{Entanglement branes and factorization in conformal field theory}",
    eprint = "1912.11201",
    archivePrefix = "arXiv",
    primaryClass = "hep-th",
    doi = "10.1103/PhysRevD.104.026012",
    journal = "Phys. Rev. D",
    volume = "104",
    number = "2",
    pages = "026012",
    year = "2021"
}

@article{Das:2025xyz,
    author ="A. Das and J. Molina-Vilaplana and P. Saura-Bastida",
    title = "Entanglement, twist fields and anomalies for non-invertible symmetries",
    journal ="To appear",
    year = "2025"
}

@article{Okada:2024qmk,
    author = "Okada, Masaki and Tachikawa, Yuji",
    title = "{Noninvertible Symmetries Act Locally by Quantum Operations}",
    eprint = "2403.20062",
    archivePrefix = "arXiv",
    primaryClass = "hep-th",
    doi = "10.1103/PhysRevLett.133.191602",
    journal = "Phys. Rev. Lett.",
    volume = "133",
    number = "19",
    pages = "191602",
    year = "2024"
}

@article{Li:2024gwx,
    author = "Li, Linhao and Huang, Rui-Zhen and Cao, Weiguang",
    title = "{Noninvertible symmetry-enriched quantum critical point}",
    eprint = "2411.19034",
    archivePrefix = "arXiv",
    primaryClass = "cond-mat.str-el",
    doi = "10.1103/mz32-k1zk",
    journal = "Phys. Rev. B",
    volume = "112",
    number = "8",
    pages = "L081113",
    year = "2025"
}

@article{PhysRevX.15.011058,
  title = {Noninvertible Symmetry-Protected Topological Order in a Group-Based Cluster State},
  author = {Fechisin, Christopher and Tantivasadakarn, Nathanan and Albert, Victor V.},
  journal = {Phys. Rev. X},
  volume = {15},
  issue = {1},
  pages = {011058},
  numpages = {41},
  year = {2025},
  month = {Mar},
  publisher = {American Physical Society},
  doi = {10.1103/PhysRevX.15.011058},
  url = {https://link.aps.org/doi/10.1103/PhysRevX.15.011058}
}

@article{Bhardwaj:2024ydc,
    author = "Bhardwaj, Lakshya and Inamura, Kansei and Tiwari, Apoorv",
    title = "{Fermionic non-invertible symmetries in (1+1)d: Gapped and gapless phases, transitions, and symmetry TFTs}",
    eprint = "2405.09754",
    archivePrefix = "arXiv",
    primaryClass = "hep-th",
    doi = "10.21468/SciPostPhys.18.6.194",
    journal = "SciPost Phys.",
    volume = "18",
    number = "6",
    pages = "194",
    year = "2025"
}

@article{Li:2025bgo,
    author = "Li, Yabo and Mitra, Aditi",
    title = "{Non-invertible symmetries out of equilibrium}",
    eprint = "2508.14213",
    archivePrefix = "arXiv",
    primaryClass = "cond-mat.str-el",
    month = "8",
    year = "2025"
}

@article{Choi:2022rfe,
    author = "Choi, Yichul and Lam, Ho Tat and Shao, Shu-Heng",
    title = "{Noninvertible Time-Reversal Symmetry}",
    eprint = "2208.04331",
    archivePrefix = "arXiv",
    primaryClass = "hep-th",
    reportNumber = "YITP-SB-2022-28, MIT-CTP/5457",
    doi = "10.1103/PhysRevLett.130.131602",
    journal = "Phys. Rev. Lett.",
    volume = "130",
    number = "13",
    pages = "131602",
    year = "2023"
}

@article{Choi:2022fgx,
    author = "Choi, Yichul and Lam, Ho Tat and Shao, Shu-Heng",
    title = "{Non-invertible Gauss law and axions}",
    eprint = "2212.04499",
    archivePrefix = "arXiv",
    primaryClass = "hep-th",
    reportNumber = "MIT-CTP/5504, YITP-SB-2022-39",
    doi = "10.1007/JHEP09(2023)067",
    journal = "JHEP",
    volume = "09",
    pages = "067",
    year = "2023"
}

@article{Lin:2023uvm,
    author = "Lin, Ying-Hsuan and Shao, Shu-Heng",
    title = "{Bootstrapping noninvertible symmetries}",
    eprint = "2302.13900",
    archivePrefix = "arXiv",
    primaryClass = "hep-th",
    reportNumber = "YITP-SB-2023-03",
    doi = "10.1103/PhysRevD.107.125025",
    journal = "Phys. Rev. D",
    volume = "107",
    number = "12",
    pages = "125025",
    year = "2023"
}

@article{Choi:2023pdp,
    author = "Choi, Yichul and Forslund, Matthew and Lam, Ho Tat and Shao, Shu-Heng",
    title = "{Quantization of Axion-Gauge Couplings and Noninvertible Higher Symmetries}",
    eprint = "2309.03937",
    archivePrefix = "arXiv",
    primaryClass = "hep-ph",
    reportNumber = "YITP-SB-2023-27, MIT-CTP/5606",
    doi = "10.1103/PhysRevLett.132.121601",
    journal = "Phys. Rev. Lett.",
    volume = "132",
    number = "12",
    pages = "121601",
    year = "2024"
}

@article{Seiberg:2024gek,
    author = "Seiberg, Nathan and Seifnashri, Sahand and Shao, Shu-Heng",
    title = "{Non-invertible symmetries and LSM-type constraints on a tensor product Hilbert space}",
    eprint = "2401.12281",
    archivePrefix = "arXiv",
    primaryClass = "cond-mat.str-el",
    reportNumber = "YITP-SB-2024-01",
    doi = "10.21468/SciPostPhys.16.6.154",
    journal = "SciPost Phys.",
    volume = "16",
    pages = "154",
    year = "2024"
}

@article{Choi:2024rjm,
    author = "Choi, Yichul and Sanghavi, Yaman and Shao, Shu-Heng and Zheng, Yunqin",
    title = "{Non-invertible and higher-form symmetries in 2+1d lattice gauge theories}",
    eprint = "2405.13105",
    archivePrefix = "arXiv",
    primaryClass = "cond-mat.str-el",
    doi = "10.21468/SciPostPhys.18.1.008",
    journal = "SciPost Phys.",
    volume = "18",
    number = "1",
    pages = "008",
    year = "2025"
}

@article{Seifnashri:2025fgd,
    author = "Seifnashri, Sahand and Shao, Shu-Heng and Yang, Xinping",
    title = "{Gauging non-invertible symmetries on the lattice}",
    eprint = "2503.02925",
    archivePrefix = "arXiv",
    primaryClass = "cond-mat.str-el",
    reportNumber = "MIT-CTP/5842, YITP-SB-2025-03",
    doi = "10.21468/SciPostPhys.19.2.063",
    journal = "SciPost Phys.",
    volume = "19",
    number = "2",
    pages = "063",
    year = "2025"
}

@article{Shao:2025mfj,
    author = "Shao, Shu-Heng and Sorce, Jonathan and Srivastava, Manu",
    title = "{Additivity, Haag duality, and non-invertible symmetries}",
    eprint = "2503.20863",
    archivePrefix = "arXiv",
    primaryClass = "hep-th",
    reportNumber = "MIT-CTP/5853, YITP-SB-2025-06",
    doi = "10.1007/JHEP08(2025)009",
    journal = "JHEP",
    volume = "08",
    pages = "009",
    year = "2025"
}

@article{Shao:2025qvf,
    author = "Shao, Shu-Heng and Zhong, Siwei",
    title = "{Where Non-Invertible Symmetries End: Twist Defects for Electromagnetic Duality}",
    eprint = "2509.21279",
    archivePrefix = "arXiv",
    primaryClass = "hep-th",
    reportNumber = "MIT-CTP/5930, YITP-SB-2025-17",
    month = "9",
    year = "2025"
}

@article{Apruzzi:2022rei,
    author = "Apruzzi, Fabio and Bah, Ibrahima and Bonetti, Federico and Schafer-Nameki, Sakura",
    title = "{Noninvertible Symmetries from Holography and Branes}",
    eprint = "2208.07373",
    archivePrefix = "arXiv",
    primaryClass = "hep-th",
    doi = "10.1103/PhysRevLett.130.121601",
    journal = "Phys. Rev. Lett.",
    volume = "130",
    number = "12",
    pages = "121601",
    year = "2023"
}

@article{Bhardwaj:2024kvy,
    author = "Bhardwaj, Lakshya and Bottini, Lea E. and Schafer-Nameki, Sakura and Tiwari, Apoorv",
    title = "{Lattice Models for Phases and Transitions with Non-Invertible Symmetries}",
    eprint = "2405.05964",
    archivePrefix = "arXiv",
    primaryClass = "cond-mat.str-el",
    month = "5",
    year = "2024"
}

@article{Bhardwaj:2023fca,
    author = "Bhardwaj, Lakshya and Bottini, Lea E. and Pajer, Daniel and Schafer-Nameki, Sakura",
    title = "{Categorical Landau Paradigm for Gapped Phases}",
    eprint = "2310.03786",
    archivePrefix = "arXiv",
    primaryClass = "cond-mat.str-el",
    doi = "10.1103/PhysRevLett.133.161601",
    journal = "Phys. Rev. Lett.",
    volume = "133",
    number = "16",
    pages = "161601",
    year = "2024"
}

@article{Bhardwaj:2024wlr,
    author = "Bhardwaj, Lakshya and Bottini, Lea E. and Schafer-Nameki, Sakura and Tiwari, Apoorv",
    title = "{Illustrating the categorical Landau paradigm in lattice models}",
    eprint = "2405.05302",
    archivePrefix = "arXiv",
    primaryClass = "cond-mat.str-el",
    doi = "10.1103/PhysRevB.111.054432",
    journal = "Phys. Rev. B",
    volume = "111",
    number = "5",
    pages = "054432",
    year = "2025"
}

@article{Thorngren:2019iar,
    author = "Thorngren, Ryan and Wang, Yifan",
    title = "{Fusion category symmetry. Part I. Anomaly in-flow and gapped phases}",
    eprint = "1912.02817",
    archivePrefix = "arXiv",
    primaryClass = "hep-th",
    reportNumber = "PUPT-2603",
    doi = "10.1007/JHEP04(2024)132",
    journal = "JHEP",
    volume = "04",
    pages = "132",
    year = "2024"
}

@article{Thorngren:2021yso,
    author = "Thorngren, Ryan and Wang, Yifan",
    title = "{Fusion category symmetry. Part II. Categoriosities at c = 1 and beyond}",
    eprint = "2106.12577",
    archivePrefix = "arXiv",
    primaryClass = "hep-th",
    doi = "10.1007/JHEP07(2024)051",
    journal = "JHEP",
    volume = "07",
    pages = "051",
    year = "2024"
}

@article{KNBalasubramanian:2025vum,
    author = "K. N. Balasubramanian, Mahesh and Buican, Matthew and Delcamp, Clement and Radhakrishnan, Rajath",
    title = "{Gauging Non-Invertible Symmetries in (2+1)d Topological Orders}",
    eprint = "2507.01142",
    archivePrefix = "arXiv",
    primaryClass = "hep-th",
    month = "7",
    year = "2025"
}

@article{Putrov:2024uor,
    author = "Putrov, Pavel and Radhakrishnan, Rajath",
    title = "{Non-anomalous non-invertible symmetries in 1+1D from gapped boundaries of SymTFTs}",
    eprint = "2405.04619",
    archivePrefix = "arXiv",
    primaryClass = "hep-th",
    month = "5",
    year = "2024"
}

@article{Belin:2013uta,
    author = "Belin, Alexandre and Hung, Ling-Yan and Maloney, Alexander and Matsuura, Shunji and Myers, Robert C. and Sierens, Todd",
    title = "{Holographic Charged Renyi Entropies}",
    eprint = "1310.4180",
    archivePrefix = "arXiv",
    primaryClass = "hep-th",
    doi = "10.1007/JHEP12(2013)059",
    journal = "JHEP",
    volume = "12",
    pages = "059",
    year = "2013"
}

@article{Goldstein:2017bua,
    author = "Goldstein, Moshe and Sela, Eran",
    title = "{Symmetry-resolved entanglement in many-body systems}",
    eprint = "1711.09418",
    archivePrefix = "arXiv",
    primaryClass = "cond-mat.stat-mech",
    doi = "10.1103/PhysRevLett.120.200602",
    journal = "Phys. Rev. Lett.",
    volume = "120",
    number = "20",
    pages = "200602",
    year = "2018"
}

@article{Zhao:2020qmn,
    author = "Zhao, Suting and Northe, Christian and Meyer, Ren\'e",
    title = "{Symmetry-resolved entanglement in AdS$_{3}$/CFT$_{2}$ coupled to U(1) Chern-Simons theory}",
    eprint = "2012.11274",
    archivePrefix = "arXiv",
    primaryClass = "hep-th",
    doi = "10.1007/JHEP07(2021)030",
    journal = "JHEP",
    volume = "07",
    pages = "030",
    year = "2021"
}

@article{Bonsignori:2019naz,
    author = "Bonsignori, Riccarda and Ruggiero, Paola and Calabrese, Pasquale",
    title = "{Symmetry resolved entanglement in free fermionic systems}",
    eprint = "1907.02084",
    archivePrefix = "arXiv",
    primaryClass = "cond-mat.stat-mech",
    doi = "10.1088/1751-8121/ab4b77",
    journal = "J. Phys. A",
    volume = "52",
    number = "47",
    pages = "475302",
    year = "2019"
}

@article{Calabrese:2009qy,
    author = "Calabrese, Pasquale and Cardy, John",
    title = "{Entanglement entropy and conformal field theory}",
    eprint = "0905.4013",
    archivePrefix = "arXiv",
    primaryClass = "cond-mat.stat-mech",
    doi = "10.1088/1751-8113/42/50/504005",
    journal = "J. Phys. A",
    volume = "42",
    pages = "504005",
    year = "2009"
}

@article{Parez:2021pgq,
    author = "Parez, Gilles and Bonsignori, Riccarda and Calabrese, Pasquale",
    title = "{Exact quench dynamics of symmetry resolved entanglement in a free fermion chain}",
    eprint = "2106.13115",
    archivePrefix = "arXiv",
    primaryClass = "cond-mat.stat-mech",
    doi = "10.1088/1742-5468/ac21d7",
    journal = "J. Stat. Mech.",
    volume = "2109",
    pages = "093102",
    year = "2021",
    note = "[Erratum: J.Stat.Mech. 2212, 129901 (2022)]"
}

@article{Sierra_2024,
      author = "Saura-Bastida, P. and Das, A. and Sierra, G. and Molina-Vilaplana, J.",
    title = "{Categorical-symmetry resolved entanglement in conformal field theory}",
    eprint = "2402.06322",
    archivePrefix = "arXiv",
    primaryClass = "hep-th",
    doi = "10.1103/PhysRevD.109.105026",
    journal = "Phys. Rev. D",
    volume = "109",
    number = "10",
    pages = "105026",
    year = "2024"
}

@article{ourPartI,
    author = "Capizzi, Luca and Castro-Alvaredo, Olalla A. and De Fazio, Cecilia and Mazzoni, Michele and Santamar\'\i{}a-Sanz, Luc\'\i{}a",
    title = "{Symmetry resolved entanglement of excited states in quantum field theory. Part I. Free theories, twist fields and qubits}",
    eprint = "2203.12556",
    archivePrefix = "arXiv",
    primaryClass = "hep-th",
    doi = "10.1007/JHEP12(2022)127",
    journal = "JHEP",
    volume = "12",
    pages = "127",
    year = "2022"
}

@article{Horvath:2021rjd,
    author = "Horvath, David X. and Calabrese, Pasquale and Castro-Alvaredo, Olalla A.",
    title = "{Branch Point Twist Field Form Factors in the sine-Gordon Model II: Composite Twist Fields and Symmetry Resolved Entanglement}",
    eprint = "2105.13982",
    archivePrefix = "arXiv",
    primaryClass = "hep-th",
    doi = "10.21468/SciPostPhys.12.3.088",
    journal = "SciPost Phys.",
    volume = "12",
    number = "3",
    pages = "088",
    year = "2022"
}

@article{FG,
    author = "Feldman, Noa and Goldstein, Moshe",
    title = "{Dynamics of Charge-Resolved Entanglement after a Local Quench}",
    eprint = "1905.10749",
    archivePrefix = "arXiv",
    primaryClass = "cond-mat.stat-mech",
    doi = "10.1103/PhysRevB.100.235146",
    journal = "Phys. Rev. B",
    volume = "100",
    number = "23",
    pages = "235146",
    year = "2019"
}

@article{Ares:2022gjb,
    author = "Ares, Filiberto and Calabrese, Pasquale and Di Giulio, Giuseppe and Murciano, Sara",
    title = "{Multi-charged moments of two intervals in conformal field theory}",
    eprint = "2206.01534",
    archivePrefix = "arXiv",
    primaryClass = "hep-th",
    doi = "10.1007/JHEP09(2022)051",
    journal = "JHEP",
    volume = "09",
    pages = "051",
    year = "2022"
}

@article{Zhao:2022wnp,
    author = "Zhao, Suting and Northe, Christian and Weisenberger, Konstantin and Meyer, Ren\'e",
    title = "{Charged moments in W$_{3}$ higher spin holography}",
    eprint = "2202.11111",
    archivePrefix = "arXiv",
    primaryClass = "hep-th",
    doi = "10.1007/JHEP05(2022)166",
    journal = "JHEP",
    volume = "05",
    pages = "166",
    year = "2022"
}

@article{Murciano:2019wdl,
    author = "Murciano, Sara and Di Giulio, Giuseppe and Calabrese, Pasquale",
    title = "{Symmetry resolved entanglement in gapped integrable systems: a corner transfer matrix approach}",
    eprint = "1911.09588",
    archivePrefix = "arXiv",
    primaryClass = "cond-mat.stat-mech",
    doi = "10.21468/SciPostPhys.8.3.046",
    journal = "SciPost Phys.",
    volume = "8",
    pages = "046",
    year = "2020"
}

@article{Gaur:2023yru,
    author = "Gaur, Himanshu and Yajnik, Urjit A.",
    title = "{Multi-charged moments and symmetry-resolved R\'enyi entropy of free compact boson for multiple disjoint intervals}",
    eprint = "2310.14186",
    archivePrefix = "arXiv",
    primaryClass = "hep-th",
    doi = "10.1007/JHEP01(2024)042",
    journal = "JHEP",
    volume = "01",
    pages = "042",
    year = "2024"
}

@article{Weisenberger:2021eby,
    author = "Weisenberger, Konstantin and Zhao, Suting and Northe, Christian and Meyer, Ren\'e",
    title = "{Symmetry-resolved entanglement for excited states and two entangling intervals in AdS$_{3}$/CFT$_{2}$}",
    eprint = "2108.09210",
    archivePrefix = "arXiv",
    primaryClass = "hep-th",
    doi = "10.1007/JHEP12(2021)104",
    journal = "JHEP",
    volume = "12",
    pages = "104",
    year = "2021"
}

@article{Murciano:2020lqq,
    author = "Murciano, Sara and Ruggiero, Paola and Calabrese, Pasquale",
    title = "{Symmetry resolved entanglement in two-dimensional systems via dimensional reduction}",
    eprint = "2003.11453",
    archivePrefix = "arXiv",
    primaryClass = "cond-mat.stat-mech",
    doi = "10.1088/1742-5468/aba1e5",
    journal = "J. Stat. Mech.",
    volume = "2008",
    pages = "083102",
    year = "2020"
}

@article{exp,
  title = "{Experimental Realization of Dicke States of up to Six Qubits for Multiparty Quantum Networking}",
  author = {Prevedel, R. and Cronenberg, G. and Tame, M. S. and Paternostro, M. and Walther, P. and Kim, M. S. and Zeilinger, A.},
  journal = {Phys. Rev. Lett.},
  volume = {103},
  issue = {2},
  pages = {020503},
  numpages = {4},
  year = {2009},
  doi = {10.1103/PhysRevLett.103.020503},
  url = {http://link.aps.org/doi/10.1103/PhysRevLett.103.020503},
  publisher = {American Physical Society}
}

@Article{Cardy,
     author    = "Cardy, J.L.",
     title     = "The central charge and universal combinations of amplitudes in
     two-dimensional theories away from criticality",
     journal   = "Phys. Rev. Lett.",
     volume    = "60",
     year      = "1988",
     pages     = "2709-2711",
     SLACcitation  = "%%CITATION = PRLTA,60,2709;%%"
}

@Article{Z,
     author    = "Zamolodchikov, A. B.",
     title     = "{Two point correlation function in scaling Lee-Yang model}",
     journal   = "Nucl. Phys.",
     volume    = "B348",
     year      = "1991",
     pages     = "619-641",
     SLACcitation  = "%%CITATION = NUPHA,B348,619;%%"
}

@Article{B,
     author    = "Bytsko, A. G.",
     title     = "{On integrable Hamiltonians for higher spin XXZ chain}",
     journal   = "J. Math. Phys.",
     volume    = "44",
     year      = "2003",
     pages     = "3698-3717",
     eprint    = "hep-th/0112163",
     SLACcitation  = "%%CITATION = HEP-TH 0112163;%%"
}

@Article{K,
     author    = "Kitanine, N.",
     title     = "{Correlation functions of the higher spin XXX chains}",
     journal   = "J. Phys.",
     volume    = "A34",
     year      = "2001",
     pages     = "8151",
     eprint    = "math-ph/0104016",
     SLACcitation  = "%%CITATION = MATH-PH 0104016;%%"
}

@article{Fossati:2023zyz,
    author = "Fossati, Michele and Ares, Filiberto and Calabrese, Pasquale",
    title = "{Symmetry-resolved entanglement in critical non-Hermitian systems}",
    eprint = "2303.05232",
    archivePrefix = "arXiv",
    primaryClass = "cond-mat.stat-mech",
    doi = "10.1103/PhysRevB.107.205153",
    journal = "Phys. Rev. B",
    volume = "107",
    number = "20",
    pages = "205153",
    year = "2023"
}

@article{Gaur:2024vdh,
    author = "Gaur, Himanshu",
    title = "{Total and Symmetry resolved Entanglement spectra in some Fermionic CFTs from the BCFT approach}",
    eprint = "2402.07557",
    archivePrefix = "arXiv",
    primaryClass = "hep-th",
    month = "2",
    year = "2024"
}

@article{Northe:2025qcv,
    author = "Northe, Christian",
    title = "{Fermion Parity Resolution of Entanglement}",
    eprint = "2509.03605",
    archivePrefix = "arXiv",
    primaryClass = "hep-th",
    month = "9",
    year = "2025"
}

@article{Fukusumi:2024cnl,
    author = "Fukusumi, Yoshiki",
    title = "{Fusion rule in conformal field theories and topological orders: A unified view of correspondence and (fractional) supersymmetry and their relation to topological holography}",
    eprint = "2405.05178",
    archivePrefix = "arXiv",
    primaryClass = "hep-th",
    month = "5",
    year = "2024"
}

@article{Fukusumi:2025xrj,
    author = "Fukusumi, Yoshiki and Kawamoto, Taishi",
    title = "{Generalizing quantum dimensions: Symmetry-based classification of local pseudo-Hermitian systems and the corresponding domain walls}",
    eprint = "2511.11059",
    archivePrefix = "arXiv",
    primaryClass = "hep-th",
    reportNumber = "YITP-25-174",
    month = "11",
    year = "2025"
}

@ARTICLE{2004math......1246C,
       author = {{Calaque}, Damien and {Etingof}, Pavel},
        title = "{Lectures on tensor categories}",
      journal = {arXiv Mathematics e-prints},
         year = 2004,
        month = jan,
          eid = {math/0401246},
        pages = {math/0401246},
          doi = {10.48550/arXiv.math/0401246},
archivePrefix = {arXiv},
       eprint = {math/0401246},
 primaryClass = {math.QA},
       adsurl = {https://ui.adsabs.harvard.edu/abs/2004math......1246C}
}

@article{Tu:2021xje,
    author = "Tu, Yi-Ting and Tzeng, Yu-Chin and Chang, Po-Yao",
    title = "{R{\'e}nyi entropies and negative central charges in non-Hermitian quantum systems}",
    eprint = "2107.13006",
    archivePrefix = "arXiv",
    primaryClass = "cond-mat.str-el",
    doi = "10.21468/SciPostPhys.12.6.194",
    journal = "SciPost Phys.",
    volume = "12",
    number = "6",
    pages = "194",
    year = "2022"
}
\bibliographystyle{jhep}

\end{document}